\documentclass[11pt]{amsart}
\usepackage[utf8]{inputenc}
\usepackage[left=1in,top=1in,right=1in,bottom=1in,head=.1in,nofoot]{geometry}

\usepackage{setspace,url,bm,amsmath} 

\usepackage{titlesec} 
\titlelabel{\thetitle.\quad} 

\usepackage{bbm}
\usepackage{latexsym}
\usepackage[round]{natbib}
\usepackage{graphicx}
\usepackage{amsmath,amssymb,amsthm, wasysym}
\usepackage{bm}
\usepackage{rotating}
\usepackage{multicol}
\usepackage{dsfont}
\usepackage{titlesec}
\usepackage{latexsym}
\usepackage{changepage}
\usepackage{multirow}
\usepackage{booktabs}
\usepackage{pbox}
\usepackage{mathrsfs}
\usepackage[title,toc, header]{appendix}
\appendixheaderoff
\usepackage{float}
\usepackage[linktoc=page,colorlinks,linkcolor=blue,citecolor=blue,urlcolor=blue]{hyperref}
\usepackage{fancyhdr}
\usepackage{setspace}
\usepackage{enumitem}
\usepackage[normalem]{ulem}
\usepackage{amsmath}
\usepackage{empheq}
\usepackage{caption}
\usepackage[margin=20pt]{subcaption}
\usepackage{textcomp}

\usepackage{xr}
\makeatletter
\newcommand*{\addFileDependency}[1]{
  \typeout{(#1)}
  \@addtofilelist{#1}
  \IfFileExists{#1}{}{\typeout{No file #1.}}
}
\makeatother
 
\newcommand*{\myexternaldocument}[1]{%
\externaldocument{#1}%
\addFileDependency{#1.tex}%
\addFileDependency{#1.aux}%
}

\myexternaldocument{RDDeAppendix}

\usepackage[table]{xcolor}

\newcommand{\GG}[1]{}

\usepackage{amsthm}
\usepackage{comment}

\theoremstyle{definition}
\newtheorem{assumption}{Assumption}
\newtheorem*{theorem*}{Theorem}

\newtheorem*{corollary*}{Corollary}

\bibpunct{(}{)}{;}{a}{}{,} 

\usepackage{etoolbox} 
\apptocmd{\sloppy}{\hbadness 10000\relax}{}{} 

\usepackage{color}
\usepackage{listings}

 \newcommand{\one}{\mathbf{1}}

\newcommand{\bX}{\mathbf{X}}
\newcommand{\bY}{\mathbf{Y}}

\newcommand{\bZ}{\mathbf{Z}}

\newcommand{\bS}{\mathbf{S}}
\newcommand{\bQ}{\mathbf{Q}}

\newcommand{\Yobs}{Y^{obs}}

\newcommand{\bG}{\bm{G}}

\newcommand{\Uset}{\mathcal{U}_{s_0}}
\newcommand{\Usetp}{\mathcal{U}_{s_0}^{+}}
\newcommand{\Usetm}{\mathcal{U}_{s_0}^{-}}

\newcommand{\balpha}{\bm{\alpha}}
\newcommand{\boeta}{\bm{\eta}}
\newcommand{\bbeta}{\bm{\beta}}
\newcommand{\bgamma}{\bm{\gamma}}
\newcommand{\bmu}{\bm{\mu}}
\newcommand{\btheta}{\bm{\theta}}

\begin{document}
\doublespacing

\title[Selecting Subpopulations in RD Designs]{\bf Selecting Subpopulations for Causal Inference in	\\[0.1in] Regression Discontinuity Designs } 

    \author[Forastiere et al.]{Laura Forastiere}  
    \address{Yale University}
\email{laura.forastiere@yale.edu}

\author[ ]{Alessandra Mattei} 
    \address{University of Florence}
\email{alessandra.mattei@unifi.it}

\author[]{Julia Pescarini} 
    \address{London School of Hygiene and Tropical Medicine}
\email{julia.pescarini1@lshtm.ac.uk}

 \author[ ]{Mauricio  Barreto } 
    \address{Universidade federal da Bahia}
\email{mauricio@ufba.fr}

  \author[]{Fabrizia Mealli}
      \address{University of Florence and European University Institute}
\email{fabrizia.mealli@eui.eu}

\begin{abstract}
The Brazil Bolsa Fam\'ilia program is a conditional cash transfer program aimed to reduce short-term poverty by direct cash transfers and to fight long-term poverty by increasing human capital among poor Brazilian people. Eligibility for Bolsa Fam\'ilia benefits depends on a cutoff rule, which classifies it as a regression discontinuity (RD) design.
Following \cite{LiMatteiMealli2015}  and \cite{BransonMealli2019}, we formally describe the Bolsa Fam\'ilia RD design as a local regular design \citep{ImbensRubin2015} within the potential outcome approach. Under this framework, causal effects can be identified and estimated on an unknown but well-defined subpopulation where the following RD assumptions hold: a local overlap assumption, a local SUTVA, and a local ignorability (unconfoundedness) assumption. 
We first discuss the potential advantages of this probabilistic framework
in settings where such assumptions are deemed plausible, over local regression methods based on continuity assumptions. The potential advantages concern the causal estimands that can be targeted, the design and the analysis, as well as the interpretation and generalizability of the results.
A critical issue of the probabilistic approach is how to identify the subpopulation for which we can draw valid causal inference. We propose to use a Bayesian model-based finite mixture approach to clustering to probabilistically classify observations into subpopulations where the RD assumptions hold and do not hold on the basis of the observed data. This approach a) allows to account for the uncertainty in the subpopulation membership, which is typically neglected; b) does not impose any constraint on the shape of the subpopulation; 
c) allows to target alternative causal estimands than the average treatment effects (ATEs); and d)  is robust to a certain degree of manipulation/selection of the forcing variable.
We apply our proposed approach to assess causal effects of the Bolsa Fam\'ilia program on leprosy incidence in 2009 for Brazilian households who registered in the Brazilian National Registry for Social Programs in 2007-2008 for the first time. We find evidence that being eligible for the program reduces the risk of leprosy.
\end{abstract}


\maketitle

\section{Introduction} 

Many treatments and interventions in medicine, public health, and social policy follow an assignment rule that agrees with a regression discontinuity (RD) design, that is, the treatment assignment is determined, at least partly, by the realized value of a variable, usually called the forcing or running variable, falling below or above a prefixed threshold or cutoff point.

In this paper, we aim to assess the causal effect of the Brazilian cash transfer program Bolsa Fam\'ilia (BF), on health, particularly on leprosy incidence.
The program eligibility can be seen as an RD design, as the
Brazilian government relied on per capita household income to select eligible families and authorize cash transfers. 
RD analyses have already been used to evaluate the BF program on a number of outcomes, e.g., fertility, schooling, labor choices \citep{superti2020effects, nilsson2013evaluation, Dourado2017, barbosa2014conditional}.

We adopt a probabilistic perspective to the design and analysis of RD designs by viewing the forcing variable  
as a random variable with a probability distribution. We formalize this perspective and describe the RD design as a regular design following  \cite{LiMatteiMealli2015} and \cite{MatteiMealli2016}, and the recent extensions proposed by \cite{BransonMealli2019}.
Differences between the proposed methodology and the existing probabilistic (local randomization) approaches are mainly on the set of assumptions used to formally describe the RD design as a regular design, the assumptions on the shape of the unknown subpopulation of units where these assumptions hold, the approach used to select this unknown subpopulation, and the estimation methods.

The core of our approach is to assume that there exists a subpopulation of units with realized values of the forcing variable above and below the threshold, where a local overlap assumption and a local Stable Unit Treatment Value Assumption (SUTVA) hold and where the forcing variable, and hence treatment assignment, is unconfounded conditional on covariates (local unconfoundedness). Throughout the paper, we refer to this set of assumptions as RD assumptions.
In the subpopulation where these assumptions hold, causal effects of interest can be identified and estimated as in a regular design.
These assumptions have two implications in the distribution of the observed data in such subpopulation: 1) conditional on the treatment status and covariates, the outcome is independent of the forcing variable (local SUTVA and local unconfoudedness), and 2) for all values of covariates, we could potentially find both treated and untreated units (local overlap).

Relying on these implications for the observed data, we use a Bayesian model-based mixture approach to probabilistic clustering to identify the subpopulation where the RD assumptions hold and to estimate causal effects for this subpopulation.
Under this approach, each unit contributes to the estimation of such effects according to their posterior probability of belonging to the subpopulation where the RD assumptions hold and the effects can be identified.
In this way, our approach relies for identification and estimation on a subpopulation that may include units with very different values of the forcing variable and where the weighting function is not imposed to be monotone and symmetric, but depends on the empirical evidence. This is different from common methods used in RD designs.
This approach also allows us to derive the posterior distribution of causal effects by
marginalizing over the uncertainty on the membership for each observation.

We use the 
proposed Bayesian approach to assess causal effects of the Bolsa Fam\'ilia program on leprosy incidence
\footnote{Brazil continues to be the second leading country in leprosy cases worldwide, with nearly 30 thousand cases being diagnosed each year. See 
\cite{mondiale2018global}.}. 
Leprosy is a rare event with only $412$ cases over $147\,399$ families in our sample, making the RD analysis particularly challenging.
We find evidence that in the Bolsa fam\'ilia study, there exists a relatively large subpopulation where the RD assumptions hold,
and in this subpopulation, there are beneficial (although not strong) effects of being eligible for the Bolsa Fam\'ilia program on the risk of leprosy.

The rest of the paper is organized as follows. In Section~\ref{sec:background}, we review recent literature on RD designs and compare our proposed framework to existing ones in terms of target causal estimands, structural and identifying assumptions, and inferential methods. In Section~\ref{sec:BF}, we provide some background on the Brazilian Bolsa Fam\'ilia Program, and in Section~\ref{sec:dataset}, we describe the dataset used here.	In Section~\ref{sec:Setup} we formally describe the Brazil's Bolsa Fam\'ilia RD design as a local unconfouded experiment by introducing the notation, the RD assumptions, and the causal estimand of interest. We also describe and discuss the RD assumptions embedding them in the literature on the probabilistic (local randomization) approach to RD designs. In Section~\ref{sec:Uset}, we first briefly review the existing approaches to the selection of suitable subpopulations for causal inference in RD designs under the local randomization framework. Then we describe our Bayesian model-based finite mixture approach and provide details on how we implement it in the Bolsa Fam\'ilia  study. In Section~\ref{sec:BFanalysis}, we present the results of the real data analysis. We conclude in Section~\ref{sec:discussion}  with some discussion.

\section{Background and contributions} \label{sec:background}
In the last two decades,  RD designs, originally introduced by \cite{ThistlethwaiteCampbell1960}, have received increasing attention in the causal inference literature from both applied and theoretical perspectives. 
Most of this literature frames RD designs in the context of the potential outcome approach to causal inference \citep{Rubin1974, Rubin1978, ImbensRubin2015}, and we also adopt this approach here. 	See \cite{LeeLemieux2010, ImbensLemieux2008} for general surveys. See also \cite{CattaneoEtAl_book2020a, CattaneoEtAl_book2020b} for a textbook discussion, and  \cite{AtheyImbens2017, CattaneoTitiunikVazquezBare:2020}, the edited volume by \cite{CattaneoEscanciano2017} and the reprint in Observational Studies \citep{ObsStudies2016} of the original paper by \cite{ThistlethwaiteCampbell1960}  with comments for more recent reviews, developments and discussions.

There is a general agreement on describing RD designs as quasi-experimental designs. The difference across the alternative existing approaches to the design and analysis of RD designs lies in the definition of the nature of the forcing variable.

Traditionally, the forcing variable is viewed as a pre-treatment covariate, and RD designs are described as irregular designs with a known but non-probabilistic assignment mechanism: the treatment assignment deterministically depends on the realized value of the forcing variable being below or above the cutoff.
In this classical perspective,
the only stochastic element of an RD design is the repeated sampling of units, and causal estimands are inherently defined for a hypothetical (almost) infinite population. Focus usually is on the treatment effect at the threshold, 
which can be identified under smoothness assumptions on the relationship between the outcome and the forcing variable, such as the continuity of conditional regression functions of the outcome given the forcing variable \citep{HTV:2001}. 
Such smoothness assumptions imply randomization of the treatment at the single threshold value \citep{BattistinRettore2008}, although randomization is not explicitly used for identification. 
The most popular methodology for drawing inference in RD designs relies on local-polynomial (non-)parametric regression methods 
within a smoothing window around the threshold
and their asymptotic proprieties
\citep{LeeLemieux2010, ImbensLemieux2008,  CattaneoEtAl_book2020a, CattaneoEtAl_book2020b, CattaneoTitiunikVazquezBare:2020}. See also \cite{CalonicoCattaneoTitiunik2014, CalonicoCattaneoTitiunik2015, CalonicoEtAl:2019} for recent developments.
An important issue arising in the use of local regression methods is the choice of the bandwidth defining this smoothing window, which determines the observations contributing to the estimation of the causal effects.
Recently theoretical developments suggest to choose the bandwidth using data-driven methods \citep{ CalonicoCattaneoFarrell2018} or Mean Square Error (MSE)-optimal criteria \citep{CalonicoCattaneoTitiunik2014,  ImbensKalyanaraman2012}.
 Three aspects of this approach are worth noting. First, because smoothing assumptions only allow the identification of causal effects at the threshold, the bandwidth does not define the subpopulation of interest for the definition of the causal estimands, but only an ``auxiliary''  subset of units	from which extrapolating information to the cutoff. 
 Second, the bandwidth defines a symmetrical subpopulation whose realized value of the forcing variable is within a distance equal to the bandwidth on either side of the discontinuity point
 \footnote{
 Asymmetrical subpopulation is also possible using two different bandwidths.
 }.
 Lastly, the uncertainty involved in this data-driven 
 choice is never incorporated in the standard errors for the estimates of interest.

A recent strand of the literature views the forcing variable as a random variable with a probability distribution rather than as a fixed covariate. 
This introduces stochasticity in the assignment mechanism in RD designs.
 Under this new probabilistic perspective, focus generally is on a subpopulation where the  RD design can be viewed as a regular design with a probabilistic (overlap) and unconfounded assignment mechanism.
This perspective enables us to overcome the longtime interpretation of RD designs as an extreme violation of the positivity or overlap assumption: because the forcing variable is seen as stochastic, we can assume that some units whose realized value of the forcing variable is observed on one side of the threshold could have been seen on the opposite side, and, thus, their probability of being treated or not treated is neither 0 nor 1, leading to a \emph{local} overlap assumption.
Notable contributions in this line of work include  \cite{CattaneoFrandsenTitiunik2015, LiMatteiMealli2015, KeeleTitiunikZubizarreta2015, MatteiMealli2016, BransonMealli2019,	SalesHansen2020}, who propose alternative ways to formalize the probabilistic assignment mechanism through local randomization assumptions.

Local probabilistic methods present distinguishing features over local regression methods:
they allow the estimation of treatment effects for all members of a subpopulation 
rather than for those at the cutoff only, making the results easier to generalize; they avoid the need for modeling assumptions on the relationship between the running variable and the outcome, and instead, place assumptions on the assignment mechanism for units near the cutoff; they allow the treatment assignment mechanism to be random rather than deterministic, so that finite population inference can be used;  they allow to easily deal with discrete running variables.

In this paper, we also adopt this probabilistic perspective,  assuming that there exists a subpopulation of units 
where the RD assumptions hold, that is, where a local overlap assumption and a local Stable Unit Treatment Value Assumption (SUTVA) hold, and where the forcing variable, and hence treatment assignment, can be seen as randomly assigned, possibly conditional on covariates.
The idea to formalize an RD design
as a design where the forcing variable is unconfounded given the covariates is also exploited in \cite{AngristRokkanen2015}.  The key difference between the framework proposed by \cite{AngristRokkanen2015}
and our local framework is about the population for which
we can draw valid causal inference: \cite{AngristRokkanen2015}  make a stronger version of unconfoundedness for the whole population, while we assume that unconfoundedness only holds for a latent subpopulation of units.
Further detailed comparisons to this and other papers are provided throughout the paper.
In a way similar to \cite{AngristRokkanen2015}, we rely on the implication of the unconfoudedness assumption that conditional on covariates and treatment status, the observed value of the forcing variable is independent of the observed outcome. While \cite{AngristRokkanen2015} rely on this implication to assess unconfoundedness and estimate effects in the whole population, we rely on it to identify our subpopulation where causal effects can be identified.

Since our approach only requires the unconfoundedness assumption to hold in a subpopulation, it is robust to non-ignorable complications, such as  strategic manipulation where units manipulate their value of the forcing variable in order to receive the treatment and this behavior is related to potential outcomes. Observations affected by non-ignorable manipulation are excluded, while those affected by ignorable manipulation, where the observed value of forcing variable only depends on observed characteristics, may be included in the target subpopulation. This selection cannot be done in a classical perspective that relies on smoothness assumptions in that the manipulation of the forcing variable by some units would violate such assumptions.

In practice, the subpopulation where the RD assumptions hold is usually unknown. 
In this paper, we deal with this issue by viewing the selection of a suitable subpopulation as an unsupervised learning problem.  We propose a Bayesian model-based finite mixture approach to clustering to classify observations into subpopulations where the RD assumptions hold and do not hold, and this is done on the basis of whether the implications of such assumptions are compatible with the observed data. 
In our framework, clustering of the units is probabilistic, 
in the sense that membership of each unit in a subpopulation is given as a probability.
Specifically, we derive the posterior distribution of the membership probabilities for each unit and
use the probability of belonging to the subpopulation where the RD assumptions hold to weight the units so that units that have a higher chance of belonging to that subpopulation contribute more to the estimation of the causal effect. Observations for whom the RD assumptions may fail are assigned smaller weights and, thus, contribute less to the inference. 
The posterior membership probability, and thus the unit's contribution to the estimation of causal effects, depends on the  
empirical evidence that units with similar characteristics exists on the other side of the threshold and that within strata of similar units under the same treatment status the forcing variable is independent of the outcome.

By weighting the observations according to the empirical evidence of meeting the RD assumptions, in our approach the subpopulation used to identify and estimate the causal effects may include observations with a realized value of
the forcing variable with any distance from the threshold, and the weighting function is not constrained to be monotone and symmetric. 
On the contrary, the existing probabilistic 
approaches focus, for convenience, on subpopulations defined by possibly asymmetric intervals around the threshold \citep{CattaneoFrandsenTitiunik2015, LiMatteiMealli2015, MatteiMealli2016, BransonMealli2019}. 
Two exceptions are \cite{KeeleTitiunikZubizarreta2015} and \cite{ricciardi2022dirichlet}. These methods do not formally require any assumption on the shape of the subpopulations, but causal inference is conducted without directly accounting for the uncertainty about a selected subpopulation.
Another exception is that of \cite{SalesHansen2020}, who consider a ``regression discontinuity donut'' specification \citep{Barreca2011SavingBR}. Our method can be indeed construed as a smooth, stochastic version of the ``donut regression discontinuity'' method, in the sense that it weights each observation depending on the posterior probability for that observation of belonging to the population where the RD assumptions hold.


Our approach closely follows \cite{LiMatteiMealli2015}, \cite{MatteiMealli2016}, and \cite{BransonMealli2019}. However, there are important differences: $(i)$ similarly to \cite{BransonMealli2019}, we make a weaker local unconfoundedness assumption that the forcing variable and the potential outcomes are conditionally independent given the covariates, rather than unconditionally\footnote{\cite{BransonMealli2019} assume that the treatment indicator rather than the forcing variable is independent of the potential outcomes given the covariates.};  $(ii)$ we do not assume that the subpopulation is defined by values of the forcing
variable falling in a symmetric interval around the threshold;
and iii) we replace Bayesian falsification tests to select the subpopulation where RD assumptions hold with a Bayesian clustering approach, which jointly identifies the subpopulation of interest and estimates causal effects in this subpopulation, accounting for its uncertainty.

Moreover, our approach presents important differences with the methodological framework developed by 
\cite{CattaneoFrandsenTitiunik2015, SalesHansen2020, EcklesEtAl2020}. \cite{CattaneoFrandsenTitiunik2015}  assume local randomization (not local unconfoundedness) and do not distinguish between random assignment and SUTVA. They then use randomization inference both to select the subpopulation where local randomization holds and as the mode of inference.
\cite{SalesHansen2020} introduce a ``residual ignorability'' assumption under which distribution-free analyses of randomized experiments are combined
with classical, wholly parametric methods for RD designs and RD specification tests. \cite{EcklesEtAl2020} see the forcing variable as a noisy measure of a latent variable, and the noise is assumed to be exogenous. This exogeneity assumption requires that the observed forcing variable and the potential outcomes are independent conditionally on the latent true forcing variable.
Under this latent unconfoundedness assumption,
average causal effects at the threshold can be identified and estimated. The exogeneity assumption introduced by \cite{EcklesEtAl2020} is a different assumption than our local unconfoundedness, and, compared to the latter, it
can be more or less plausible in real settings and has different inferential implications, allowing the identification of different causal estimands.

To summarize, our Bayesian model-based mixture approach to clustering has important advantages.
First, it does not impose any constraint on the shape of the subpopulation but allows the subpopulation to include observations with a realized value of the forcing variable with any distance from the threshold, as long as the RD assumptions are met.
Second, it allows us to account for the uncertainty about the subpopulation membership, which is typically neglected. 
Third, it is robust with respect to strategic (non-ignorable) manipulations that may lead to the failure of the unconfoundedness assumption, in the sense that we probabilistically ``pull out''  observations for whom the forcing variable may be confounded by assigning them smaller weights.
Fourth, our approach can be used as a design phase before the application of any type of analysis for any causal estimand. 
Fifth, for the selection of the subpopulation where the RD assumptions hold, our approach avoids relying on falsification tests  \citep{CattaneoFrandsenTitiunik2015, LiMatteiMealli2015, MatteiMealli2016, LicariMattei2020, BransonMealli2019},
which are highly dependent on the sample size and are not straightforward when local randomization is replaced by local unconfoudedness, as we do here \citep{BransonMealli2019}. 
Finally, our approach allows dealing with different estimands from the average treatment effect, like e.g., relative risks for relatively rare outcomes,  where standard estimators might not work well. Drawbacks of our approach are discussed throughout the paper.

	\section{The Brazilian Bolsa Fam\'ilia Program}
\label{sec:BF}

The Bolsa Fam\'ilia Program (BF) is a social welfare program of the Brazilian government that started in 2003 and is still ongoing.
The program has reached around 13 million families, more than 50 million people, a major portion of the country's low-income population.
Its primary objectives are to reduce short-term poverty through direct cash transfers and
to fight long-term poverty by increasing human capital among poor Brazilian people.
The Bolsa Fam\'ilia program is a conditional cash transfer program, that is, benefits are paid over time to beneficiaries only conditional on their investments in health and education. 
%
Eligibility to the Bolsa Fam\'ilia program is conditional on being registered in the Brazilian National Registry for Social Programs (Cadastro Unico or CadUnico) and having a monthly per capita below the poverty line.
In 2008, this threshold was 120 Brazilian reals (BRL)
\footnote{Note that $1$ BRL $= 0.4321$ USD in 12/31/2008. See https://it.investing.com/currencies/usd-brl-historical-data}. 
Beneficiaries can then receive 
variable benefits that depend on the number of children and their ages
\citep{Brazil2008, Brazil2014, WWP2014}.
In addition to these variable benefits, 
those with per capita income below a lower threshold (60 BRL in 2008)  and categorized as living in `extreme poverty' were eligible for a basic grant
(see Section~A.1 in the e-Appendix for further details).

Because of the eligibility criterion based on household per capita income, we adopt the RD design framework. As the RD threshold, we use the higher threshold of 120 BRL, which distinguishes between ineligible and potentially eligible families. Those whose monthly per capita income was lower than 120 BRL were potentially eligible for BF benefits but could still not receive them if their income was above the lower threshold of 60 BRL and they did not have children or pregnant women or because of delays in the application approval. In principle, this makes the Bolsa Fam\'ilia program a fuzzy RD design.
Nevertheless, here, we focus on the intention to treat effect of eligibility (having a monthly per capita income below 120 BRL) rather than on the effect of the actual receipt of the benefits.
This allows us to avoid making further assumptions, such as exclusion restrictions, that would be required to  estimate the effects of receiving the benefits
\footnote{
 We could have applied our method to the lower threshold of 60 BRL. However, by doing so, we would have estimated the effect of the eligibility to the basic benefits in addition to the variable benefits, instead of the overall effect of eligibility to the program. 
}. 

\section{The Brazil's Bolsa Fam\'ilia Dataset} \label{sec:dataset}
We analyze the Brazilian Bolsa Fam\'ilia program using
a subset of the 100 Million Brazilian Cohort, including the $N=147\,399$ families who registered in CadUnico in $2007-08$ for the first time and have a monthly per capita household income in $2008$ not greater than $300$ BRL\footnote{We chose the limit of $300$ BRL because some bumping in the distribution of the forcing variable was observed above $300$ BRL, possibly due to salaries being determined by state contracts. Excluding observations above $300$ BRL allows us to ease model convergences and improve the stability of results. Nevertheless, we have also conducted the analysis using all the data as a robustness check. In this analysis, very few units, around $150 (0.19\%)$, with per capita household income larger than $300$ BRL are included in the subpopulation denoted $\mathcal{U}_{s_{0}}$ and results (available on request to the authors) lead to the same substantive conclusions.}. See Section~A.2 in the e-Appendix for some details on the 100 Million Brazilian Cohort.

Let $\mathcal{U}$
denote the set of families indexed by $i \in\{1, \ldots, N\}$. 
Let $S_i$  denote the forcing variable, here, family $i$'s monthly per capita household income in Brazilian Reals (BRL) in $2008$. Let $Z_i \in \{0, 1\}$ indicate  the eligibility status; $Z_i$ is a deterministic function of monthly per capita income $S_i$:  $Z_i =\mathbb{I}\{S_i \leq s_0\}=\mathbb{I}\{S_i \leq 120\}$.  
In our sample,  $138\,220$ ($93.8\%$) families are eligible with monthly per capita household income not greater than 120 BRL ($S_i\leq 120$) and $9\,179$ ($6.2\%$) families are not eligible with monthly per capita household income greater than 120 BRL ($S_i> 120$).

\begin{figure}[t]
\begin{center}
\begin{subfigure}{0.4\textwidth}
\hspace{-1cm}
\includegraphics[width=8cm]{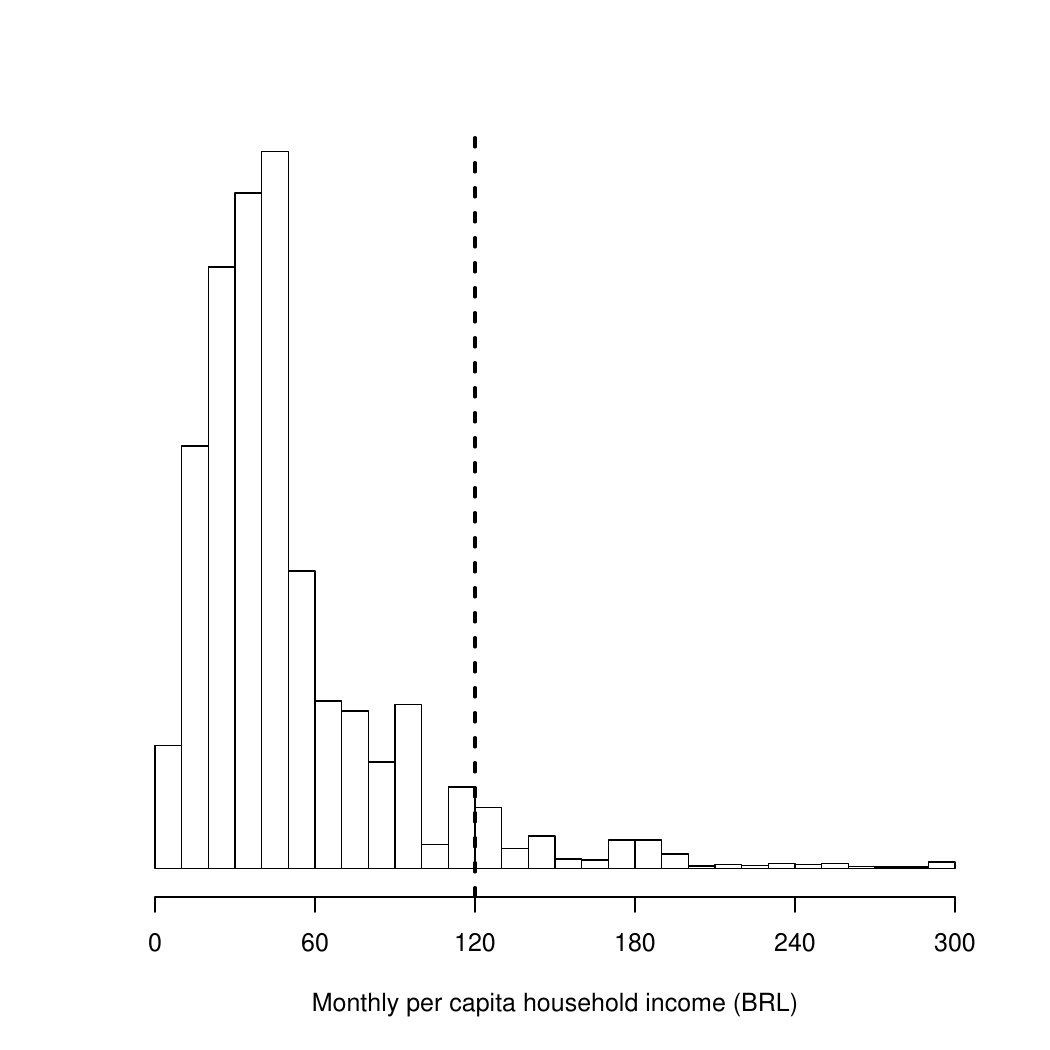}  
\caption{}\label{Fig1:S}
\end{subfigure} 
\begin{subfigure}{0.5\textwidth}
\hspace{-1cm}
$$
\begin{array}{lccc}
\\
\\
\\
\\
\\
\multicolumn{4}{c}{\hbox{Per- capita household income } (S_i).}\\
\hline
\vspace{-0.25cm}\\
\hbox{Statistic} & \hbox{All }  & Z_i=0 & Z_i=1  \\
(\hbox{Sample size})&(147\,399)&(9\,179)&(138\,220)\\
\hline
\hbox{Min}    &  \,\,\,0.0  & 120.2 & \,\,\,0.0    \\
Q_1           &  28.0  &   130.0 & 26.7 \\
\hbox{Median} &  45.0  &   156.7 & 40.0\\
\hbox{Mean}   &  53.4  &  168.6  & 45.8   \\
Q_3           &  60.0 & 190.0  & 58.9 \\
\hbox{Max}    &  \!\!\!300.0  & 300.0  & \!\!\!120.0    \\
\hbox{SD }  &  39.8  & 42.8& 25.1    \\
\hline
\\
\\
\\
\\
\end{array}
$$	
\caption{}\label{tab1}
\end{subfigure} 
\end{center}
\caption{Bolsa Fam\'ilia Study: Descriptive statistics of the forcing variable: Monthly per capita household income (BRL) in 2008. (a) Histogram of monthly per capita household income. (b) Summary statistics of monthly per capita household income by eligibility status.} \label{tab1Fig1}
\end{figure}

%

%
Figure~\ref{tab1Fig1} presents the histogram of the empirical distribution of monthly per capita household income for the whole sample and summary statistics of monthly per capita household income by eligibility status,  $Z_i$. 
As we can see, 
the empirical distribution of monthly per capita household income is skewed to the right: 
more than $95\%$ of families have a value of monthly per capita household income lower than 130 BRL, and more than $99\%$ of families have a value of monthly per capita household income not greater than 200 BRL. The median $S_i$ is $40$ BRL for eligible families and $156.7$ BRL  for ineligible families.

In addition to the forcing variable, and thus, the eligibility status, for each family $i$, we observe: a binary outcome $Y_i$, equal to 1 if at least a leprosy case in family $i$ occurs in 2009, and $0$ otherwise, and a vector of $p=24$ covariates, $\bX_i$, including information on the household structure, living and economic conditions of the family,	and household head's characteristics.

\begin{figure}[t]
	\begin{center}
		\includegraphics[width=8cm ]{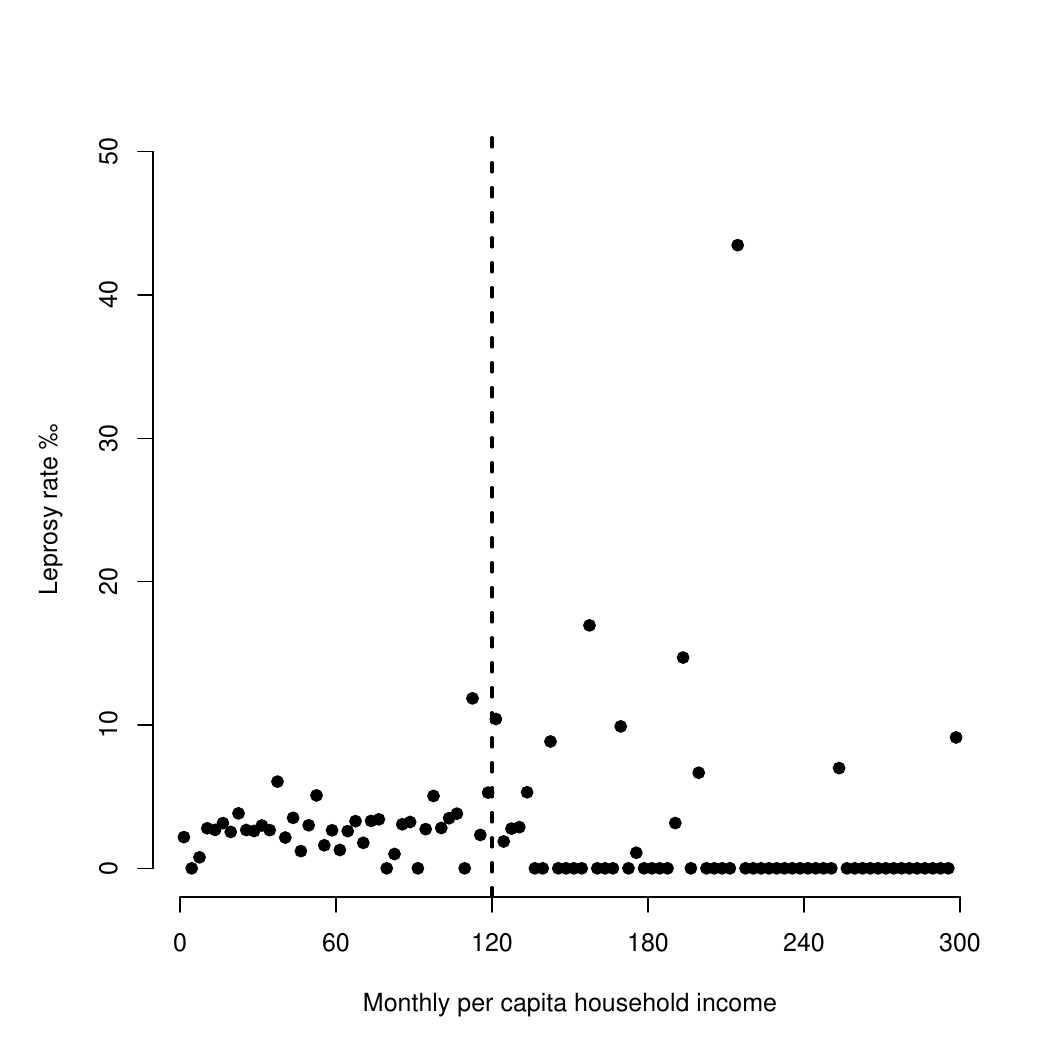} 
    	\end{center}
	\caption{Leprosy rate (per mil) as a function of the forcing variable (monthly per capita household income).} \label{Fig2:Y}
\end{figure}

Table A1 
in the e-Appendix presents some summary statistics for the sample, classified by eligibility.
We notice systematic differences in background characteristics between eligible and ineligible families.
Eligible families are, on average, younger and larger than ineligible families; they comprise a larger number of children and are in worse living and economic conditions. 
Moreover,  the proportion of unemployed household heads is higher for eligible families than for ineligible families. The overall leprosy rate is $2.80 \permil$, and it is slightly higher	among eligible families: 
$2.80 \permil$ versus $2.72 \permil$. 
A graphical analysis of the outcome	by forcing variable is shown in Figure~\ref{Fig2:Y}, where we display leprosy rate (per mil) as a function of monthly per capita household income.
Since leprosy is a rare outcome, it makes it difficult to 
use the standard graphical diagnostics for RD designs \cite[e.g.,][]{ImbensLemieux2008} \footnote{Note that such a graphical analysis is also very sensitive to the choice of the bin width.}.

\section{The Brazil's Bolsa Fam\'ilia RD Design as Local 
Unconfounded Experiment} \label{sec:Setup}

In RD designs under the 
probabilistic framework, the forcing variable is stochastic and can be seen as the assignment variable. Therefore, under the potential outcomes approach \citep{Rubin1974, Rubin1978},  potential outcomes need to be 
postulated as being indexed by the forcing variable.

Throughout, we will maintain the common Stable Unit Treatment Value Assumption  \cite[SUTVA;][]{Rubin1980}, which implies that there is no interference, in the sense that the potential leprosy outcomes for a family cannot be affected by the value of monthly per capita household income  (and thus, by the eligibility status) of other families. 
Conducting the analysis at the household level mitigates the possible violation of the no-interference assumption\footnote{If interference were still present across households, \cite{Aronow2021TheRD} showed that under the local randomization framework, an estimator neglecting interference and targeted to the causal treatment effect in a subpopulation above and below the threshold is unbiased for the average direct effect in the same subpopulation, marginalized over the treatment assignments.}.
Under this assumption, let $Y_i({s})$ denote the potential leprosy status that would be observed for the $i-$th  family under a monthly per capita household income equal to ${s} \in \mathbb{R}_+$.

The 
probabilistic framework we adopt is based on three key assumptions that we now introduce and discuss in the context of the Bolsa Fam\'ilia study. 

The first assumption is a \textit{local overlap} assumption: it requires that there exists
a subpopulation 
comprising eligible and ineligible families for which an overlap assumption holds, namely, families that have a non-zero conditional probability given the covariates of having 
a value of 
$S$ falling on both sides of the threshold. 
Formally, 
\begin{assumption} \label{ass:loverlap}\textbf{(Local Overlap)}.
	There exists a subset of units, $\Uset \subseteq \mathcal{U}$, such that for each $i \in \Uset$, $0<\Pr(S_i \leq s_0 \mid \bX_i) < 1$, and thus, also  $0<\Pr(S_i > s_0 \mid \bX_i)<1$.
\end{assumption}
Assumption \ref{ass:loverlap}  implies that eligible (or ineligible) families in $\Uset$ with an observed value of the monthly per capita household income below (or above) the threshold, given their observed characteristics $\bX_i$, may have been observed with a possibly different value of the monthly per capita household income making them ineligible (or eligible) for the BF program. 
We can hypothesize that this local assumption is more likely to hold for families with an observed $S$ close to the threshold. However, as we will see in the analysis, this can also be true for those whose monthly per capita household income was observed relatively far from the threshold.

For families belonging to the subpopulation $\Uset$ we also adopt a modified Stable Unit Treatment Value Assumption  \cite[SUTVA;][]{Rubin1980}  specific to RD
settings:
\begin{assumption} \label{ass:lsutva} \textbf{(Local RD-SUTVA)}.
	For each  $i \in \Uset$, consider two eligibility statuses
	$z^{'} = \one(S_i=s' \leq s_0)$ and $z^{''} = \one(S_i=s^{''} \leq s_0)$, with possibly $s^{'} \neq s^{''}$.
	$ \hbox{If }  z^{'} = z^{''} \hbox{ then }    Y_i({s}^{'})=Y_i({s}^{''})$.
\end{assumption}
Assumption \ref{ass:lsutva}  implies that the potential leprosy outcomes depend on monthly per capita household income solely through the eligibility status, $Z_i$, but not directly, so that values of monthly per capita household income leading to the same eligibility status define the same potential leprosy outcome. 
Assumption \ref{ass:lsutva}  allows us to write $Y_i(s)$ as $Y_i(z)$ for each unit $i \in \Uset$, and thus,  under local RD-SUTVA for each family $i$ within $\Uset$ there exist only two potential outcomes, $Y_i(0)$ and $Y_i(1)$: they are the values of the leprosy indicator if family $i$ had a value of monthly per capita household income falling above and below the threshold of $s_0=120$ BRL, respectively.

Under local overlap and local RD-SUTVA (Assumptions \ref{ass:loverlap} and \ref{ass:lsutva}), causal effects are defined as comparisons of these
two potential outcomes for a common set of units in the subpopulation $\Uset$. 
They are \textit{local} causal effects in that they are causal effects for units belonging to the subpopulation $\Uset$. For our Bolsa Fam\'ilia study, we focus on the finite sample causal relative risk:
\begin{equation}\label{eq:rr} 
	RR_{\Uset} \equiv 
	\dfrac{ \sum_{i: i \in \Uset} Y_i(1)\Big/ N_{\Uset} }{ \sum_{i: i \in \Uset} Y_i(0)\Big/ N_{\Uset}  },  
\end{equation}
where $N_{\Uset}$ is the number of units in $\Uset$. For rare outcomes, such as leprosy, the causal relative risk is generally more informative than the causal risk difference.
Note that $RR_{\Uset}$ is well-defined only if there is at least one unit in $\Uset$ with $Y_i(0)=1$. Also, because we do not generally observe which units belong to the subpopulation $\Uset$, the number of units in $\Uset$, $N_{\Uset}$, is unknown.

We formalize the concept of an RD design as a  locally unconfounded experiment or as a locally regular design
invoking the following assumption: 
\begin{assumption} \label{ass:lunc} \textbf{Local Unconfondedness (LU)}. For each  $i \in \Uset$,
	$$
	p\left(S_i \mid Y_i(0), Y_i(1), \bX_i \right) =p\left(S_i  \mid \bX_i  \right),
	$$
 where $p(\cdot \mid \cdot)$ denote a conditional probability density function.
\end{assumption}
\noindent Assumption \ref{ass:lunc} implies that for each family $i \in \Uset$, 
$$ \Pr\left(S_i \leq s_0 \mid Y_i(0), Y_i(1), \bX_i \right) =\Pr\left(S_i \leq s_0 \mid \bX_i  \right)=\Pr\left(Z_i =1 \mid \bX_i  \right),$$ 
which amounts to state that within the subpopulation $\Uset$, families with the same values of 
$\bX_i$
 have the same probability of being observed under 
 the threshold value of the forcing variable, and, in turn, of being eligible or not for the Bolsa Fam\'ilia program. That is, in $\Uset$, eligibility is as good as random conditional on covariates and does not depend on endogenous factors.
Given that the vector of covariates $\bX_i$ includes
characteristics of the household structure, living and economic conditions,	and household head's  socio-demographic characteristics (see Table~A1 
in the e-Appendix), it seems plausible that there exist some households with the same characteristics whose monthly  per capita income  in 2008 fell below or above the threshold due to random factors or events not related to leprosy  
\footnote{
Note that indexing the potential outcomes by $S$ allows us to distinguish between local RD-SUTVA and local unconfoundedness, which are similar but separate assumptions, as first noticed by \cite{LiMatteiMealli2015}.}. 
Assumptions \ref{ass:loverlap}-\ref{ass:lunc} are local assumptions in the sense that they are assumed to hold only for a (latent) subpopulation of units. Therefore, we are postulating that there may exist both units for which these assumptions hold as 
well as units for which some of these assumptions do not hold.  In particular,  
unconfoundedness may hold for some units for which the individual-level assignment probabilities can vary with the unit's covariate values but are free from the potential outcomes, and not for others, for which the individual-level probabilities are a function of
the potential outcomes.

As we highlight in Section~\ref{sec:background}, \cite{AngristRokkanen2015} also use the concept of unconfoundedness in RD settings; they describe the RD design as a ``global'' unconfounded experiment by assuming unconfoundedness of the forcing variable conditionally on covariates for the whole population. Thus, \cite{AngristRokkanen2015} make a stronger version of unconfoundedness about the whole population to identify overall average causal effects. We, instead, assume that unconfoundedness only holds for a subpopulation of units, and our inference targets the identification of this subpopulation, which is generally unknown, and estimation of local causal effects for units belonging to it.

\cite{AngristRokkanen2015} first note that in RD settings, unconfoundedness implies that 
the forcing variable is independent of the observed potential outcome conditional on the eligibility
status and covariates. They use this implication, which is testable given that it involves only observed quantities, as
a falsification test for unconfoundedness. Under our RD assumptions, this implication holds locally, that is, for units in $\Uset$. Formally, under Assumption \ref{ass:loverlap}-\ref{ass:lunc}, for each $i \in \Uset$,
\begin{equation} \label{eq:limplication}
    p(S_i \mid Z_i=z, \Yobs_i, \bX_i) = p(S_i \mid Z_i=z,  \bX_i) \qquad z=0,1. 
\end{equation}
See Section~B.1 in the e-Appendix for the proof of Equation~\ref{eq:limplication}. 
The approach that we propose here for the selection of $\Uset$ relies precisely on this implication:
the Bayesian model-based clustering approach that we use to identify $\Uset$ amounts to identifying the set of units for whom 
Equation~\ref{eq:limplication} holds.

Under Assumptions \ref{ass:loverlap}-\ref{ass:lunc}, if the subpopulation $\Uset$ were observed, $RR_{\Uset}$ in Equation \eqref{eq:rr} would be identified from the observed data. Formally,  under Assumptions \ref{ass:loverlap}-\ref{ass:lunc}, 
we have
\begin{eqnarray*}
	\Pr\left(Y_i(z)=1 \mid i \in \Uset\right) &=&	\mathbb{E}\left[Y_i(z) \mid i \in \Uset\right] \\&=&
	\mathbb{E}_X\left[\mathbb{E}\left[Y_i(z) \mid \bX_i; i \in \Uset\right]\right]\\&=&	\mathbb{E}_X\left[\mathbb{E}\left[Y_i(z) \mid \bX_i, Z_i=z; i \in \Uset\right]\right]\\&=&
	\mathbb{E}_X\left[\mathbb{E}\left[\Yobs_i \mid \bX_i, Z_i=z; i \in \Uset\right]\right], 
\end{eqnarray*}
where the first equation holds by definition, the second equality follows from the law of iterated expectation, the third equation follows from Assumption \ref{ass:lunc}, and the last equality follows from Assumption~\ref{ass:lsutva}. Therefore, in principle, we could exploit this result implied by Assumptions \ref{ass:loverlap}-\ref{ass:lunc} to derive simple estimators of causal effects for families in $\Uset$. However, note that under Assumptions \ref{ass:loverlap}-\ref{ass:lunc}, $\Uset$ exists and  comprises all units for which Assumptions \ref{ass:loverlap}-\ref{ass:lunc} hold, but is not observed.



Assumptions \ref{ass:loverlap}-\ref{ass:lunc} present
similarities and differences 
with alternative probabilistic 
approaches \citep{LiMatteiMealli2015,MatteiMealli2016,BransonMealli2019,CattaneoFrandsenTitiunik2015,SalesHansen2020,EcklesEtAl2020}, that were already discussed in Section~\ref{sec:background}.

\section{Selection of the Subpopulation  $\Uset$}
\label{sec:Uset}

In practice,  the 
subpopulation $\Uset$,  where the RD assumptions hold, is unknown. Therefore, a critical issue of the approach is how to choose $\Uset$.

\subsection{Selection of Subpopulations  $\Uset$: State of the Art}	\label{subsec:UsetA}
As discussed briefly in Section~\ref{sec:background}, the existing probabilistic approaches in the literature 
exploit the implications of the assumptions on the hypothesized   assignment mechanism.
Under the assumption that the subpopulation has a rectangular shape, comprising units with a realized value of the forcing variable falling in a (possibly symmetric) interval around the threshold: $\Uset= \{i: S_i \in[s_0-h^{L}, s_0+h^{U}]\}$, \cite{CattaneoFrandsenTitiunik2015, LiMatteiMealli2015, MatteiMealli2016, LicariMattei2020, BransonMealli2019} propose to use falsification tests for selecting the bandwidth $h$ that exploits the local nature of the invoked randomization / unconfoundedness assumption. 
For instance, under local randomization, the covariates 
should be well balanced in the two subsamples defined by $Z_i$ in $\Uset$, and thus any test of the null	hypothesis of no difference in the distribution (or mean) of covariates between eligible and ineligible should fail to reject the null. 


The falsification test approach is relatively simple to implement, but it has two important pitfalls. First, it generally relies on the assumption that 
the subpopulation we are looking for has a rectangular shape. Although this assumption offers computational advantages, restricting the set of candidate subpopulations, it may be without grounds or be difficult to justify from a substantive perspective. Second, causal inference is drawn conditionally on the results of the falsification tests without incorporating the uncertainty about the selected subpopulation.

Alternative methods include the penalized matching framework proposed by \cite{KeeleTitiunikZubizarreta2015}, the Dirichlet process mixture model approach proposed by \cite{ricciardi2022dirichlet}, and the  ``regression discontinuity donut'' specification proposed by \cite{SalesHansen2020}.  These methods do not require any assumption on the shape of the subpopulation,  
but causal inference is again conducted without	directly accounting for the uncertainty in the selection of the	subpopulation.

\subsection{A Bayesian Model-Based Finite Mixture Approach to the Selection of Subpopulations  $\Uset$}
\label{subsec:Usetapproach}
The data challenges raised by the Bolsa Fam\'ilia study and the pitfalls of the existing approaches prompted us to develop a new approach to the selection of the suitable subpopulation in RD designs. We propose a Bayesian model-based finite mixture approach to clustering to classify observations into subpopulations where the RD assumptions hold and do not hold on the basis of the observed data.

The key insight underlying our approach is to  view the families in the   Bolsa Fam\'ilia  study as coming from  three subpopulations:
\begin{enumerate}
	\item  $\Uset$: the subpopulation of families where the RD assumptions hold, for which the support of the forcing variable (monthly per capita household income) includes $s_0$ and values  above and below the threshold, $s_0$;
	\item $\Usetm$:  the subpopulation of families who 
 have a realized value of the forcing variable not greater than the threshold, $s_0$, but do not belong to $\Uset$ given that some of the RD assumptions may fail to hold; 
	\item $\Usetp$: the subpopulation of families  who 
 have a realized value of the forcing variable above the threshold, $s_0$, but do not belong to $\Uset$ given that some of the RD assumptions may fail to hold. 
\end{enumerate}
The subpopulations $\Uset$, $\Usetp$ and $\Usetm$ define a partition of the whole population, $\mathcal{U}$: $\mathcal{U}= \Uset \cup \Usetm \cup \Usetp$, and $\Uset \cap \Usetm = 
\Uset \cap \Usetp= \Usetm \cap \Usetp= \emptyset$, and thus, each family in the study must belong to one of the three subpopulations.
A clear definition of the characteristics of each subpopulation is key to our approach. 
For families who belong to $\Uset$, the RD assumptions must hold and thus, Equation~\ref{eq:limplication} must hold. 
For families who do not belong to $\Uset$, the local overlap assumption may be untenable -- in the sense that those families may have a zero probability of	being assigned to either eligibility statutes -- and/or there may be a relationship between the forcing variable and potential outcomes, monthly per capita household income and presence of leprosy cases in the family,  implying that either local RD-SUTVA (Assumption \ref{ass:lsutva}) or local unconfoundedness   (Assumption \ref{ass:lunc}) is questionable.
For families in $\Usetm \cup \Usetp$, the failure of local RD-SUTVA affects the definition of the potential outcomes, which must be indexed by the forcing variable; and the failure of local unconfoundedness implies that the assignment mechanism depends on the potential outcomes.

We use these features characterizing the three subpopulations as input for our Bayesian model-based finite mixture approach for classifying families into the three subpopulations.
Specifically, we view the joint conditional distribution of the forcing variable and the potential outcomes	given the observed covariates as a three-component finite mixture distribution with unknown mixture proportions \cite[e.g.,][]{TitteringtonSmithMarkov:1985, McLachlanBasford:1988}. The three components correspond to the three subpopulations, $\Uset$, $\Usetp$, and $\Usetm$:	we characterize them to match their definition with respect to the RD assumptions 
and the value of the forcing variable falling below or above the threshold.	Formally, we specify the following mixture model:
{
\allowdisplaybreaks
\begin{eqnarray} \label{eq:mix}
	\lefteqn{p\left(S_i, \{Y_i({s})\}_{{s}\in \mathbb{R}_+} \mid \bX_i; \btheta \right) 	
	}\nonumber \\&=& \label{eq:mixa}
	\pi_i(\Uset; \btheta) \,  p(S_i,  \{Y_i({s})\}_{{s}\in \mathbb{R}_+}   \mid \bX_i, i \in \Uset; \btheta) +\\&& \nonumber
	\pi_i(\Usetm; \btheta) \, p(S_i, \{Y_i({s})\}_{{s}\in \mathbb{R}_+} \mid   \bX_i; i \in \Usetm;\btheta) + \\&& \nonumber
	\pi_i(\Usetp; \btheta) \,  p(S_i ,\{Y_i({s})\}_{{s}\in \mathbb{R}_+}  \mid  \bX_i, i \in \Usetp; \btheta)\\&=&  \label{eq:mixb}
	\pi_i(\Uset; \balpha) \,  p(S_i \mid \bX_i, i \in \Uset; \boeta) \, p(Y_i(0), Y_i(1)  \mid \bX_i, i \in \Uset; \bgamma)+\\&&\nonumber
	\pi_i(\Usetm; \balpha) \, p(S_i \mid \bX_i; i \in \Usetm; \boeta^{-}) \, p(\{Y_i(s)\}_{s\in \mathbb{R}_+} \mid S_i, \bX_i; i \in \Usetm; \bgamma^{-}) +\\&&\nonumber
	\pi_i(\Usetp; \balpha) \,  p(S_i \mid \bX_i, i \in \Usetp; \boeta^{+})\, p(\{Y_i(s)\}_{s\in \mathbb{R}_+} \mid S_i, \bX_i, i \in \Usetp; \bgamma^{+}),
\end{eqnarray}
}
where $\pi_i(\Uset; \balpha)=Pr(i \in \Uset \mid   \bX_i; \balpha)\geq 0$,	$\pi_i(\Usetm; \balpha) =Pr(i \in \Usetm \mid   \bX_i; \balpha)\geq 0$ and $ \pi_i(\Usetp; \balpha)=Pr(i \in \Usetp \mid   \bX_i; \balpha)\geq 0$ 	are the mixing probabilities, with $\pi_i(\Usetm; \balpha)+\pi_i(\Uset; \balpha)+\pi_i(\Usetp; \balpha)=1$; $(\boeta^{-}, \bgamma^{-})$, $(\boeta, \bgamma)$ and $(\boeta^{+}, \bgamma^{+})$ are parameter vectors defining each mixture component, and	$\btheta = \left(\balpha, \boeta^{-}, \bgamma^{-},  \boeta, \bgamma, \boeta^{+}, \bgamma^{+}\right)$	is the complete set of parameters specifying the mixture. 

Equation~\eqref{eq:mixa} formally describes the joint distribution of the forcing variable, $S_i$, and the potential outcomes,  $\{Y_i({s})\}_{{s}\in \mathbb{R}_+}$, as a three mixture distribution, and Equation~\eqref{eq:mixb} follows from imposing the RD assumptions for units in $\Uset$ and allowing for 
violations of those assumptions for units who do not belong to $\Uset$. Specifically, the mixture component corresponding to the subpopulation $\Uset$ in Equation~\eqref{eq:mixb} is specified to reflect the RD assumptions: under local overlap (Assumption~\ref{ass:loverlap}), first,   local RD-SUTVA (Assumption \ref{ass:lsutva}) implies that for each family in $\Uset$ there exist only two potential outcomes, $Y_i(0)$ and $Y_i(1)$, the two potential outcomes corresponding to the two eligibility statuses; second, local unconfoundedness (Assumption \ref{ass:lunc}) implies that the forcing variable and the potential outcomes are conditionally independent given the covariates so that the joint distribution of the forcing variable and the potential outcomes factorizes into the product of the marginal distribution of the forcing variable and the marginal distribution of potential outcomes. 

The specification of the mixture components describing the joint conditional distribution of the forcing variable and the potential outcomes in the subpopulations $\Usetm$ and $\Usetp$ reflects possible violations of local RD-SUTVA and/or local unconfoundedness: potential outcomes are defined as a function of the values of the forcing variable, and the forcing variable and the potential outcomes may be not independent, even conditional on the covariates. 
A possible threat to unconfoundedness is the presence of some manipulation of the forcing variable, when characteristics leading to manipulation are correlated with potential outcomes even after conditioning on the observed covariates. 
Observations where this occur would belong to $\Usetm$ or $\Usetp$.

After specifying a parametric form for each probability distribution involved in the mixture, we propose to use a Bayesian approach to fit the mixture model. Posterior inference of the parameters can be obtained using Gibbs sampling with a data augmentation step to impute the missing subpopulation membership for	each unit \cite[][]{DieboltRober:1994, RichardsonGreen:1997}.	Specifically, we derive the posterior distribution of the causal estimand in Equation \eqref{eq:rr}  using  an MCMC algorithm with data augmentation \citep{TannerWong:1987}, where at each iteration $\ell$, $\ell=1, \ldots, L$, of the MCMC algorithm: $(1)$ we impute the missing subpopulation membership for each family using a data
augmentation step; $(2)$ we update the model parameters using Gibbs sampling methods; and $(3)$ for each family classified in $\Uset^\ell$, we draw the missing potential outcome,	$Y_i^{\ell}=Z_i Y_i(0) +(1-Z_i)Y_i(1)$,	from its posterior predictive distribution and calculate the causal relative risk ratio:
$$
RR^{\ell}_{\Uset}=	\dfrac{ \sum_{i: i \in \Uset^\ell} [Z_i Y_i + (1-Z_i) Y_i^{\ell}(1)]\Big/ N_{\Uset^{\ell}} }{ \sum_{i: i \in \Uset^{\ell}} [(1-Z_i) Y_i + Z_i Y_i^{\ell}(0)]\Big/ N_{\Uset^{\ell}}  },
$$	
where $Y_i=Z_i Y_i(1) +(1-Z_i)Y_i(0)$ is the observed potential outcome for $i \in \Uset^{\ell}$.  See Section~B.4 in the e-Appendix for further details. 

In this framework,  each unit's contribution to the posterior distribution of the causal estimand of interest depends on the posterior probability of that unit belonging to the subpopulation $\Uset$. 
The unit-level posterior probabilities of agreeing with the RD assumptions, i.e., of belonging to $\Uset$ (as	determined by the relationship between the outcome and the forcing variable conditional on the covariates), will act as weights in the final posterior inference. Observations with a higher posterior probability of being members of $\Uset$ will be assigned a higher weight and, thus,  will provide more information to the posterior inference of $RR_{\Uset}$.
Observations exhibiting characteristics that may undercut the plausibility of the RD assumptions will be assigned a smaller weight, and thus, will contribute less to the inference of $RR_{\Uset}$.

By down-weighting observations where there is less evidence that the unconfoundedness holds, 
our method is robust with respect to manipulations that lead to failure of unconfoundedness.
Note that this feature of our method can be viewed as a smooth, stochastic version of the ``donut regression discontinuity'' method \citep{Barreca2011SavingBR, SalesHansen2020}; it can also be viewed as a stochastic version of the mixture model for confounding contamination proposed by \cite{BonviniKennedy2021}.
We will return to this when discussing our case study in Section \ref{sec:BFanalysis}.

\subsection{Specification of the Finite Mixture Model in the Bolsa Fam\'ilia Study} 
In the Bolsa Fam\'ilia study, to model the mixing probabilities, we adopt two conditional probit
models, defined using latent variables $G^\ast_i(-)$ and 	$G^\ast_i(+)$, for
whether family $i$ belongs to the subpopulation $\Usetm$ or $\Usetp$:
$$	\pi_i(\Usetm) = \Pr(G^\ast_i(-) \leq 0) \qquad \hbox{and} \qquad
\pi_i(\Usetp) = \Pr(G^\ast_i(-) > 0 \, \mathrm{and}  \, G^\ast_i(+) \leq 0),
$$
where
$$G^\ast_i(-) = \alpha_0^{-} +   \mathbf{X}_i'\balpha^{-}_X + \epsilon_{i}^{-} \qquad \text{and}\qquad G^\ast_i(+) = \alpha_{0}^{+} +   \mathbf{X}_i'\balpha^{+}_X + \epsilon_{i}^{+},  $$
with $\epsilon_{i}^{-} \sim N(0,1)$ and  $\epsilon_{i}^+ \sim N(0,1)$, independently. Clearly $\pi_i(\Uset) =1- 	\pi_i(\Usetm)-	\pi_i(\Usetp)$.
For the forcing variable, we specify log-normal models.  For the subpopulations $\Usetm$ and $\Usetp$, we specify truncated log-normal models, respectively,  bounded from above and from below at $\log(s_0)$ with the location parameters linear in the covariates with subpopulation-specific parameters and with subpopulation-specific variances.   For the subpopulations $\Uset$, we specify a log-normal model with the mean linear in the covariates with subpopulation-specific parameters and subpopulation-specific variance. Formally,
\begin{eqnarray*}
	\log(S_i) \mid \bX_i, i \in \Usetm  &\sim& TN_{(-\infty, \log(s_0)]}\left(\beta_0^{-}+   \mathbf{X}_i'\bbeta^{-}_X; \sigma^2_{-} \right),\\
	\log(S_i) \mid \bX_i, i \in \Usetp &\sim& TN_{(\log(s_0),\infty)}\left(\beta_0^{+}+   \mathbf{X}_i'\bbeta^{+}_X; \sigma^2_{+} \right), \\
	\log(S_i)  \mid \bX_i, i \in \Uset  &\sim& N\left(\beta_0 +   \mathbf{X}_i'\bbeta_X; \sigma^2 \right).
\end{eqnarray*}
It is worth noting that this model specification implies that the probability of having a value of the forcing variable on the other side of the threshold is zero for units in $\Usetm$ and $\Usetp$, that is, for these units the overlap assumption (Assumption \ref{ass:loverlap}) does not hold.

Because our outcome is 	dichotomous, we assume that the marginal  distributions of the outcome take the form of generalized linear Bernoulli models with a probit link:
\begin{eqnarray*}
	\Pr(Y_i(s) =1 \mid  S_i=s, \bX_i, i \in \Usetm) &	=& \Phi\left(\gamma_{0}^{-} +  \tilde{s} \gamma_{1}^{-}   + \mathbf{X}_i'\bgamma^{-}_X \right),\\
	\Pr(Y_i(s) =1 \mid  S_i=s, \bX_i, i \in \Usetp) &	=& \Phi\left(\gamma_{0}^{+} + \tilde{s} \gamma_{1}^{+}   + \mathbf{X}_i'\bgamma^{+}_X \right),\\
	{\Pr(Y_i(z) =1 | \bX_i, i \in \Uset)} &	{=}& {\Phi\left(\gamma_{0,z} +   \mathbf{X}_i'\bgamma_{X,z} \right), \qquad z=0,1,}
\end{eqnarray*}
where $\Phi(\cdot)$ denotes the cumulative distribution function of a standard Normal distribution, and $\tilde{s}$ are the standardized values of the forcing variable. 
For parsimony and for gaining information across groups,  we impose 
equality of the slope coefficients in the outcome regressions:  $\bgamma_{X}^{-}=\bgamma_{X}^{+} =\bgamma_{X,z=0}=\bgamma_{X,z=1} \equiv\bgamma_{X}$.

Bayesian inference on finite sample causal estimands for a subset of units in the study, such as the local causal relative risk in Equation~\eqref{eq:rr} we are interested in, follows from predictive Bayesian inference and generally involves parameters describing the association between potential outcomes \cite[e.g.][]{ImbensRubin1997, ImbensRubin2015}. Nevertheless,   the association parameters do not enter the likelihood function: because we never simultaneously observe all the potential outcomes for any unit, the data contain no information about the association between the potential outcomes. Therefore, the posterior distribution of the association parameters will be identical to their prior distribution if we assume that they are a priori independent of the other parameters. 
Throughout the paper, we assume that the unobserved potential outcomes are independent of the observed potential outcome conditional on the covariates and parameters.
Therefore, the above model assumptions completely specify the mixture model in Equation \eqref{eq:mix}\footnote{Relaxing the conditional independence assumption usually leads to posterior distributions of the causal estimands centered on approximately the same values but a larger posterior variability \cite[e.g.,][]{li2023bayesian}.}.
The full parameter vector is $\btheta = \{(\alpha_0^{-}, \balpha^{-}_X),(\alpha_0^{+}, \balpha^{+}_X),$ $(\beta_0^{-},  \bbeta^{-}_X, \sigma^2_{-}),$ $(\beta_0^{+},  \bbeta^{+}_X,\sigma^2_{+}),$ $(\beta_0,  \bbeta_X, \sigma^2),$ $(\gamma_0^{-}, \gamma_1^{-}),$ $(\gamma_0^{+}, \gamma_1^{+}),$ $\gamma_{0,z=0},$ $\gamma_{0,z=1},$ $\bgamma_X)\}$, which includes $6\times p + 14=6\times 24 + 14=158$ parameters. 

Bayesian inference is conducted under the assumption that parameters are a priori independent and using  multivariate normal prior distributions for the regression coefficients and 
inverse-chi square distributions for the variances of the models for the forcing variable.
Specifically, the prior distributions for the mixing probabilities are multivariate normal distributions with covariance-variance matrices equal to scalar matrices with equal-valued elements along the diagonal set at $1$. The  mean vectors are vectors of zeros with the exception of the first element of the mean vector of the coefficients of the probit submodel for $\Usetm$ membership, which is set equal to  $\Phi^{-1}(2/3) \sqrt{1+\overline{\bX}'\overline{\bX}} $, where 
$\overline{\bX}$ is the mean vector of the covariates. These prior specifications result in approximately setting the prior probability of belonging to each subpopulation at 1/3 for each individual (see Figure~A3(a) 
in the e-Appendix), which reflects our a priori ignorance about such probabilities. For the regression coefficients of the models for the outcome, we use as prior distributions multivariate normal distributions with mean vector zero and covariance-variance matrices equal to scalar matrices with equal-valued elements along the diagonal set at $1$.
For the  truncated log-normal models and the log-normal model of the forcing variable, we specify similar prior distributions for the regression coefficients (although the diagonal of the set covariance-variance matrices is set at $100$)
and  inverse   chi-squared distributions with degrees of freedom set at $3$ and scale parameter  set at $1/3$ for the variances (See Section~B.3 
in the e-Appendix for details). 

\subsection{Simulation Study under Challenging Scenarios}
We conduct extensive simulation studies to investigate how our approach works even in challenging scenarios
and to  shed light on the sources of identification of the subpopulation membership. 
We also investigate 
whether our procedure is able to recognize when the subpopulation $\Uset$ is empty. 
The simulation results convey three essential messages. First, our   Bayesian model-mixture approach works very well in extreme scenarios where there are no units in $\Uset$, i.e., $\Uset$ is empty: it provides extremely small posterior means for the proportion of units in $\Uset$ with very high precision, successfully suggesting that the subpopulation $\Uset$ is empty. Second,  the stronger the association between the covariate and forcing variable, the better the performance of the Bayesian model-based approach to clustering is. Third, our approach leads to more accurate and more efficient estimates of the causal estimands in studies where the dependence of subpopulation membership on the forcing variable	is stronger.
See Section~D 
in the e-Appendix for an in-depth description of the simulation designs and results.

\section{Design and Analysis of the Bolsa Fam\'ilia Study} \label{sec:BFanalysis}

We apply the  Bayesian model-based finite mixture approach described in Section~\ref{sec:Uset} for both the design and the analysis of the Bolsa Fam\'ilia program to assess its effect on leprosy incidence.

A preliminary discussion is deserved. In RD designs, concerns about the validity of the RD assumptions may arise if the forcing variable is susceptible to manipulation. Because Bolsa Fam\'ilia applicants know the eligibility criteria, there is a concern that they might attempt to report a lower income in order to end up below the threshold and receive the Bolsa Fam\'ilia benefits. If this is the case, the eligible families just below the threshold may include those who have manipulated their income, and thus, they may have somewhat different characteristics from the ineligible families just above the threshold who have not reported their true income, invalidating the basic RD design. Empirically, we can get some insight on the presence of a manipulation by inspecting the density of the forcing variable. 
In the presence of manipulation,
we would expect to see a discontinuity in the density of $S_i$ at the cutoff point.
The histogram of the forcing variable in Figure~\ref{Fig1:S} suggests that the distribution of monthly	per capita household income is quite smooth around the threshold,	although there is a small jump. 
\citet{firpo2014evidence} applied the McCrary test \citep{mccrary2008manipulation}  and found evidence of a discontinuity, and interpreted it as suggestive evidence that individuals manipulate their income by voluntarily reducing their labor supply in order to become eligible for the program. We also find some evidence of a discontinuity in the density of the forcing variable at the threshold by applying the McCrary test to our sample: the estimated log difference in the densities of the forcing variable at the threshold is $0.252$ with $s.e = 0.028$, which lead to a $p-$value $= 0.000$,  suggesting to reject the null hypothesis of continuity. 
A discontinuity in the marginal density of the forcing variable may indeed suggest the presence of manipulation of the forcing variable.
However, although for some families this may have been a strategic manipulation, leading to the forcing variable being confounded even after conditioning on covariates, for others the manipulation mechanism may have been ignorable, preserving unconfoundedness.
As discussed in Section
\ref{subsec:Usetapproach}, our approach
is able to deal with an ignorable manipulation and
is robust with respect
to non-ignorable manipulation by probabilistically down-weighing families for which our RD assumptions do not hold.

\subsection{Results} Table~\ref{tab:prop} shows the median and 95\%  highest density interval of the posterior distributions of the mixing probabilities and of the number of families by eligibility status in $\Uset$. At the posterior median there are $77\,643$ ($52.7\%$) families classified in the subpopulation $\Uset$.
The posterior median of the number of families who are eligible and not eligible to receive   Bolsa Fam\'ilia benefits are, respectively,  $75\,372$	 and $2\,269$. 
The estimated proportions of families in $\Usetm$ and $\Usetp$ are $42.6\%$ and $4.7\%$, respectively  (see also Section~D in the e-Appendix). 

\begin{table*}	\caption{Bolsa Fam\'ilia study: 
Mixture-model Bayesian analysis. Summary statistics of the posterior distributions of the mixing probabilities and of the number of families in $\Uset$ by eligibility status.}\label{tab:prop} 
\begin{center}$
\begin{array}{lrrr}
\hline
\vspace{-0.35cm}\\
& & \multicolumn{2}{c}{95\% \hbox{ HDI}}\\
\hbox{Estimand } &   \hbox{Median} &   \hbox{Lower bound} & \hbox{Upper bound} \\ 
\vspace{-0.35cm}\\
\hline
\vspace{-0.3cm}\\
\pi(\Usetm)  & 0.426 & 0.424 & 0.429\\ 
\vspace{-0.2cm}\\
\pi(\Usetp) & 0.047 & 0.046 & 0.047 \\
\vspace{-0.2cm}\\
\pi(\Uset)   & 0.527 & 0.524 & 0.529 \\ 
\\
N_{\Uset}   & 77\,643 & 77\,246  & 78\,038 \\ 
\\
\sum_{i \in \Uset} (1-Z_i)& 2\,269 & 2\,197 & 2\,343 \\  
\vspace{-0.2cm}\\
\sum_{i \in \Uset} Z_i  &  75\, 372& 74\,996  & 75\,746\\ 
\vspace{-0.35cm}\\
\hline
\end{array}
$
	\end{center}
\end{table*}

\begin{figure}[t]
\begin{center}
\begin{subfigure}[b]{0.485\textwidth}
\centering
\includegraphics[width=\textwidth]{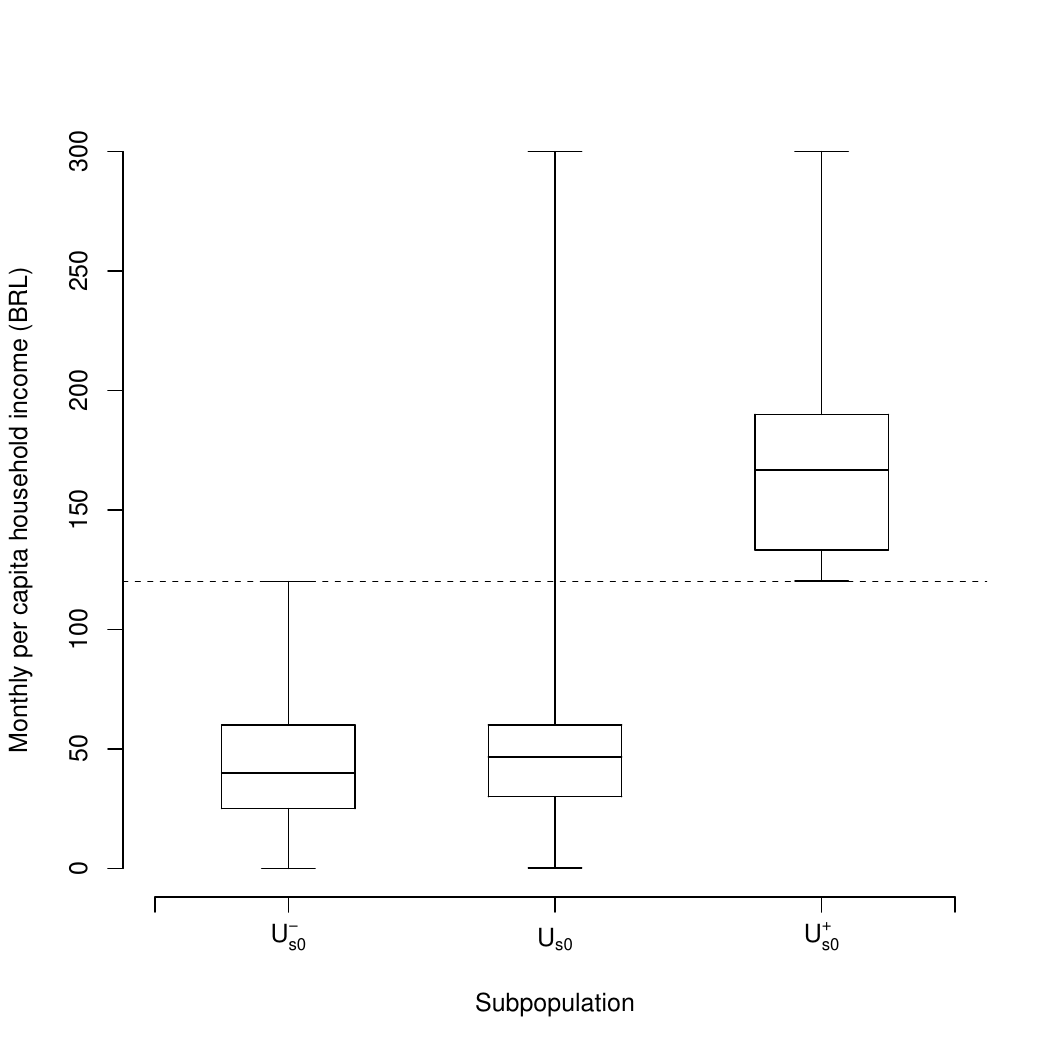}
\caption{Posterior median of the boxplots of monthly per capita household income by subpopulation (horizontal dash line at the threshold $s_0=120$).} 
\label{fig:PostBoxplotS}
\end{subfigure}
\hfill
\begin{subfigure}[b]{0.485\textwidth}
\centering
\includegraphics[width=\textwidth]{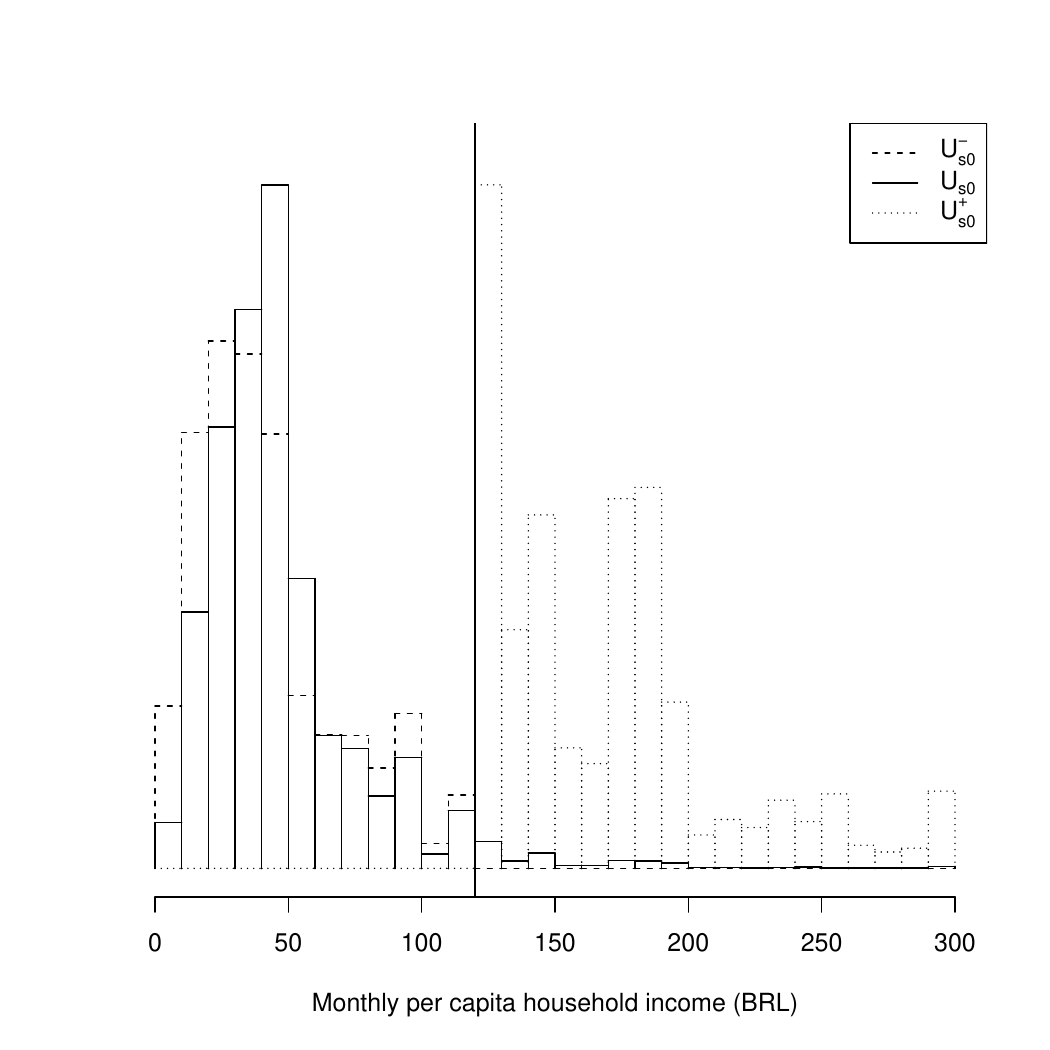}
\caption{Posterior median of the histogram of monthly per capita household income by subpopulation (vertical solid line at the threshold $s_0=120$).}
\label{fig:PostHistS}
\end{subfigure}
\end{center}
\caption{Bolsa Fam\'ilia study: Mixture-model Bayesian analysis. 	Posterior medians of the distribution of monthly per capita household income by subpopulation.  } \label{fig:PostS}
\end{figure}

Figure~\ref{fig:PostS} shows the posterior median of the distribution of the forcing variable, monthly per capita household income, for the sample classified by subpopulation membership: Figure~\ref{fig:PostBoxplotS} shows the boxplots constructed using the posterior median of the minimum, the first quartile, the median, the third quartile and the maximum of the forcing variable in each subpopulation, and Figure~\ref{fig:PostHistS} shows the histograms constructed using the posterior median of the densities of the forcing variable in each subpopulation.
As we can see in Figures~\ref{fig:PostBoxplotS} and \ref{fig:PostHistS}, 
the posterior median of the third quartile of monthly per capita household income for families in $\Uset$ is just $60$ BLR, and more than 96\% of families in $\Uset$ have realized values of monthly per capita household income falling below the threshold. Except for the long right tail and lower densities for values of $S_i \leq 30$, the distribution of the forcing variable for families in $\Uset$ is very similar to the distribution of the forcing variable for families in $\Usetm$. 

Table~\ref{tab:pUsetS} shows the posterior median and standard deviation of the probability of belonging to $\Uset$ by values of monthly per capita household income. 
The probability of belonging to $\Uset$ is higher than $0.416$  for families with monthly per capita household income below the threshold and reaches its maximum, equal to $0.628$, for families with monthly per capita household income falling in the interval $(30, 60]$. Families with monthly per capita household income above the threshold of 120 BLR have a generally decreasing probability of belonging to $\Uset$, with the posterior median of the probability of belonging to $\Uset$ being lower than $0.20$ for families with monthly per capita household income greater than $180$ BLR.
In summary, these results suggest that poorer families have a higher probability of being classified as members of the subpopulation $\Uset$: on average, $\Uset$ comprises even extremely poor families with values of monthly per capita household income below the threshold, and most of the ineligible families in $\Uset$ have values of $S$ relatively close to the threshold.
The results in Table~\ref{tab:pUsetS}, along with the high posterior probability of belonging to $\Uset$ (see Table~\ref{tab:prop}), also provide insights into the failure of the McCrary test. 
Specifically, Table~\ref{tab:pUsetS}  shows that the posterior median of the probability of belonging to
$\Uset$ is $0.494$ for families with a realized value of the forcing variable just below the threshold (between $110$ and $120$ BLR) and is $0.306$ for families with a realized value of the forcing variable just above the threshold (between $120$ and $130$ BLR).
Therefore, there are families with a realized value of the forcing variable falling in a neighborhood of the threshold for which the posterior probability of belonging to
$\Uset$ is small; that is, for these families, our model shows evidence that the forcing variable is confounded with the potential outcomes, even conditional on the covariates.
The presence of these families may explain the failure of the McCrary test.

\begin{table*}\caption{Bolsa Fam\'ilia study:  Mixture-model Bayesian analysis.  Posterior median and SD of the probability of belonging to $\Uset$ by monthly per capita household income.} \label{tab:pUsetS}
\begin{center}$
\begin{array}{rcc}
\hline
\multicolumn{1}{c}{S \hbox{ values} }& \multicolumn{1}{c}{\hbox{Median}} & \multicolumn{1}{c}{\hbox{SD}} \\ 
\hline
[0,  30]   & 0.449 & 0.001 \\ 
(30, 60]   & 0.628 & 0.002 \\ 
(60, 90]   & 0.523 & 0.003 \\ 
(90,  100] & 0.459 & 0.004\\
(100,  110] & 0.416 & 0.009\\
(110,  120] & 0.494 & 0.005\\
(120,  130] & 0.306 & 0.007\\
(130,  140] & 0.259 & 0.011\\
(140, 150] & 0.327 & 0.009\\ 
(150, 180] & 0.206 & 0.007 \\ 
(180, 240] & 0.192 & 0.006 \\ 
(240, 300] & 0.174 & 0.009 \\ 
\hline
\end{array}
$
\end{center}
\end{table*}

Table~A12 in the e-Appendix 
shows the posterior median of the sample means and standard deviations of the covariates by subpopulation membership. As we can see in Table~A12 
families belonging to the subpopulation $\Uset$  have, in general, background characteristics relatively similar to families who belong to the subpopulation $\Usetm$. The major difference between these two subpopulations concerns expenditures: families in $\Uset$ have, on average, much higher expenditures, and almost no families in $\Uset$ have zero expenditures.
Families belonging to $\Uset$ are younger, larger, comprise a higher number of children, and are more likely to comprise vulnerable people (namely,    pregnant women, breastfeeding women, or disabled people) than families belonging to $\Usetp$. Moreover, families in $\Uset$ are in worse living and economic conditions than families in $\Usetp$, and their household head is more likely to be unemployed than the household head of families in $\Usetp$.
Table~A13 in the e-Appendix 
shows the posterior median of sample averages and standard
deviations by eligibility status and some measure of covariate balance for families in $\Uset$. Table~A13 shows evidence that covariates are not well balanced between eligible and ineligible families in $\Uset$, and thus, we have conducted the analysis under our RD assumptions conditioning on the pre-treatment variables.

\begin{figure}[t]
	\begin{center}
		\includegraphics[width=7cm]{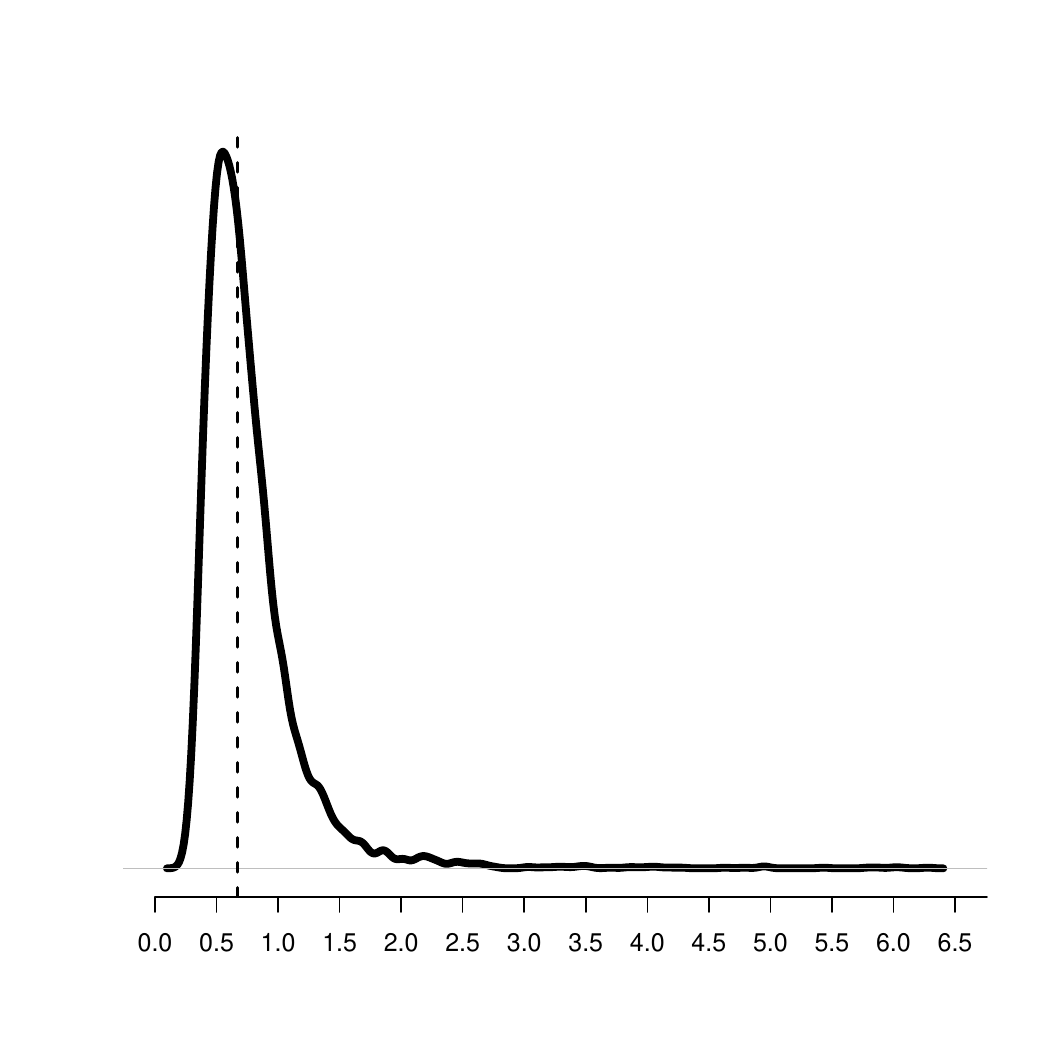} 
	\end{center}
	\caption{Bolsa Fam\'ilia study: Mixture-model Bayesian analysis. Posterior distributions of the finite sample causal relative risk (Dashed line = Posterior median). } \label{fig:mbares} 
\end{figure}

Figure~\ref{fig:mbares} and  Table~\ref{tab:mbares2} show the posterior distribution of the causal relative risk in Equation~\eqref{eq:rr} and some summary statistics of it. 
The posterior median of the causal relative risk is equal to $0.671$,  and the $95\%$  highest density interval is rather wide, including values from $0.273$ to $1.444$. The posterior probability that the causal relative risk is less than 1 is approximately $83.0\%$, but the posterior distribution of $RR_{\Uset}$ is skewed with a long right tail. 
Therefore, there is evidence that being eligible for Bolsa Fam\'ilia program based on per capita income reduces the risk of leprosy (see Section~F in the e-Appendix for results on the causal risk difference).  
We can hypothesize that this effect may be due in part to the improvement of living conditions, nutrition, education, and  health-care access resulting directly from BF benefits, and in part to the increase in vaccination and regular check-ups resulting from the conditions imposed to receive the benefits 
\citep{Pescarini2020}.
However, it is worth highlighting that our analysis informs us only about the effect of eligibility. 
In order to estimate the effect of the actual receipt of the Bolsa Fam\'ilia program, we would have to deal with complications arising in fuzzy RD designs. Here, we prefer to avoid dealing with such complications, which may mask the main contribution of our research work.

\begin{table*}	\caption{Bolsa Fam\'ilia study: Mixture-model Bayesian approach.		Summary statistics of the posterior distributions of the finite sample causal relative risk, $RR_{\Uset}$.} \label{tab:mbares2} 
\begin{center}$
	\begin{array}{ccccc}
		\hline
		\vspace{-0.35cm}\\
		&\multicolumn{3}{c}{95\% \hbox{ HDI}}  \\ 
		\vspace{-0.35cm}\\
		\cline{2-4}
		\vspace{-0.35cm}\\
		\hbox{Median}  & \hbox{Lower bound}& \hbox{Upper bound}\%& \hbox{(width)} &Pr\left(RR_{\Uset}  < 1\right)\\
		\vspace{-0.35cm}\\	\hline
		\vspace{-0.35cm}\\
		0.671   & 0.273  & 1.444 &(1.171)&  0.830\\ 
		\vspace{-0.35cm}\\
		\hline
	\end{array}
	$
\end{center}
\end{table*}

It is worth remarking that we derive the posterior distribution of the causal relative risk, $RR_{\Uset}$, shown in Figure~\ref{fig:mbares} and summarized in Table~\ref{tab:mbares2},  by
marginalizing over the uncertainty in whether each observation is a member of an unknown subpopulation $\Uset$. Each family's contribution to the posterior distribution of $RR_{\Uset}$  differs depending on the posterior probability of that family's inclusion in $\Uset$.  
Although closely related, this differs from the sensitivity analysis method introduced by \cite{BonviniKennedy2021}.  
 
\subsection{Model check}
We evaluate the influence of the model assumptions through model checking using posterior predictive checks. 
We compute a posterior predictive $p-$value (PPPV) for various discrepancy measures to assess whether the model- defined by the prior distributions and the likelihood - can successfully replicate the classification of the units in the subpopulation $\Uset$ and adequately preserve features of signal, noise, and signal-to-noise ratio of the outcome distribution in the subpopulation $\Uset$ (see  Section~G in the e-Appendix for details).
Results from these checks, displayed in Table~A15 in the e-Appendix, provide no evidence 
of systematic discrepancies between the observed and replicated data. 
Also, Figure~A6 in the e-Appendix shows that the proportion of correct outcome prediction in each replicated dataset is very high.%

\subsection{Sensitivity checks}
In order to investigate the behavior of the proposed Bayesian mixture model approach, we conduct several additional analyses. 
First, we investigate the characteristics of subpopulations chosen by applying bandwidth selectors for local polynomial RD point estimators based on continuity assumptions of the conditional distribution functions of the potential outcome given the forcing variable. Specifically, we exploit the data-driven bandwidth selectors based on MSE-optimal criteria proposed by \cite{CalonicoCattaneoTitiunik2014} using uniform and triangular kernel functions to construct local-polynomial estimators of order one and two.  We use two different MSE-optimal bandwidth selectors: one for below and one for above the threshold.	The four selected  MSE-optimal subpopulations are shown in Table~A16 in the e-Appendix. 
The sample size of the selected subpopulations, and especially the number of eligible families falling in these subpopulations (namely the left bandwidth), is strongly affected by both the kernel function and the order of the local-polynomial estimator. 
Moreover,  we find that some covariates are generally not well balanced between eligibility groups in the four subpopulations, and the evidence that the two eligibility groups are apart gets stronger as the selected subpopulation gets larger
(see Tables~A17-A20 and Figure~A7 in the e-Appendix).
%
%
These results suggest that standard analyses based on continuity assumptions, where local polynomial estimators with no covariate adjustment are used, might lead to misleading results 
(see  Section~K in the e-Appendix for results based on continuity assumptions). 
Recently, \cite{CalonicoEtAl:2019} proposed to augment standard local polynomial regression estimators to allow for covariate adjustment,
although inference is not provided for relative risks.


We also 
conduct a model-based causal Bayesian analysis within each of the four MSE-optimal subpopulations under Assumptions \ref{ass:loverlap}-\ref{ass:lunc}, and compare the results with those obtained using the mixture-model Bayesian approach we propose. 
For the two potential outcomes, $Y_i(0)$ and $Y_i(1)$, we specify the same models used in the mixture-model Bayesian analysis for families classified in an MSE-optimal subpopulation, $\Uset$, namely,  probit regression models with an intercept per eligibility group and equal slope coefficients between eligibility groups: $Pr(Y_i(z)=1 \mid \bX_i, i \in \Uset)= \Phi(\gamma_0,z+ \bX_i' \bgamma_{X})$, $z=0,1$. The same prior assumptions and distributions are also used (see Section~J in the e-Appendix for details). 
It is worth highlighting that this analysis does not account for the uncertainty in the subpopulation membership, as it is conducted conditionally on the pre-selected MSE-optimal subpopulation.

Table~\ref{tab:baUset} shows summary statistics of the posterior distributions of the causal relative risk in Equation~\eqref{eq:rr} in each of the four  MSE-optimal subpopulations 
(see Table~A21 in the e-Appendix for results on the causal risk difference). 
The posterior medians of the causal relative risk are greater than $1$, but the 95\% posterior credible intervals are rather wide, covering $1$. The posterior probability that the causal relative risk effect is less than $1$, that is, that eligibility for Bolsa Fam\'ilia benefits decreases leprosy rate, ranges between $18.4\%$ and $26.0\%$.   
Nevertheless, drawing causal conclusions from these results requires some care. First, the plausibility of the RD  Assumptions for the four  MSE-optimal subpopulations may be arguable because they are not defined to find subpopulations where Assumptions \ref{ass:loverlap}-\ref{ass:lunc} hold,  but to find an optimal balance between precision and bias at the threshold for local polynomial estimators. 
Moreover, neglecting uncertainty in subpopulation membership makes inference conclusions less credible.  

Although results from both the Bayesian analysis for specific MSE-optimal subpopulations 
and the mixture-model Bayesian approach 
do not allow us to derive firm conclusions about the effectiveness of the Bolsa Fam\'ilia program due to large posterior variability, we judge results based on the Bayesian mixture model approach more plausible, because they rely on observations exhibiting the most empirical basis for causal inference. Indeed the Bayesian mixture model approach leads to a posterior probability that the causal relative risk is less than one that is much greater than the Bayesian analysis based on specific MSE-optimal subpopulations, thus providing evidence that being eligible for the  Bolsa Fam\'ilia program is beneficial. This is in line with other studies on the health effects of the  Bolsa Fam\'ilia program \citep{Pescarini2020}.


\begin{table*}	\caption{Bolsa Fam\'ilia study: 
Summary statistics of the posterior distributions of the finite sample causal relative risk, $RR_{\Uset}$, for specific MSE-optimal subpopulations.}\label{tab:baUset}
\begin{center}
	$
  \hspace*{-2cm}
\begin{array}{l c ccccc}
\hline
\vspace{-0.35cm}\\
&&&\multicolumn{3}{c}{95\% \hbox{ HDI}}  \\ 
\cline{4-6}
\vspace{-0.35cm}\\
\hbox{Selector parameters} &		\hbox{Suppopulation: }  \Uset  & \hbox{Median}  & \hbox{LB} & \hbox{UB}& \hbox{(width)} &Pr\left(RR_{\Uset}  < 1\right)\\
\vspace{-0.35cm}\\
\hline
\vspace{-0.35cm}\\
\hbox{Uniform kernel }  (p=1) &	\{i \in \mathcal{U}:  43.5 \leq S_i\leq 163.9\}  & 1.196 &   0.646 & 2.021  & (1.375) & 0.256\\ 
\vspace{-0.15cm}\\
\hbox{Triangular kernel  } (p=1)   &		\{i \in \mathcal{U}:  39.0 \leq S_i\leq 174.5\} & 1.202 &  0.627 & 2.000&(1.373) &0.260\\
\vspace{-0.15cm}\\
\hbox{Uniform kernel } (p=2) &			\{i \in \mathcal{U}: \ \ 0.0 \leq S_i\leq 166.0\}&1.278   & 0.675 & 2.144 &(1.469) &0.184\\  
\vspace{-0.15cm}\\
\hbox{Triangular kernel } (p=2)  &		\{i \in \mathcal{U}: \ \ 0.0 \leq S_i\leq 165.6\}& 1.281 & 0.699 & 2.221 &(1.522)&0.194\\ 
\vspace{-0.35cm}\\
\hline
\end{array}
 \hspace*{-2cm}
$
\end{center}
\end{table*}

\section{Discussion} \label{sec:discussion}

In RD studies, selecting a subpopulation of units to use for drawing valid causal inference is a pervasive problem that presents inherent challenges. 
Specifically, in the innovative framework we have used to formally describe the Bolsa Fam\'ilia RD design as a local 
unconfounded experiment, 
causal inference concerns a group of families
where  a set of RD assumptions (Assumptions~\ref{ass:loverlap}, ~\ref{ass:lsutva}, ~\ref{ass:lunc}) hold.  
In practice, 
the subpopulation where these assumptions hold is unknown.

We have proposed a principled method to deal with this issue
with some distinct features:
it allows to propagate uncertainty in the subpopulation membership into the inferences on the causal effects;
the procedure works without imposing any constraints on the shape of the subpopulation.
Our approach also allows us to easily investigate the distributions of the forcing variable and of the baseline characteristics within each latent subgroup of units. 
This information is extremely valuable, providing key insights on the importance of adjusting for covariates in drawing inference on causal effects. In our application, we find evidence that covariates are not well balanced between eligible and ineligible families selected for causal inference, and thus, we have conducted the analysis under the local unconfoundedness assumption (Assumption \ref{ass:lunc}) conditioning on the pre-treatment variables, which is relatively simple in our Bayesian approach.

Additional strengths make the proposed Bayesian model-based finite mixture approach an appealing framework for the design and analysis of RD studies. 
It is   scalable to high-dimensional settings, 
and it performs well even when focus is on causal estimands different from average causal effects, as the causal relative risk for 
the leprosy indicator in the Bolsa Fam\`ilia study.  
The proposed approach can also be viewed as a Bayesian sensitivity analysis to the RD assumptions, where the proportion of units for which we can draw valid causal inference (or, equivalently, its complement to one) is the key sensitivity parameter. 
From this perspective,  our approach can be construed as a stochastic version of the mixture model for confounding recently proposed by  \cite{BonviniKennedy2021}. 
Rather than varying the proportion of confounded units as a sensitivity parameter,  our procedure stochastically selects observations in $\Uset$, for which assumptions hold.
Ultimately, we regard membership in the hypothetical subpopulation $\Uset$	as unknown for each observation and derive the posterior distribution of the causal effect by marginalizing over the uncertainty in whether each observation is a member. Thus, the resulting posterior distribution of the causal effect, where the contribution of each unit depends on the posterior probability of that unit's inclusion in $\Uset$, incorporates the uncertainty in $\Uset$ membership.

Our approach of prioritizing inclusion in $\Uset$ of observations for which the observed data suggest a reasonably high likelihood that local RD-SUTVA and local unconfoundedness hold allows us to also elegantly deal with issues arising from the manipulation of the forcing variable. Manipulation of the forcing variable is a potential threat to the validity of the RD design if it makes the unconfoundedness assumption untenable, but in our Bayesian mixture model-based approach, units for which local RD-SUTVA and/or local unconfoundedness doubtfully hold on an empirical basis will have a low probability of being members of the subpopulation $\Uset$, and thus, their contribution to the posterior distribution of the causal effect, $RR_{\Uset}$, will be small.

Although in this paper we opt for a Bayesian approach to inference, combining the ``design'' and ``analysis'' phases, 
we can also postpone the  ``analysis'' phase and use the proposed approach as an innovative and insightful method to select suitable subpopulations for causal inference in RD designs.  Specifically, we can view the selection of suitable subpopulations as a missing data problem and use the proposed approach to multiply impute subpopulation membership,   creating a set of completed membership datasets. 
For each of the complete membership datasets, we can use units classified in $\Uset$  using methods under either the design-based framework or the traditional framework.
Under the design-based framework, for each of the complete membership datasets, 
we can  draw inference on the causal effects for units in $\Uset$ using any mode of causal inference under the treatment assignment mechanism described by the invoked local unconfoundedness assumption, including design-based methods \cite[][]{LiMatteiMealli2015, MatteiMealli2016, BransonMealli2019, LicariMattei2020}.
In the traditional framework, under assumptions of continuity of conditional regression functions (or conditional distribution functions) of the outcomes given the forcing variable, for each of the complete membership datasets, units classified in the subpopulation $\Uset$	can either be directly used as an ``auxiliary'' subset of units from which  information is extrapolated to the cutoff, or be viewed as a properly selected pool from which  the ``auxiliary'' subset of units is constructed through the choice of a bandwidth defining a smoothing window around the threshold, using data-driven methods \cite[][]{CalonicoCattaneoFarrell2018} or Mean Square Error (MSE)-optimal criteria \cite[][]{ImbensKalyanaraman2012, CalonicoCattaneoTitiunik2014}.
Then, we can combine the complete-data inferences to form one inference that properly reflects uncertainty on subpopulation membership and possibly sampling variability \cite[][]{Rubin1987}. See Section~H in the e-Appendix.

Despite these appealing features, our approach may also raise questions about its performance and the sources of identification of subpopulation membership.  
We investigate these issues by conducting extensive simulation studies, where we find that our approach works well even in challenging scenarios.

We apply the Bayesian model-based mixture approach in the Bolsa Fam\`ilia study under specific parametric assumptions and relatively common prior distributions.
Nevertheless, our framework is general and may be implemented under alternative, possibly more flexible, model structures, including Bayesian non- or semi-parametric models \citep{li2023bayesian, linero2023and} for both the forcing variable and the outcome.
Moreover, we specify the prior for the subpopulation membership model to reflect our a priori
ignorance about the probability that any individual belongs to
each subpopulation by approximately setting each of their
prior probabilities at 1/3. Data are strongly informative about the subpopulation membership, leading to posterior distributions for the membership probabilities that are very different from the prior. 
It might be worthwhile to investigate, in future work, alternative prior distributions that depend on some measure of the distance of the realized value of the forcing variable from the threshold.

In the Bolsa fam\'ilia study, results from our Bayesian model-based mixture approach indicate that the subpopulation where the RD assumptions hold comprises 
	more than 50\% of our sample families. 
	For families classified as members of this subpopulation, we find evidence that being eligible for the Bolsa Fam\'ilia program reduces the risk of leprosy: although the posterior distribution of the causal relative risk 
   is skewed with a long right tail,    its posterior median is less than one  (approximately equal to $0.671$), and the posterior probability that it is less than one is high (approximately 83\%).

In this paper, we have analyzed the effect of Bolsa Fam\'ilia eligibility as a sharp RD design.
The extension of our approach to fuzzy RD designs is relatively straightforward and will be pursued by our future research agenda.

\clearpage

\begin{appendices}

The Appendix consists of ten Sections A-K.	In Section~A, we provide additional information on the Bolsa Fam\'ilia program and the dataset we use in the analysis. In Section~B, we provide computational details.  In Section~C, we describe the simulation studies.
In Sections~D-K, additional results on the Bolsa Fam\'ilia study are reported.
In Section~D, we compare the prior and the posterior distributions of the individual mixing probabilities. 
In Section~E, we investigate covariate distributions by subpopulation membership. 
In Section~F, we show results on the causal risk difference based on our Bayesian mixture model approach.
In Section~G, we provide details on the posterior predictive checks we use for model checking. 
In Section~H, we provide details on how to combine complete-membership inferences using multiply imputed subpopulation membership.
 In Section~I, we investigate covariate balance in specific MSE-optimal sub-populations. In Section~J, we provide computational details on the Bayesian approach applied on a MSE-optimal subpopulation. Finally, in Section~K, we show results based on  methods relying onthe continuity assumption.
 
 \clearpage

\section{Bolsa Fam\'ilia Program} \label{sec:BFP}
\subsection{Eligibility, Benefits, and Conditions} \label{sec:BFEligibility}

In order to have access to the Bolsa Fam\'ilia benefits, families must first register in the Brazilian National Registry for Social Programs (Cadastro Unico or CadUnico), which is a social registry established in July 2001  to facilitate the selection of beneficiaries for social assistance programs run by the Brazilian federal government, through the identification and socio-economic characterization of low-income Brazilian households.
Then, for a family to be eligible for Bolsa Fam\'ilia, it must have a monthly per capita income below the poverty line (R\$ 120 in 2008).  
However, the amount of benefits depends on the income per capita being under the extreme poverty line (R\$ 60 in 2008) or above and on the family composition. 

All families in extreme poverty receive a Basic Benefit (BB) (R\$ 62 in 2008), while families living in poverty (as opposed to extreme poverty) do not receive this benefit. The  Variable  Benefit (BV) (R\$  20 in 2008) is for families in poverty and extreme poverty whose composition includes women who are pregnant or lactating and/or children and adolescents up to 15 years of age.  The  Variable Youth Benefit (BVJ) (R\$  33 in 2008) is for families in poverty and extreme poverty whose composition includes youths  16 - 17  years of age.  The family of an adolescent beneficiary continues to regularly receive the payment benefit until the month of December of the year when the adolescent will turn 18.
Each family can receive up to a limited number of these benefits (3 for the BV and 2 for the BVJ in 2008).
The two thresholds for the definition of extreme poverty and poverty, the amount of these benefits, and the maximum number each family can receive varied over the years.

There are also schooling and health conditions attached to the payments. In terms of schooling, program participation requires school enrollment for children and teenagers and 85\% attendance for children between 6 and 15, and 75\% attendance for adolescents between 16 and 17. 
Regarding health, conditions include fulfillment of a vaccination and growth and development calendar for children under 7, prenatal care for pregnant women, and health monitoring for lactating women.
In principle, if one of these conditions is not met, the corresponding benefit is not given.
There are four consequences of noncompliance with conditions that progressively become more severe:  (1)  notification, (2)  blockage, (3) suspension, and  (4)  cancellation.
These consequences are related to the benefits that correspond to the unfulfilled condition.
Nevertheless, the Basic Benefit (BB) given to families in extreme poverty is suspended only when the income criterion is no longer met.
Each beneficiary receives a debit card, which is charged up every month unless the recipient has not met the necessary conditions, in which case (and after a couple of warnings) the payment is suspended.

\subsection{The Bolsa Fam\'ilia Dataset} \label{sec:BFdataset}
 
The 100 Million Brazilian Cohort is a large-scale linked cohort that aims to evaluate the impact of Bolsa Fam\'ilia and other social programs on health outcomes in Brazil \citep{Barreto2019TheCF}. For the analysis of the Bolsa Fam\'ilia  program, it linked data from i) the Brazilian National Registry for Social Programs Cadastro Único (CadUnico), which includes sociodemographic variables for the head of the family, household living conditions, and per capita income; ii) the Bolsa Fam\'ilia program Payroll Database, which includes information on Bolsa Fam\'ilia payments and verified conditions; iii) the Brazilian Notifiable Disease Registry (SINAN), which includes health outcome data. CadUnico and Bolsa Fam\'ilia  program data sets are deterministically linked using the Social Identification Number, whereas the SINAN data set is linked using the CIDACS-RL tool (https://gitHub.com/gcgbarbosa/cidacs-rl), which performs a two-step deterministic and probabilistic linkage based on five individual-level identifiers. In the first step, entries are deterministically linked. In the second step, entries that are not linked deterministically are then linked based on a similarity score	\citep{pita2018accuracy, ali2019administrative, CIDACS-RL}.
 
Table~\ref{tab2:stat} presents some summary statistics of the background variables and the outcome variable (leprosy) for the sub-sample we use in our analyses, which comprises   $N=147\,399$ families with monthly per capita household income in 2008 not greater than 300 BRL. For each variable, Table~\ref{tab2:stat} reports the grand mean calculated on the whole sample and the group means defined by the eligibility status, $Z_i$.

\begin{table*}\caption{Bolsa Fam\'ilia study: Summary Statistics.} \label{tab2:stat}
$$
\begin{array}{l rrr}
\hline 
& \hbox{Grand}   & \multicolumn{2}{c}{\hbox{Means}}\\
\hbox{Variable} & \hbox{Means} &  Z_i=0&Z_i=1\\
\multicolumn{1}{r}{\hbox{(Sample size)}}  & (147\,399)&(9\,179)&(138\,220)\\
\hline
\vspace{-0.3cm}\\
\multicolumn{4}{l}{\hbox{\textit{Household structure}}}\\
\vspace{-0.4cm}\\
\hbox{Min age}& 10.93 & 19.97 & 10.32 \\ 
\hbox{Mean age}&22.60 & 33.26 & 21.89 \\ 
\hbox{Household size} & 2.99 & 2.58 & 3.02 \\ 
\hbox{N. Children} & 1.34 & 0.72 & 1.38 \\ 
\hbox{N. Adults}& 1.60 & 1.55 & 1.60 \\ 
\hbox{Children not at school}& 0.04 & 0.02 & 0.04 \\ 
\hbox{Presence of weak people}& 0.22 & 0.16 & 0.23 \\ 
\vspace{-0.3cm}\\
\multicolumn{4}{l}{\hbox{\textit{Living and economic conditions}}}\\
\vspace{-0.4cm}\\
\hbox{Rural} & 0.39 & 0.22 & 0.40 \\ 
\hbox{Apartment}& 0.95 & 0.96 & 0.95 \\ 
\hbox{Home ownership: Homeowner} & 0.59 & 0.67 & 0.58 \\ 
\hbox{No rooms pc}& 1.60 & 2.04 & 1.57 \\ 
\hbox{House of bricks/row dirt }& 0.91 & 0.95 & 0.91 \\ 
\hbox{Water treatment} & 0.79 & 0.85 & 0.78 \\ 
\hbox{Water supply}& 0.63 & 0.77 & 0.63 \\ 
\hbox{Lighting}& 0.79 & 0.91 & 0.79 \\
\hbox{Bathroom fixture}& 0.61 & 0.47 & 0.62 \\ 
\hbox{Waste treatment}& 0.64  & 0.81 & 0.63 \\ 
\hbox{Zero PC expenditure}& 0.21 & 0.16 & 0.22 \\ 
\hbox{Log PC expenditure} & 2.95 & 3.91 & 2.89 \\ 
\hbox{Other programs}  &0.06 & 0.06 & 0.06 \\ 
\vspace{-0.3cm}\\
\multicolumn{4}{l}{\hbox{\textit{Household head's characteristics}}}\\
\vspace{-0.4cm}\\
\hbox{Male}& 0.86 & 0.83 & 0.87 \\ 
\hbox{Race: White}& 0.88 & 0.86 & 0.88 \\ 
\hbox{Primary/Middle Education} & 0.47 & 0.45 & 0.47 \\ 
\hbox{Occupation: Unemployed}&  0.49 & 0.37 & 0.49 \\ 
\vspace{-0.3cm}\\
\multicolumn{4}{l}{\hbox{\textit{Outcome variable}}}\\
\vspace{-0.4cm}\\
\hbox{Leprosy } (\permil)&     2.80  & 2.72&   2.80   \\
\hline
\end{array}
$$
\end{table*}

\section{A Bayesian Model-Based Finite Mixture Approach to the Selection of Subpopulations  $\Uset$: Computational Details} \label{sec:MixtureModel}

\subsection{Proposition.}  Suppose the RD assumptions (Assumptions 1-3) hold. Then, for all $i \in \Uset$, 
$$p(S_i \mid Z_i=z, Y^{obs}_i, \mathbf{X}_i)=p(S_i \mid Z_i=z, \mathbf{X}_i),$$
for $z=0,1$.
\medskip

\noindent \textbf{Proof}. For all $i \in \Uset$ and for all $A \subseteq \mathbb{R}$
\begin{eqnarray}
\Pr(S_i \in A \mid Z_i=1, Y_i, \mathbf{X}_i) &=& \Pr(S_i \in A \mid Z_i=1, Y_i(1), \mathbf{X}_i)\\
&=& \dfrac{\Pr(S_i \in A, Z_i=1 \mid  Y_i(1),\mathbf{X}_i)}{\Pr(Z_i=1 \mid Y_i(1),\mathbf{X}_i) )}\\
&=& \dfrac{\Pr(S_i \in A, S_i \leq s_0 \mid  Y_i(1),\mathbf{X}_i)}{\Pr(S_i\leq s_0 \mid Y_i(1),\mathbf{X}_i) )}\\
&=& \dfrac{\Pr(S_i \in A  \cap [0, s_0] \mid  Y_i(1),\mathbf{X}_i)}{\Pr(S_i\in [0,s_0] \mid Y_i(1),\mathbf{X}_i) }\\
&=& \dfrac{\Pr(S_i \in A \cap [0,s_0] \mid  \mathbf{X}_i)}{\Pr(S_i\in [0,s_0] \mid \mathbf{X}_i) }\\
&=& \Pr(S_i \in A \cap [0,s_0]  \mid S_i\in [0,s_0], \mathbf{X}_i)\\
&=&  \Pr(S_i \in A  \mid S_i\in [0,s_0], \mathbf{X}_i)\\
&=&  \Pr(S_i \in A  \mid Z_i=1, \mathbf{X}_i)
\end{eqnarray}		
where $(1)$ follows from the local RD-SUTVA,  $(2)$ follows from the definition of conditional distributions, the $(3)$ follows because $Z_i$ is a function of $S_i$: $Z_i= \mathbb{I}\{S_i \leq s_0\}$, $(4)$ holds by the definition of intersection of sets, $(5)$ follows from local unconfoundedness,  $(6)$ and $(7)$ follow from the definition of conditional distributions and simple set operations, and  $(8)$ holds by definition of $Z_i$.
Therefore, we have that 
$$p(S_i \mid Z_i=1, Y_i, \mathbf{X}_i)=p(S_i \mid Z_i=1, \mathbf{X}_i),$$
and similarly, $p(S_i \mid Z_i=0, Y_i, \mathbf{X}_i)=p(S_i \mid Z_i=0,  \mathbf{X}_i)$.

\subsection{Likelihood Functions} \label{sec:Mixturelikelihood}
We can write the  observed likelihood function in terms of the observed data as follows:
\begin{eqnarray*}
	\lefteqn{\mathscr{L}\left(\btheta\mid  \bX, \bS, \bZ, \bY\right)=
		\mathscr{L}\left(\balpha, \boeta^{-}, \bgamma^{-},  \boeta, \bgamma_0, \bgamma_1, \boeta^{+}, \bgamma^{+} \mid  \bX, \bS, \bZ, \bY \right)\propto}\\&
	\prod\limits_{i: Z_i=0}& \big[\pi_i(\Uset; \balpha) \,  p(S_i \mid \bX_i, i \in \Uset; \boeta) \, p(Y_i \mid \bX_i, i \in \Uset; \bgamma_0)+ \\&&
	\,\, \pi_i(\Usetp; \balpha) \, p(S_i \mid \bX_i; i \in \Usetp; \boeta^{+})  \, p(Y_i \mid S_i, \bX_i; i \in \Usetp, \bgamma^{+})\big]\times\\&
	\prod\limits_{i: Z_i=1}& \big[\pi_i(\Uset; \balpha) \,  p(S_i \mid \bX_i, i \in \Uset; \boeta) \, p(Y_i \mid \bX_i, i \in \Uset; \bgamma_1)+ \\&&
	\,\, \pi_i(\Usetm; \balpha) \, p(S_i \mid \bX_i; i \in \Usetm; \boeta^{-})  \, p(Y_i \mid S_i, \bX_i; i \in \Usetm; \bgamma^{-})\big]
\end{eqnarray*}

Let $G_i$ denote the subpopulation membership for unit $i$: $G_i\in \{\Usetm, \Uset, \Usetp\}$ and let $\bG$ be the $N-$ dimensional vector with $i-th$ element equal to $G_i$.
The subpopulation complete-data likelihood function, based on observing  $\bX, \bS, \bZ, \bY$ as well as the subpopulation membership is
\begin{eqnarray*}
\lefteqn{\mathscr{L}\left(\btheta\mid  \bX, \bS, \bZ, \bY, \bG\right)=}\\&&
\lefteqn{
\mathscr{L}\left(\balpha, \boeta^{-}, \bgamma^{-},  \boeta, \bgamma_0, \bgamma_1, \boeta^{+}, \bgamma^{+} \mid  \bX, \bS, \bZ, \bY,\bG \right)\propto}\\&&
\prod\limits_{i \in \Uset: Z_i=0} \pi_i(\Uset; \balpha) \,  p(S_i \mid \bX_i, i \in \Uset; \boeta) \, p(Y_i \mid \bX_i, i \in \Uset; \bgamma_0) \times \\&&
\prod\limits_{i\in \Usetp: Z_i=0}  \pi_i(\Usetp; \balpha) \, p(S_i \mid \bX_i; i \in \Usetp; \boeta^{+})  \, p(Y_i \mid S_i, \bX_i; i \in \Usetp, \bgamma^{+})\times\\&&
\prod\limits_{i \in \Uset: Z_i=1} \pi_i(\Uset; \balpha) \,  p(S_i \mid \bX_i, i \in \Uset; \boeta) \, p(Y_i \mid \bX_i, i \in \Uset; \bgamma_1)\times \\&&
\prod\limits_{i \in \Usetm: Z_i=1}  \pi_i(\Usetm; \balpha) \, p(S_i \mid \bX_i; i \in \Usetm; \boeta^{-})  \, p(Y_i \mid S_i, \bX_i; i \in \Usetm; \bgamma^{-})
\end{eqnarray*}

In the Bolsa Fam\'ilia study, we have
$\balpha=(\balpha^{-},   \balpha^{+})$ with $\balpha^{-}=(\alpha_0^{-}, \balpha_X^{-})$ and $\balpha^{+}= (\alpha_0^{+}, \balpha_X^{+})$; 
$\boeta=(\beta_0, \bbeta_X, \sigma^2)$; 
$\boeta^{-}=(\beta_0^{-}, \bbeta_X^{-}, \sigma^2_{-})$;
$\boeta^{+}=(\beta_0^{+}, \bbeta_X^{+}, \sigma^2_{+})$; 
$\bgamma_{0}=(\gamma_{0,0}, \bgamma_{X,0})$;
$\bgamma_{1}=(\gamma_{0,1}, \bgamma_{X,1})$;
$\bgamma^{-}=(\gamma_0^{-}, \gamma_1^{-}, \bgamma_X^{-})$; and
$\bgamma^{+}=(\gamma_0^{+}, \gamma_1^{+}, \bgamma_X^{+})$, and we impose $\bgamma_X^{-}=  \bgamma_X^{+}=\bgamma_{X,0} =  \bgamma_{X,1} \equiv  \bgamma_{X}$. Let $f(\cdot; \mu, \sigma^2)$ denote the pdf of a Normal distribution with mean $\mu$ and variance $\sigma^2$ and let $\Phi(\cdot)$ denote the cdf of a standard Normal distribution. Then, the observed-data likelihood is
\begin{eqnarray*}
	\lefteqn{\mathscr{L}\left(\btheta\mid  \bX, \bS, \bZ, \bY\right)=}\\
	\lefteqn{ 
		\mathscr{L}\left( \balpha^{-},   \balpha^{+}, \beta_0, \bbeta_X, \sigma^2,
		\beta_0^{-}, \bbeta_X^{-}, \sigma^2_{-},
		\beta_0^{+}, \bbeta_X^{+}, \sigma^2_{+},
		\gamma_{0,0}, \gamma_{0,1}, 
		\gamma_0^{-}, \gamma_1^{-}, 
		\gamma_0^{+}, \gamma_1^{+}, \bgamma_X, 
		\mid  \bX, \bS, \bZ, \bY \right)\propto}\\& 
	\prod\limits_{i: Z_i=0}&  \bigg[
	\left(1-\pi_i(\Usetp;  \balpha^{+}) -
	\pi_i(\Usetm;  \balpha^{-})\right) f\left(\log(S_i); \beta_0+ \bX_i' \bbeta_X, \sigma^2\right) \\& & \quad \Phi\left(\gamma_{0,0}+ \bX_i'\bgamma_{X}\right)^{Y_i}
	\left[1-\Phi\left(\gamma_{0,0}+ \bX_i'\bgamma_{X}\right)\right]^{1-Y_i}+ \\&& 
	\quad \pi_i(\Usetp; \balpha^{+}) 
	\dfrac{f\left(\log(S_i); \beta_0^{+} + \bX_i' \bbeta_X^{+}, \sigma^2_{+}\right)}{1-\Phi\left((\log(s_0) -\beta_0^{+} - \bX_i' \bbeta_X^{+})/\sqrt{\sigma^2_{+}}\right)}\\&&\quad \Phi\left(\gamma_{0}^{+}+\widetilde{S}_i\gamma_{1}^{+}+ \bX_i'\bgamma_{X}\right)^{Y_i}
	\left[1-\Phi\left(\gamma_{0}^{+}+\widetilde{S}_i\gamma_{1}^{+}+ \bX_i'\bgamma_{X}\right)\right]^{1-Y_i}
	\bigg]\times\\&
	\prod\limits_{i: Z_i=1}&  \bigg[
	\left(1-\pi_i(\Usetp; \balpha^{+}) -
	\pi_i(\Usetm;   \balpha^{-})\right) f\left(\log(S_i); \beta_0+ \bX_i' \bbeta_X, \sigma^2\right) \\&&\quad \Phi\left(\gamma_{0,1}+ \bX_i'\bgamma_{X}\right)^{Y_i}
	\left[1-\Phi\left(\gamma_{0,1}+ \bX_i'\bgamma_{X}\right)\right]^{1-Y_i}+ \\&&\quad  \pi_i(\Usetm;   \balpha^{-})  
	\dfrac{f\left(\log(S_i); \beta_0^{-} + \bX_i' \bbeta_X^{-}, \sigma^2_{-}\right)}{\Phi\left( (\log(s_0) - \beta_0^{-} - \bX_i' \bbeta_X^{-})/\sqrt{\sigma^2_{-}}\right)  }\\&&\quad  \Phi\left(\gamma_{0}^{-}+\widetilde{S}_i\gamma_{1}^{-}+ \bX_i'\bgamma_{X}\right)^{Y_i}
	\left[1-\Phi\left(\gamma_{0}^{-}+\widetilde{S}_i\gamma_{1}^{-}+ \bX_i'\bgamma_{X}\right)\right]^{1-Y_i}
	\bigg]
\end{eqnarray*}
where 
$$
\widetilde{S}_i= \dfrac{S_i - \overline{S}}{\sqrt{S^2_S}}= \dfrac{S_i - \sum_{i=1}^N S_i / N}{\sqrt{  \sum_{i=1}^N (S_i-\overline{S})^2 / (N-1)} }
$$
\noindent The subpopulation-complete data likelihood is
{\normalsize
	\begin{eqnarray*}
		\lefteqn{ \mathscr{L}_c\left(\btheta\mid  \bX, \bS, \bZ, \bY, \bG\right)\propto}\\
		\lefteqn{\mathscr{L}_c\left(\balpha^{-},  \balpha^{+}, \beta_0, \bbeta_X, \sigma^2, \beta_0^{-}, \bbeta_X^{-}, \sigma^2_{-}, \beta_0^{+}, \bbeta_X^{+}, \sigma^2_{+}, \gamma_{0,0}, \gamma_{0,1}, 
			\gamma_0^{+}, \gamma_1^{+}, \bgamma_X
			\mid  \bX, \bS, \bZ, \bY, \bG\right)\propto}\\&
		\prod\limits_{i \in \Usetp}&
		\pi_i(\Usetp;   \balpha^{+}) 
		\dfrac{f\left(\log(S_i); \beta_0^{+} + \bX_i' \bbeta_X^{+}, \sigma^2_{+}\right)}{1-\Phi\left((\log(s_0) -\beta_0^{+} - \bX_i' \bbeta_X^{+})/\sqrt{\sigma^2_{+}}\right)}\\&&  \Phi\left(\gamma_{0}^{+}+\widetilde{S}_i\gamma_{1}^{+}+ \bX_i'\bgamma_{X}\right)^{Y_i}
		\left[1-\Phi\left(\gamma_{0}^{+}+\widetilde{S}_i\gamma_{1}^{+}+ \bX_i'\bgamma_{X}\right)\right]^{1-Y_i}
		\times\\&
		\prod\limits_{i \in \Uset: Z_i=0}&
		\left(1-\pi_i(\Usetp;   \balpha^{+}) -
		\pi_i(\Usetm;  \balpha^{-})\right) f\left(\log(S_i); \beta_0+ \bX_i' \bbeta_X, \sigma^2\right) \\&&  \Phi\left(\gamma_{0,0}+ \bX_i'\bgamma_{X}\right)^{Y_i}
		\left[1-\Phi\left(\gamma_{0,0}+ \bX_i'\bgamma_{X}\right)\right]^{1-Y_i}
		\times \\&
		\prod\limits_{i \in \Uset: Z_i=1}   &
		\left(1-\pi_i(\Usetp;  \balpha^{+}) -
		\pi_i(\Usetm;  \balpha^{-})\right) f\left(\log(S_i); \beta_0+ \bX_i' \bbeta_X, \sigma^2\right)\\&&  \Phi\left(\gamma_{0,1}+ \bX_i'\bgamma_{X}\right)^{Y_i}
		\left[1-\Phi\left(\gamma_{0,1}+ \bX_i'\bgamma_{X}\right)\right]^{1-Y_i} \times\\&
		\prod\limits_{i \in \Usetm} & \pi_i(\Usetm;  \balpha^{-}) 
		\dfrac{f\left(\log(S_i); \beta_0^{-} + \bX_i' \bbeta_X^{-}, \sigma^2_{-}\right)}{\Phi\left( (\log(s_0) - \beta_0^{-} - \bX_i' \bbeta_X^{-})/\sqrt{\sigma^2_{-}}\right)  }\\&&  \Phi\left(\gamma_{0}^{-}+\widetilde{S}_i\gamma_{1}^{-}+ \bX_i'\bgamma_{X}\right)^{Y_i}
		\left[1-\Phi\left(\gamma_{0}^{-}+\widetilde{S}_i\gamma_{1}^{-}+ \bX_i'\bgamma_{X}\right)\right]^{1-Y_i}
	\end{eqnarray*}
}

\subsection{Prior Distributions} \label{sec:Mixturepriors}

In the Bolsa Fam\'ilia study, we assume that parameters are a priori independent and use 
proper prior distributions. Specifically, the prior distributions for the model for the mixing probabilities are $\balpha^{-} \equiv \left(\alpha_0^{-}, \balpha^{-}_X\right) \sim N_{p+1}\left(
\Phi^{-1}(2/3) \sqrt{1+\overline{X}'\overline{X}},\boldsymbol{0}_p),\right.$ $\left. \sigma^{2}_{\balpha^{-}} \mathbf{I}_{p+1}\right)$, and $\balpha^{+} \equiv \left(\alpha_0^{+}, \balpha^{+}_X\right) \sim N_{p+1}(\boldsymbol{0}, \sigma^{2}_{\balpha^{+}} \mathbf{I}_{p+1})$ independently, where $\Phi(\cdot)$ is the CDF of a standard Normal distribution, $\overline{X}$ is the mean vector of the covariates, and $\sigma^{2}_{\balpha^{-}}$ and $\sigma^{2}_{\balpha^{+}}$ are hyperparameters set at $1$. The prior distributions for the parameters of the models for the forcing variable  are  $\bbeta^{-} \equiv \left(\beta_0^{-},  \bbeta^{-}_X\right)  \sim N_{p+1}(\boldsymbol{0}, \sigma^{2}_{\bbeta^{-}} \mathbf{I}_{p+1})$,  $\bbeta^{+} \equiv \left(\beta_0^{+}, \bbeta^{+}_X\right)  \sim N_{p+1}(\boldsymbol{0}, \sigma^{2}_{\bbeta^{+}} \mathbf{I}_{p+1})$,   $\bbeta  \equiv \left(\beta_0,\bbeta_X\right)  \sim N_{p+1}(\boldsymbol{0}, \sigma^{2}_{\bbeta} \mathbf{I}_{p+1})$, where $\sigma^{2}_{\bbeta^{-}}$, $\sigma^{2}_{\bbeta^{+}}$ and $\sigma^{2}_{\bbeta}$  are hyperparameters set at $100$, and $\sigma^2_{-} \sim \mathrm{inv}-\chi^2(\nu_{-}, s^2_{-})$,  $\sigma^2_{+} \sim \mathrm{inv}-\chi^2(\nu_{+}, s^2_{+})$, and $\sigma^2 \sim \mathrm{inv}-\chi^2(\nu, s^2)$, where $\mathrm{inv}-\chi^2$ refers to the distribution of the inverse of a chi-squared random variable. We set  $\nu_{-}=\nu_{+}=\nu=3$ and $s^2_{-}=s^2_{+}=s^2=1/3$.
The prior distributions for the parameters of the outcome models are: $\bgamma^{-} \equiv \left(\gamma_0^{-}, \gamma_1^{-}\right) \sim N_{2}(\boldsymbol{0}, \sigma^{2}_{\bgamma^{-}} \mathbf{I}_{2})$, 
$\bgamma^{+} \equiv \left(\gamma_0^{+}, \gamma_1^{+} \right) \sim N_{2}(\boldsymbol{0}, \sigma^{2}_{\bgamma^{+}} \mathbf{I}_{2})$,  $\gamma_{0,z}  \sim N(0, \sigma^{2}_{\gamma_{0,z}})$, $z=0,1$, and      $\bgamma_X \sim N_{p}(\boldsymbol{0}, \sigma^{2}_{\bgamma_X} \mathbf{I}_{p})$,  
where $\sigma^{2}_{\bgamma^{-}}$,  $\sigma^{2}_{\bgamma^{+}}$  $\sigma^{2}_{\gamma_{0,z}}$, $z=0,1$, and  $\sigma^{2}_{\bgamma_X}$ are hyperparameters set at $1$.

\subsection{MCMC Algorithm with Data Augmentation for the Bolsa Fam\'ilia study}
\label{sec:MixtureMCMC}
Our MCMC Algorithm with Data Augmentation iteratively simulates $\btheta$ and $\bG$ given each other and the observed data.

Let $(\btheta, \bG)$ denote the current state of the chain, with
\begin{align*}
\btheta = \left[\balpha^{-}, \balpha^{+},(\beta_0, \bbeta_X, \sigma^2), (\beta_0^{-}, \bbeta_X^{-}, \sigma^2_{-}),
(\beta_0^{+}, \bbeta_X^{+}, \sigma^2_{+}), \gamma_{0,0}, \gamma_{0,1},   (\gamma_0^{-}, \gamma_1^{-}), (\gamma_0^{+}, \gamma_1^{+}), \bgamma_{X}\right]
\end{align*}

Given the parameter $\btheta$ and observed data, $\bX, \bS, \bZ, \bY$, we first draw the missing subpopulation membership indicator $G_i$ for all $i$. Then, given the imputed subpopulation complete data, $\bX, \bS, \bZ, \bY$, $\bG$, we draw the following sub-vectors of $\btheta$ in sequence, conditional on all others:
$\balpha^{-}$, $\balpha^{+}$, $(\beta_0, \bbeta_X)$,  $\sigma^2$, $(\beta_0^{-}, \bbeta_X^{-})$, $\sigma^2_{-}$,
$(\beta_0^{+}, \bbeta_X^{+})$, $\sigma^2_{+}$, $\gamma_{0,0}$, $\gamma_{0,1}$,  $(\gamma_0^{-}, \gamma_1^{-})$, $(\gamma_0^{+}, \gamma_1^{+})$, $\bgamma_{X}$.

Our MCMC algorithm uses the data augmentation and Gibbs sampling described in \cite{albert1993bayesian} for deriving the posterior distributions of the parameters of the probit regression models for the outcome; the data augmentation and Gibbs sampling for the posterior distributions of the parameters of the membership probabilities; and the Gibbs sampling based on the inverse distribution function method described in \cite{griffiths2002gibb} for the posterior distribution of the parameters of the truncated normal models for the forcing variable in $\Usetm$ and $\Usetm$. Finally, we use the standard Gibbs sampling for the parameters of the normal model with semi-conjugate priors for the posterior distribution of the parameters of the normal model for the forcing variable in $\Uset$. Below are the computational details.

{
\allowdisplaybreaks
\begin{enumerate}
	\item Draw the missing subpopulation membership indicator $G_i$ for all $i$ according to  
	$\Pr(G_i \mid \bX_i, S_i, Z_i=0, Y_i; \btheta)$: 
 \smallskip
	\begin{enumerate}
		\item[$-$] For families with $Z_i=0$
		\begin{eqnarray*}
			\lefteqn{  \Pr(G_i= \Usetm \mid \bX_i, S_i, Z_i=0, Y_i; \btheta)  = 0}\\
			\\
			\lefteqn{  \Pr(G_i= \Uset \mid \bX_i, S_i, Z_i=0, Y_i; \btheta)  = }\\&& 
			\Bigg[\left(1-\pi_i(\Usetp;   \balpha^{+}) -
			\pi_i(\Usetm;  \balpha^{-})\right) f\left(\log(S_i); \beta_0+ \bX_i' \bbeta_X, \sigma^2\right) \\&& \quad \Phi\left(\gamma_{0,0}+ \bX_i'\bgamma_{X}\right)^{Y_i}
			\left[1-\Phi\left(\gamma_{0,0}+ \bX_i'\bgamma_{X}\right)\right]^{1-Y_i}\Bigg] \Bigg/\\&& 
			\Bigg[\left(1-\pi_i(\Usetp;   \balpha^{+}) -
			\pi_i(\Usetm;  \balpha^{-})\right) f\left(\log(S_i); \beta_0+ \bX_i' \bbeta_X, \sigma^2\right) \\&&\quad   \Phi\left(\gamma_{0,0}+ \bX_i'\bgamma_{X}\right)^{Y_i}
			\left[1-\Phi\left(\gamma_{0,0}+ \bX_i'\bgamma_{X}\right)\right]^{1-Y_i} +\\&&\quad
			\pi_i(\Usetp;   \balpha^{+}) 
			\dfrac{f\left(\log(S_i); \beta_0^{+} + \bX_i' \bbeta_X^{+}, \sigma^2_{+}\right)}{1-\Phi\left( (\log(s_0) - \beta_0^{+} - \bX_i' \bbeta_X^{+})/\sqrt{\sigma^2_{+}}\right)  }\\&&  \quad  \Phi\left(\gamma_{0}^{+}+\widetilde{S}_i\gamma_{1}^{+}+ \bX_i'\bgamma_{X}\right)^{Y_i}
			\left[1-\Phi\left(\gamma_{0}^{+}+\widetilde{S}_i\gamma_{1}^{+}+ \bX_i'\bgamma_{X}\right)\right]^{1-Y_i}\Bigg]
		\end{eqnarray*}
		\begin{eqnarray*}
			\lefteqn{ \Pr(G_i= \Usetp \mid \bX_i, S_i, Z_i=0, Y_i; \btheta)  = }\\&& 
			\Bigg[ \pi_i(\Usetp;   \balpha^{+}) 
			\dfrac{f\left(\log(S_i); \beta_0^{+} + \bX_i' \bbeta_X^{+}, \sigma^2_{+}\right)}{1-\Phi\left( (\log(s_0) - \beta_0^{+} - \bX_i' \bbeta_X^{+})/\sqrt{\sigma^2_{+}}\right)  }\\&&  \quad  \Phi\left(\gamma_{0}^{+}+\widetilde{S}_i\gamma_{1}^{+}+ \bX_i'\bgamma_{X}\right)^{Y_i}
			\left[1-\Phi\left(\gamma_{0}^{+}+\widetilde{S}_i\gamma_{1}^{+}+ \bX_i'\bgamma_{X}\right)\right]^{1-Y_i}\Bigg] \Bigg/\\&& 
			\Bigg[\left(1-\pi_i(\Usetp;   \balpha^{+}) -
			\pi_i(\Usetm;  \balpha^{-})\right) f\left(\log(S_i); \beta_0+ \bX_i' \bbeta_X, \sigma^2\right) \\&&\quad   \Phi\left(\gamma_{0,0}+ \bX_i'\bgamma_{X}\right)^{Y_i}
			\left[1-\Phi\left(\gamma_{0,0}+ \bX_i'\bgamma_{X}\right)\right]^{1-Y_i} +\\&&\quad
			\pi_i(\Usetp;   \balpha^{+}) 
			\dfrac{f\left(\log(S_i); \beta_0^{+} + \bX_i' \bbeta_X^{+}, \sigma^2_{+}\right)}{1-\Phi\left( (\log(s_0) - \beta_0^{+} - \bX_i' \bbeta_X^{+})/\sqrt{\sigma^2_{+}}\right)  }\\&&  \quad  \Phi\left(\gamma_{0}^{+}+\widetilde{S}_i\gamma_{1}^{+}+ \bX_i'\bgamma_{X}\right)^{Y_i}
			\left[1-\Phi\left(\gamma_{0}^{+}+\widetilde{S}_i\gamma_{1}^{+}+ \bX_i'\bgamma_{X}\right)\right]^{1-Y_i}\Bigg]
		\end{eqnarray*}
		
		\item[$-$] For families with $Z_i=1$
		\begin{eqnarray*}
			\lefteqn{  \Pr(G_i= \Usetm \mid \bX_i, S_i, Z_i=1, Y_i; \btheta)  = }\\&& 
			\Bigg[ \pi_i(\Usetm;   \balpha^{-}) 
			\dfrac{f\left(\log(S_i); \beta_0^{-} + \bX_i' \bbeta_X^{-}, \sigma^2_{-}\right)}{\Phi\left((\log(s_0)- \beta_0^{-} - \bX_i' \bbeta_X^{-})/\sqrt{\sigma^2_{-}}\right)
   }\\&&  \quad  \Phi\left(\gamma_{0}^{-}+\widetilde{S}_i\gamma_{1}^{-}+ \bX_i'\bgamma_{X}\right)^{Y_i}
			\left[1-\Phi\left(\gamma_{0}^{-}+\widetilde{S}_i\gamma_{1}^{-}+ \bX_i'\bgamma_{X}\right)\right]^{1-Y_i}\Bigg] \Bigg/\\&& 
			\Bigg[\left(1-\pi_i(\Usetp;   \balpha^{+}) -
			\pi_i(\Usetm;  \balpha^{-})\right) f\left(\log(S_i); \beta_0+ \bX_i' \bbeta_X, \sigma^2\right) \\&&\quad   \Phi\left(\gamma_{0,1}+ \bX_i'\bgamma_{X}\right)^{Y_i}
			\left[1-\Phi\left(\gamma_{0,1}+ \bX_i'\bgamma_{X}\right)\right]^{1-Y_i} +\\&&\quad
			\pi_i(\Usetm;   \balpha^{-}) 
			\dfrac{f\left(\log(S_i); \beta_0^{-} + \bX_i' \bbeta_X^{-}, \sigma^2_{-}\right)}{\Phi\left((\log(s_0)- \beta_0^{-} - \bX_i' \bbeta_X^{-})/\sqrt{\sigma^2_{-}}\right)}\\&&  \quad  \Phi\left(\gamma_{0}^{-}+ \widetilde{S}_i\gamma_{1}^{-}+ \bX_i'\bgamma_{X}\right)^{Y_i}
			\left[1-\Phi\left(\gamma_{0}^{-}+\widetilde{S}_i\gamma_{1}^{-}+ \bX_i'\bgamma_{X}\right)\right]^{1-Y_i}\Bigg]
		\end{eqnarray*}
		\begin{eqnarray*}
			\lefteqn{  \Pr(G_i= \Uset \mid \bX_i, S_i, Z_i=1, Y_i; \btheta)  = }\\&& 
			\Bigg[\left(1-\pi_i(\Usetp;   \balpha^{+}) -
			\pi_i(\Usetm;  \balpha^{-})\right) f\left(\log(S_i); \beta_0+ \bX_i' \bbeta_X, \sigma^2\right) \\&& \quad \Phi\left(\gamma_{0,1}+ \bX_i'\bgamma_{X}\right)^{Y_i}
			\left[1-\Phi\left(\gamma_{0,1}+ \bX_i'\bgamma_{X}\right)\right]^{1-Y_i}\Bigg] \Bigg/\\&& 
			\Bigg[\left(1-\pi_i(\Usetp;   \balpha^{+}) -
			\pi_i(\Usetm;  \balpha^{-})\right) f\left(\log(S_i); \beta_0+ \bX_i' \bbeta_X, \sigma^2\right) \\&&\quad   \Phi\left(\gamma_{0,1}+ \bX_i'\bgamma_{X}\right)^{Y_i}
			\left[1-\Phi\left(\gamma_{0,1}+ \bX_i'\bgamma_{X}\right)\right]^{1-Y_i} +\\&&\quad
			\pi_i(\Usetm;   \balpha^{-}) 
\dfrac{f\left(\log(S_i); \beta_0^{-} + \bX_i' \bbeta_X^{-}, \sigma^2_{-}\right)}{\Phi\left((\log(s_0)- \beta_0^{-} - \bX_i' \bbeta_X^{-})/\sqrt{\sigma^2_{-}}\right)}			
   \\&&  \quad  \Phi\left(\gamma_{0}^{-}+\widetilde{S}_i\gamma_{1}^{-}+ \bX_i'\bgamma_{X}\right)^{Y_i}
			\left[1-\Phi\left(\gamma_{0}^{-}+\widetilde{S}_i\gamma_{1}^{-}+ \bX_i'\bgamma_{X}\right)\right]^{1-Y_i}\Bigg]
			\\
			\\
			\lefteqn{ \Pr(G_i= \Usetp \mid \bX_i, S_i, Z_i=1, Y_i; \btheta)  = 0}\\
		\end{eqnarray*}
		
	\end{enumerate}

	\item Sample the coefficients $\balpha^{-}$ and $\balpha^{+}$
	\begin{enumerate}
		\item Sample the latent variables $G_i^\ast(-)$ and $G_i^\ast(+)$:  Sample the latent variable $G_i^\ast(-)$
		from $N(\alpha_0^{-} + \bX_i'\balpha_X^{-}, 1)$ truncated
		to $(-\infty, 0]$ if $G_i = \Usetm$ and to $(0, +\infty)$ if $G_i\neq  \Usetm$; sample the latent variable $G_i^\ast(+)$
		from $N(\alpha_0^{+} + \bX_i'\balpha_X^{+}, 1)$ truncated
		to $(-\infty, 0]$ if $G_i = \Usetp$ and to $(0, +\infty)$ if $G_i\neq  \Usetp$;
		\item Sample the coefficients $\balpha^{-}$ from
		$N_{p+1}\left(\bmu(\balpha^{-}), \Sigma(\balpha^{-})\right)$, where
		$$
		\Sigma(\balpha^{-}) = \left[\dfrac{1}{\sigma^2_{\balpha^{-}}}\mathbf{I}_{p+1} + [\boldsymbol{1},\bX]'[\boldsymbol{1},\bX] \right]^{-1} $$
		and $$
		\bmu(\balpha^{-})=\Sigma(\balpha^{-})  \left[
		\dfrac{1}{\sigma^2_{\balpha^{-}}}\mathbf{I}_{p+1} \boldsymbol{0} + [\boldsymbol{1},\bX]'G_i^\ast(-)\right]=
		\Sigma(\alpha^{-})   [\boldsymbol{1},\bX]'\bG^\ast(-) 
		$$
		\item Sample the coefficients $\balpha^{+}$ from
		$N_{p+1}\left(\bmu(\balpha^{+}), \Sigma(\balpha^{+})\right)$, where
		$$
		\Sigma(\balpha^{+}) = \left[\dfrac{1}{\sigma^2_{\balpha^{+}}}\mathbf{I}_{p+1} + [\boldsymbol{1},\bX]_{\Usetp \cup \Uset}'[\boldsymbol{1},\bX]_{\Usetp \cup \Uset} \right]^{-1} $$
		and 
		$$
		\bmu(\balpha^{+})=\Sigma(\balpha^{+})  \left[
		\dfrac{1}{\sigma^2_{\balpha^{+}}}\mathbf{I}_{p+1} \boldsymbol{0} + [\boldsymbol{1},\bX]_{\Usetp \cup \Uset}'\bG^\ast(+)_{\Usetp \cup \Uset}\right]=
		\Sigma(\balpha^{+})   [\boldsymbol{1},\bX]_{\Usetp \cup \Uset}'\bG^\ast(+)_{\Usetp \cup \Uset}
		$$
	\end{enumerate}
	\item Sample the coefficients $\bbeta^{-} = (\beta_0^{-}, \bbeta^{-}_{X})$ and the scale parameter $\sigma^2_{-}$

 \begin{enumerate}
		\item Compute the values of the latent variable $Q^{-}_{i}$ for all $i \in \Usetm$:
  $$
Q^{-}_{i} = \beta_0^{-}+\bX_i'\bbeta^{-}_{X} + \sqrt{\sigma^2_{-}}
\Phi^{-1}\left(\dfrac{\Phi\left(\frac{\log(S_i) - \beta_0^{-}-\bX_i'\bbeta^{-}_{X}}{\sqrt{\sigma^2_{-}}}\right)}{
\Phi\left(\frac{\log(s_0) - \beta_0^{-}-\bX_i'\bbeta^{-}_{X}}{\sqrt{\sigma^2_{-}}}\right)}\right) \qquad i \in \Usetm
 $$
 Define $\bQ^{-}_{\Usetm} = [Q_i^{-}]_{i \in \Usetm}$
 \item Sample the coefficients $\bbeta^{-} = (\beta_0^{-}, \bbeta^{-}_{X})$  from $N_{p+1}\left(\bmu(\bbeta^{-}), \Sigma(\bbeta^{-})\right)$ where
	$$
	\Sigma(\bbeta^{-}) = \left[\dfrac{1}{\sigma^2_{\bbeta^{-}}}\mathbf{I}_{p+1} + \dfrac{1}{\sigma^2_{-}}[\boldsymbol{1},\bX]_{\Usetm}'[\boldsymbol{1},\bX]_{\Usetm} \right]^{-1} $$
	and $$
	\bmu(\bbeta^{-})=\Sigma(\bbeta^{-})  \left[
	\dfrac{1}{\sigma^2_{\bbeta^{-}}}\mathbf{I}_{p+1} \boldsymbol{0} + \dfrac{1}{\sigma^2_{-}}[\boldsymbol{1},\bX]_{\Usetm}'\bQ^{-}_{\Usetm}\right]= \Sigma(\bbeta^{-})   \dfrac{1}{\sigma^2_{-}}[\boldsymbol{1},\bX]_{\Usetm}'\bQ^{-}_{\Usetm} 
	$$ 
	\item Sample  $\sigma^2_{-}$ from inv$-\chi^2\left(\nu(\sigma^2_{-}), s^2(\sigma^2_{-})\right)$, where
	$$
	\nu(\sigma^2_{-}) = \nu_{-} + \sum_{i=1}^N \one\{G_i=\Usetm\}
	$$
	and
	$$
	s^2(\sigma^2_{-}) = \dfrac{1}{\nu(\sigma^2_{-}) }
	\left[\sum_{i \in \Usetm} \left(Q^{-}_i - \beta_0^{-} - \bX_i'\bbeta_{X}^{-}\right)^2 + \nu_{-} s^2_{-}\right]
	$$
\end{enumerate}	
\item  Sample the coefficients $\bbeta^{+} = (\beta_0^{+}, \bbeta^{+}_{X})$ and the scale parameter $\sigma^2_{+}$

 \begin{enumerate}
 \item  Compute the values of the latent variable $Q^{+}_{i}$ for all $i \in \Usetp$:
  $$
Q^{+}_{i} = \beta_0^{+}+\bX_i'\bbeta^{+}_{X} + \sqrt{\sigma^2_{+}}
\Phi^{-1}\left(\dfrac{\Phi\left(\frac{\log(S_i) - \beta_0^{+}-\bX_i'\bbeta^{+}_{X}}{\sqrt{\sigma^2_{+}}}\right)-\Phi\left(\frac{\log(s_0) - \beta_0^{+}-\bX_i'\bbeta^{+}_{X}}{\sqrt{\sigma^2_{+}}}\right)}{1-
\Phi\left(\frac{\log(s_0) - \beta_0^{+}-\bX_i'\bbeta^{+}_{X}}{\sqrt{\sigma^2_{+}}}\right)}\right) \qquad i \in \Usetp
 $$
 Define $\bQ^{+}_{\Usetp} = [Q_i^{+}]_{i \in \Usetp}$
	\item Sample the coefficients $\bbeta^{+} = (\beta_0^{+}, \bbeta^{+}_{X})$ from $N_{p+1}\left(\bmu(\bbeta^{+}), \Sigma(\bbeta^{+})\right)$ where
	$$
	\Sigma(\bbeta^{+}) = \left[\dfrac{1}{\sigma^2_{\bbeta^{+}}}\mathbf{I}_{p+1} + \dfrac{1}{\sigma^2_{+}}[\boldsymbol{1},\bX]_{\Usetp}'[\boldsymbol{1},\bX]_{\Usetp} \right]^{-1} $$
	and $$
	\bmu(\bbeta^{+})=\Sigma(\bbeta^{+})  \left[
	\dfrac{1}{\sigma^2_{\bbeta^{+}}}\mathbf{I}_{p+1} \boldsymbol{0} + \dfrac{1}{\sigma^2_{+}}[\boldsymbol{1},\bX]_{\Usetp}' \bQ^{+}_{\Usetp}\right]= 
	\Sigma(\bbeta^{+})  \dfrac{1}{\sigma^2_{+}}[\boldsymbol{1},\bX]_{\Usetp}' \bQ^{+}_{\Usetp} 
	$$ 
	\item Sample the variance $\sigma^2_{+}$ from inv$-\chi^2\left(\nu(\sigma^2_{+}), s^2(\sigma^2_{+})\right)$, where
	$$
	\nu(\sigma^2_{+}) = \nu_{+} + \sum_{i=1}^N \one\{G_i=\Usetp\}
	$$
	and
	$$
	s^2(\sigma^2_{+}) = \dfrac{1}{\nu(\sigma^2_{+}) }
	\left[\sum_{i \in \Usetp} \left(\bQ^{+}_i - \beta_0^{+} - \bX_i'\bbeta_{X}^{+}\right)^2 + \nu_{+} s^2_{+}\right]
	$$
	\end{enumerate}

	\item Sample the coefficients $\bbeta = (\beta_0, \bbeta_{X})$ from $N_{p+1}\left(\bmu(\bbeta), \Sigma(\bbeta)\right)$ where
	$$
	\Sigma(\bbeta) = \left[\dfrac{1}{\sigma^2_{\bbeta}}\mathbf{I}_{p+1} + \dfrac{1}{\sigma^2}[\boldsymbol{1},\bX]_{\Uset}'[\boldsymbol{1},\bX]_{\Uset} \right]^{-1} $$
	and $$
	\bmu(\bbeta)=\Sigma(\bbeta)  \left[
	\dfrac{1}{\sigma^2_{\bbeta}}\mathbf{I}_{p+1} \boldsymbol{0} + \dfrac{1}{\sigma^2}[\boldsymbol{1},\bX]_{\Uset}' \log(\bS)_{\Uset}\right]=   \Sigma(\bbeta)  \dfrac{1}{\sigma^2}[\boldsymbol{1},\bX]_{\Uset}' \log(\bS)_{\Uset} 
	$$ 
	\item Sample the variance $\sigma^2$ from inv$-\chi^2\left(\nu(\sigma^2), s^2(\sigma^2)\right)$, where
	$$
	\nu(\sigma^2) = \nu + \sum_{i=1}^N \one\{G_i=\Uset\}
	$$
	and
	$$
	s^2(\sigma^2) = \dfrac{1}{\nu(\sigma^2) }
	\left[\sum_{i \in \Uset} \left(\log(S_i) - \beta_0 - \bX_i'\bbeta_{X}\right)^2 + \nu s^2\right]
	$$
	
	\item  Sample the coefficients $\bgamma^{-}=(\gamma_0^{-}, \gamma_1^{-}), \bgamma^{+}=(\gamma_0^{+}, \gamma_1^{+}), \gamma_{0,0}, \gamma_{0,1}$  

\begin{enumerate}
\item  Sample the latent variable $\bY^\ast$

\begin{enumerate}
\item  For $i \in \Usetm$, sample the latent variable $Y_i^\ast$
from $N(\gamma_0^{-} +\gamma_1^{-} \widetilde{S}_i +  \bX_i'\bgamma_X, 1)$ truncated to $[0, +\infty)$ if $Y_i=1$ and to $(-\infty, 0]$ if $Y_i = 0$;
\item  For $i \in \Usetp$, sample the latent variable $Y_i^\ast$
from $N(\gamma_0^{+} +\gamma_1^{+} \widetilde{S}_i +  \bX_i'\bgamma_X, 1)$ truncated to $[0, +\infty)$ if $Y_i=1$ and to $(-\infty, 0]$ if $Y_i = 0$;
\item  For $i \in \Uset$ with $Z_i=0$, sample the latent variable $Y_i^\ast$
from $N(\gamma_{0,0} + \bX_i'\bgamma_X, 1)$ truncated to $[0, +\infty)$ if $Y_i=1$ and to $(-\infty, 0]$ if $Y_i = 0$;
\item  For $i \in \Uset$ with $Z_i=1$, sample the latent variable $Y_i^\ast$
from $N(\gamma_{0,1} + \bX_i'\bgamma_X, 1)$ truncated to $[0, +\infty)$ if $Y_i=1$ and to $(-\infty, 0]$ if $Y_i = 0$;
\end{enumerate} 
		
		\item    Sample the coefficients $\bgamma^{-}=(\gamma_0^{-}, \gamma^{-}_{1})$ from $N_{2}\left(\bmu(\bgamma^{-}), \Sigma(\bgamma^{-})\right)$ where
		$$
		\Sigma(\bgamma^{-}) = \left[\dfrac{1}{\sigma^2_{\bgamma^{-}}}\mathbf{I}_{2} +  [\boldsymbol{1},\bS]_{\Usetm}'[\boldsymbol{1},\bS]_{\Usetm} \right]^{-1} $$
		and $$
		\bmu(\bgamma^{-})=\Sigma(\bgamma^{-})  \left[
		\dfrac{1}{\sigma^2_{\bgamma^{-}}}\mathbf{I}_{2} \boldsymbol{0} +  [\boldsymbol{1},\bS]_{\Usetm}' [\bY^\ast - \bX \bgamma_X]_{\Usetm}\right]=  \Sigma(\bgamma^{-})     [\boldsymbol{1},\bS]_{\Usetm}' [\bY^\ast - \bX \bgamma_X]_{\Usetm} 
		$$ 
		
		\item    Sample the coefficients $\bgamma^{+}=(\gamma_0^{+}, \gamma^{+}_{1})$ from $N_{2}\left(\bmu(\bgamma^{+}), \Sigma(\bgamma^{+})\right)$ where
		$$
		\Sigma(\bgamma^{+}) = \left[\dfrac{1}{\sigma^2_{\bgamma^{+}}}\mathbf{I}_{2} +  [\boldsymbol{1},\bS]_{\Usetm}'[\boldsymbol{1},\bS]_{\Usetp} \right]^{-1} $$
		and $$
		\bmu(\bgamma^{+})=\Sigma(\bgamma^{+})  \left[
		\dfrac{1}{\sigma^2_{\bgamma^{+}}}\mathbf{I}_{2} \boldsymbol{0} +  [\boldsymbol{1},\bS]_{\Usetp}' [\bY^\ast - \bX \bgamma_X]_{\Usetp}\right]=  \Sigma(\bgamma^{+})     [\boldsymbol{1},\bS]_{\Usetp}' [\bY^\ast - \bX \bgamma_X]_{\Usetp} 
		$$ 
		
		\item    Sample the coefficient $\gamma_{0,0}$ from $N\left(\mu(\gamma_{0,0}), \sigma^2(\gamma_{0,0})\right)$ where
		$$
		\sigma^2(\gamma_{0,0}) =  \left[\dfrac{1}{\sigma^2_{\gamma_{0,0}}} + \sum_{i \in \Uset} (1-Z_i) \right]^{-1} $$
		and $$
		\mu(\gamma_{0,0})= \sigma^2(\gamma_{0,0}) \left[
		\dfrac{0}{\sigma^2_{\gamma_{0,0}}}+
		\sum_{i \in\Uset}(Y_i^{\ast}- \bX_i' \bgamma_X) (1-Z_i)\right]=   \sigma^2(\gamma_{0,0}) 
		\sum_{i \in\Uset}(Y_i^{\ast}- \bX_i' \bgamma_X) (1-Z_i)   
		$$ 
		
		\item    Sample the coefficient $\gamma_{0,1}$ from $N\left(\mu(\gamma_{0,1}), \sigma^2(\gamma_{0,1})\right)$ where
		$$
		\sigma^2(\gamma_{0,1}) =  \left[\dfrac{1}{\sigma^2_{\gamma_{0,1}}} + \sum_{i \in \Uset} Z_i \right]^{-1} $$
		and $$
		\mu(\gamma_{0,1})= \sigma^2(\gamma_{0,1}) \left[
		\dfrac{0}{\sigma^2_{\gamma_{0,1}}}+
		\sum_{i \in\Uset}(Y_i^{\ast}- \bX_i' \bgamma_X) Z_i\right]=   \sigma^2(\gamma_{0,1}) 
		\sum_{i \in\Uset}(Y_i^{\ast}- \bX_i' \bgamma_X) Z_i   
		$$ 
	\end{enumerate}

	\item  Sample the coefficients $\bgamma_X$  
	
	\begin{enumerate}
		\item  Sample the latent variable $\bY^\ast$
		
		\begin{enumerate}
			\item  For $i \in \Usetm$, sample the latent variable $Y_i^\ast$
			from $N(\gamma_0^{-} +\gamma_1^{-} \widetilde{S}_i +  \bX_i'\bgamma_X, 1)$ truncated to $[0, +\infty)$ if $Y_i=1$ and to $(-\infty, 0]$ if $Y_i = 0$;
			\item  For $i \in \Usetp$, sample the latent variable $Y_i^\ast$
			from $N(\gamma_0^{+} +\gamma_1^{+} \widetilde{S}_i +  \bX_i'\bgamma_X, 1)$ truncated to $[0, +\infty)$ if $Y_i=1$ and to $(-\infty, 0]$ if $Y_i = 0$;
			\item  For $i \in \Uset$ with $Z_i=0$, sample the latent variable $Y_i^\ast$
			from $N(\gamma_{0,0} + \bX_i'\bgamma_X, 1)$ truncated to $[0, +\infty)$ if $Y_i=1$ and to $(-\infty, 0]$ if $Y_i = 0$;
			\item  For $i \in \Uset$ with $Z_i=1$, sample the latent variable $Y_i^\ast$
			from $N(\gamma_{0,1} + \bX_i'\bgamma_X, 1)$ truncated to $[0, +\infty)$ if $Y_i=1$ and to $(-\infty, 0]$ if $Y_i = 0$;
		\end{enumerate} 
		\item Let  $\widetilde{Y}^\ast$ be a $N-$dimensional vector with $i$th element equal to
		$$\widetilde{Y}_i^\ast= \begin{cases}
			Y_i^\ast - \gamma_{0}^{-} - \gamma_{1}^{-} \widetilde{S}_i & \hbox{if } G_i= \Usetm\\
			Y_i^\ast - \gamma_{0}^{+} - \gamma_{1}^{+} \widetilde{S}_i& \hbox{if } G_i= \Usetp\\
			Y_i^\ast - \gamma_{0,0} & \hbox{if } G_i= \Uset, Z_i=0\\
			Y_i^\ast - \gamma_{0,1} & \hbox{if } G_i= \Uset, Z_i=1
		\end{cases} 
		$$
		
		\item    Sample the coefficients $\bgamma_X$ from $N_{p}\left(\bmu(\bgamma_X), \Sigma(\bgamma_X)\right)$ where
		$$
		\Sigma(\bgamma_X) = \left[\dfrac{1}{\sigma^2_{\bgamma_X}}\mathbf{I}_{p} +  \bX'\bX \right]^{-1} $$
		and 
		$$
		\bmu(\bgamma_X)=\Sigma(\bgamma_X)  \left[
		\dfrac{1}{\sigma^2_{\bgamma_X}}\mathbf{I}_{p} \boldsymbol{0} +  \bX'\widetilde{Y}^\ast \right]= \Sigma(\bgamma_X)   \bX'\widetilde{Y}^\ast.  
		$$ 
	\end{enumerate}
\end{enumerate}
}

\section{Simulation Studies} \label{sec:sim}

We conduct extensive simulation studies to investigate how our approach works in challenging scenarios. The primary aim of our simulation studies is to shed light on the sources of identification of subpopulation membership. 
In our Bayesian model-based finite mixture approach to clustering, the covariates and the outcome model play a crucial role in classifying units into the three subpopulations: $\Uset$, $\Usetm$, and $\Usetp$.
Units belonging to the subpopulation $\Uset$ should have characteristics such that their probability to fall on either side of the threshold is sufficiently far away from zero and one. 
In the subpopulation $\Uset$, local-SUTVA (Assumption 2) and local unconfoundedness (Assumption 3) hold, and thus, potential outcomes and the forcing variable are structurally and statistically independent for units belonging to $\Uset$. For units belonging to either  $\Usetm$ or $\Usetp$, local-SUTVA or local unconfoundedness does not hold, implying that potential outcomes depend on the forcing variable.


\subsection*{Data Generating Processes} \label{sec:dgp}

We consider three simulation settings. The first setting focuses on  RD studies where there are no units in $\Uset$; the second setting aims to investigate the role of covariates for classifying units into the three subpopulations, $\Uset$, $\Usetm$, and $\Usetp$; and the third setting explores the role of the dependency between the subpopulation membership and the forcing variable.
We consider a large sample size $N=10\,000$ to avoid sampling variability issues, and we set $s_0=0$.

\subsubsection*{Data Generating Process: Setting 1}
The first simulation setting is an extreme scenario; we consider an RD study where there are no units in $\Uset$, i.e., $\Uset$ is empty. A key concern is whether our procedure is able to recognize that the subpopulation is empty or whether, instead, it wrongly creates a ``fake'' subpopulation $\Uset$, selecting observations such that the dependency between the outcome and the forcing variable is broken.

Under this first setting, data are generated using the following process:
$$
\begin{array}{ccc}
\multicolumn{3}{c}{\pi_i(\Usetm) = \pi, \qquad 
	\pi_i(\Usetp) = 1-\pi, \qquad	\pi_i(\Uset) =0;
}\\
\\
S_i \mid   i \in \Usetm \sim TN\left(\beta_0^{-}; \sigma^2_{-}; -\infty, s_0 \right),&\quad &S_i \mid   i \in \Usetp \sim TN\left(\beta_0^{+}; \sigma^2_{+}; s_0, \infty \right); \\
\\
Y_i(s) \mid    i \in \Usetm \sim  N\left(\gamma_{0}^{-} + (s-s_0) \gamma_{1}^{-};    \tau^2_{-} \right),&\quad& 
Y_i(s) \mid   i \in \Usetp  \sim  N\left(\gamma_{0}^{+} + (s-s_0) \gamma_{1}^{+}; \tau^2_{+} \right),
\end{array}
$$
where $TN(\mu, \sigma^2, a, b)$ denotes a truncated normal distribution with mean equal to $\mu$ and variance equal to $\sigma^2$ before truncation, truncated on the interval $[a, b]$.
We set  
$$
\begin{array}{c}
\pi=0.75, \quad \beta_0^{-}=-2, \quad  \sigma^2_{-}=1, \quad  \beta_0^{+}=2.25, \quad \sigma^2_{+}=1, \quad 
\gamma_{0}^{-}= 1.75, \quad \gamma_{0}^{+}= 0.75.
\end{array}
$$
We consider $3 \times 2$ scenarios by varying $\gamma_{1}^{-}$ and $\gamma_{1}^{+}$, which describe the strength of association between the outcome and the forcing variable, and $\tau^2_{-}$ and $\tau^2_{+}$, the conditional outcome variability given the forcing variable. Specifically, we set 
$$
\gamma_{1}^{-}=1.5; \gamma_{1}^{+}=1.2,
\qquad \qquad  \gamma_{1}^{-}=1.0; \gamma_{1}^{+}=0.6,
\qquad \qquad \gamma_{1}^{-}=0.50; \gamma_{1}^{+}=0.30
$$
to mimic studies with a strong, medium, and weak association between the outcome and the forcing variable and 
$$
\tau^2_{-}=\tau^2_{+}=0.50^2, \qquad \qquad  \tau^2_{-}=\tau^2_{+}=0.75^2
$$
to mimic studies where the outcome variability conditional on the forcing variable is low and high. Table~\ref{tab:rho.s.y} shows the correlation coefficients between the forcing variable and the outcome in the $3 \times 2$ scenarios.

\begin{table}\caption{Simulation setting 1. Correlation coefficients between the forcing variable and the outcome in $\Usetm$ and $\Usetm$, $\rho_{S,Y}^{-}$ and $\rho_{S,Y}^{+}$, in the $3 \times 2$ scenarios.} \label{tab:rho.s.y}
$$
\begin{array}{c|cc}
	\multicolumn{3}{c}{\tau^2_{-}=\tau^2_{+} = 0.50^2}\\
	
	{(\gamma_1^{-},\gamma_1^{+})}&\rho_{S,Y}^{-}& \rho_{S,Y}^{+}\\
	\hline
	(1.5, 1.2) &  0.95 &0.92\\
	(1.0, 0.6) & 0.89 & 0.77 \\
	(0.5, 0.3)&  0.71 & 0.51 \\
\end{array} \qquad \qquad
\begin{array}{c|cc}
	\multicolumn{3}{c}{\tau^2_{-}=\tau^2_{+} = 0.75^2}\\	 
	{(\gamma_1^{-},\gamma_1^{+})}&\rho_{S,Y}^{-}& \rho_{S,Y}^{+}\\
	\hline
	(1.5, 1.2) &  0.89 & 0.85 \\
	(1.0, 0.6) & 0.80 & 0.63\\
	(0.5, 0.3)&  0.56& 0.37 \\
\end{array}
$$
\end{table}

\subsubsection*{Data Generating Process: Setting 2}
Under the second setting, data are generated using the following process. We first generate a covariate from a standard Normal distribution and standardize it using the sample mean and the sample variance. Let $X_i$ denote the standardized covariate.  Then, we generate
$$G^\ast_i(-) \sim N\left( \alpha_0^{-} +   \alpha^{-}_X X_i, 1\right) \quad \hbox{and} \quad G^\ast_i(+)\sim N\left(  \alpha_{0}^{+} +   \alpha^{+}_X  X_i,1\right),$$ independently, and we calculate the subpopulation membership probabilities as follows:
\begin{eqnarray*}
&	\pi_i(\Usetm) = \Pr(G^\ast_i(-) \leq 0), \qquad 
\pi_i(\Usetp) = \Pr(G^\ast_i(-) > 0 \, \mathrm{and}  \, G^\ast_i(+) \leq 0),&\\
&	\pi_i(\Uset) =1- 	\pi_i(\Usetm)-	\pi_i(\Usetp). \qquad & 
\end{eqnarray*}
Finally, we generate the forcing variable and the outcome using the following models:
$$
\begin{array}{cl}
S_i \mid   i \in \Usetm \sim & TN\left(\beta_0^{-}+\beta^{-}_X X_i ; \sigma^2_{-}; -\infty, s_0 \right), \\ 
\vspace{-0.3cm}\\
S_i \mid   i \in \Usetp \sim &TN\left(\beta_0^{+} +\beta^{-}_X X_i; \sigma^2_{+}; s_0, \infty \right), \\
\vspace{-0.3cm}\\
	S_i \mid   i \in \Uset \sim &N\left(\beta_0+\beta_X X_i ; \sigma^2 \right);
\end{array}
$$
and
$$
\begin{array}{ccc}
Y_i(s) \mid    i \in \Usetm \sim  N\left(\gamma_{0}^{-} + (s-s_0) \gamma_{1}^{-}+\gamma^{-}_X X_i;    \tau^2_{-} \right),\\
\vspace{-0.3cm}\\
Y_i(s) \mid   i \in \Usetp  \sim  N\left(\gamma_{0}^{+} + (s-s_0) \gamma_{1}^{+}+\gamma^{+}_X X_i; \tau^2_{+} \right), 
\end{array}
$$
$$
\begin{array}{ccc}
Y_i(0) \mid    i \in \Uset \sim  N\left(\gamma_{0,z=0} +   \gamma_{X, z=0} X_i;    \tau^2_{z=0} \right),
\\
\vspace{-0.3cm}\\
Y_i(1) \mid    i \in \Uset \sim  N\left(\gamma_{0,z=1} +   \gamma_{X, z=1} X_i;    \tau^2_{z=1} \right).
\end{array}
$$

We consider eight simulation scenarios, varying: the dependence of the subpopulation membership on the covariate  ($\alpha_X^{-}$ and $\alpha_X^{+}$); the association between the forcing variable and the covariate in each subpopulation, $\Usetm$, $\Usetp$ and $\Uset$ ($\beta_X^{-}$, $\beta_X^{+}$, and $\beta_X$); and the conditional outcome variability given the forcing variable and the covariate in $\Usetm$ and $\Usetp$ ($\tau^2_{-}$ and $\tau^2_{+}$)  and given the covariate in $\Uset$ ($\tau^2$). Table~\ref{tab:sim2} shows the true parameter values. 
\begin{table}\caption{Simulation setting 2. True parameter values. Parameters that do not vary across simulation scenarios are shown only for the first scenario.} \label{tab:sim2}
$$
\begin{array}{c|rr rr rr rr}
	\hline 
	& \multicolumn{8}{c}{\hbox{Simulation scenario}}\\
	\hbox{Parameters} & 1 & 2 & 3 & 4 & 5 & 6 & 7 & 8\\
	\hline
	\alpha_0^{-}  & 0.75 \\
	\alpha_X^{-}  & 0.80 &  0.80 & 0.80 &  0.80 & 0.40 &  0.40 & 0.840 &  0.40\\
	\\
	\alpha_0^{+}  &  0.75\\
	\alpha_X^{+}  & -0.40 & -0.40& -0.40 & -0.40& -0.20 & -0.20& -0.20 & -0.20\\
	\\
	\beta_0^{-}  & -2.00\\
	\beta_X^{-}  & 1.50 & 1.50 	   & 0.75 & 0.75 & 1.50 & 1.50 	   & 0.75 & 0.75  \\
	\sigma^2_{-} & 1.00 & \\
	\\           
	\beta_0^{+} & 2.25\\
	\beta_X^{+} & 1.25& 1.25 & 0.70& 0.70& 1.25& 1.25 & 0.70& 0.70\\
	\sigma^2_{+} & 1.00\\
	\\
	\beta_0   & -1.00\\
	\beta_X  & 1.75& 1.75 & 0.80& 0.80 & 1.75& 1.75 & 0.80& 0.80\\
	\sigma^2 & 1.00\\
	\\
	\gamma_0^{-} & 1.75\\
	\gamma_1^{-} & 1.00\\
	\gamma_X^{-} & 0.75\\
	\tau^2_{-} & 0.50^2& 0.75^2& 0.50^2& 0.75^2& 0.50^2& 0.75^2& 0.50^2& 0.75^2\\
	\\
	\gamma_0^{+} & 0.75\\
	\gamma_1^{+} & 0.70\\
	\gamma_X^{+} & 0.75\\
	\tau^2_{+} & 0.50^2& 0.75^2& 0.50^2& 0.75^2& 0.50^2& 0.75^2& 0.50^2& 0.75^2\\
	\\
	\gamma_{0,z=0} & 0.50\\
	\gamma_{X,z=0} & 0.80\\
	\tau^2_{z=0} & 0.50^2& 0.75^2& 0.50^2& 0.75^2& 0.50^2& 0.75^2& 0.50^2& 0.75^2\\
	\\
	\gamma_{0,z=1} & 1.50\\
	\gamma_{X,z=1} & 0.80\\
	\tau^2_{z=1} & 0.50^2& 0.75^2& 0.50^2& 0.75^2& 0.50^2& 0.75^2& 0.50^2& 0.75^2\\
	\hline
\end{array}
$$
\end{table}

\begin{figure}
\begin{center}
	\begin{tabular}{ccc}
		Scenarios 1-4 &$\qquad$&	Scenarios 5-8\\
		Strong dependence between&$\qquad$& 
		Weak dependence  between \\
		subpopulation membership and $X_i$ &$\qquad$& 	subpopulation membership and $X_i$\\
		$\alpha_X^{-}=0.8$ and $\alpha_X^{+}=-0.4$ &$\qquad$&	$\alpha_X^{-}=0.4$ and $\alpha_X^{+}=-0.2$ \\
		\includegraphics[width=6.5cm]{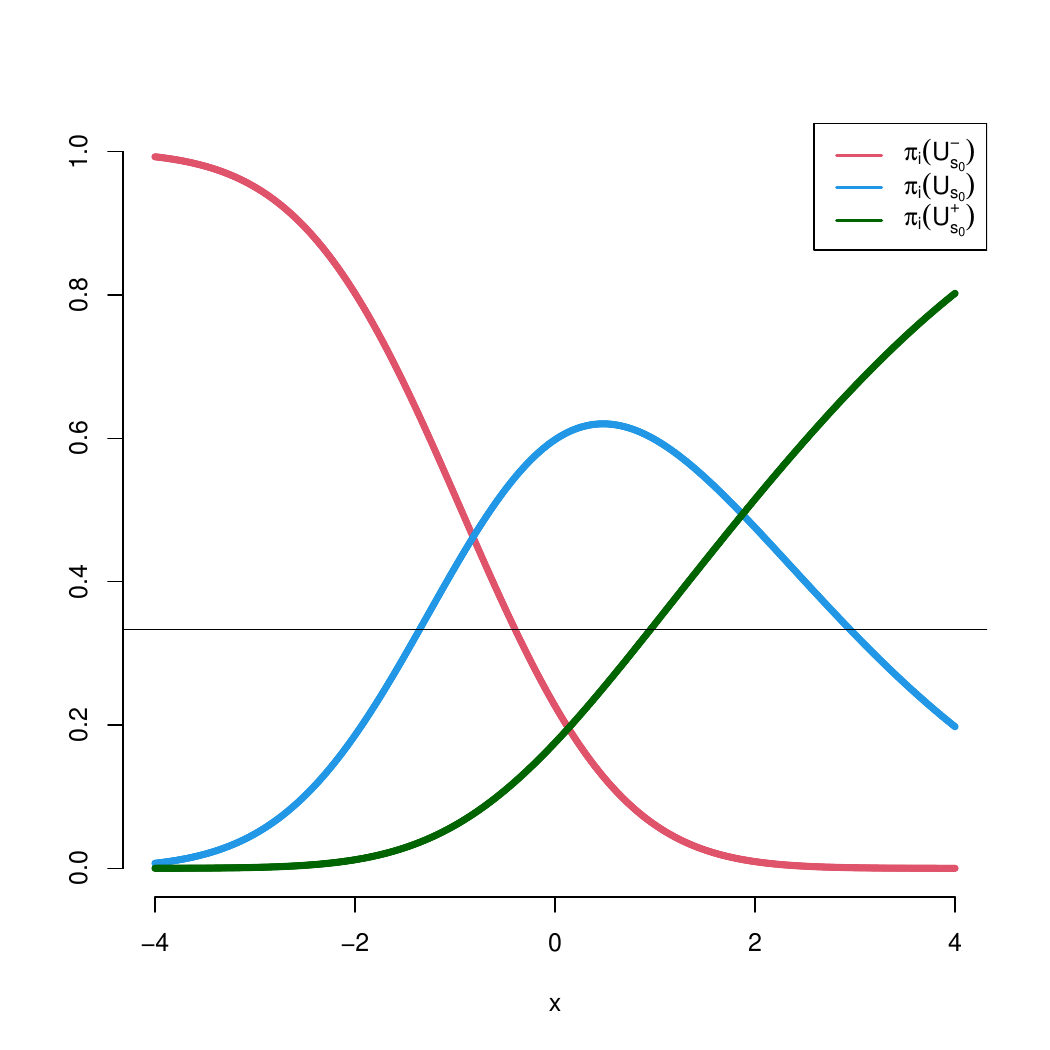} &$\qquad$& \includegraphics[width=6.5cm]{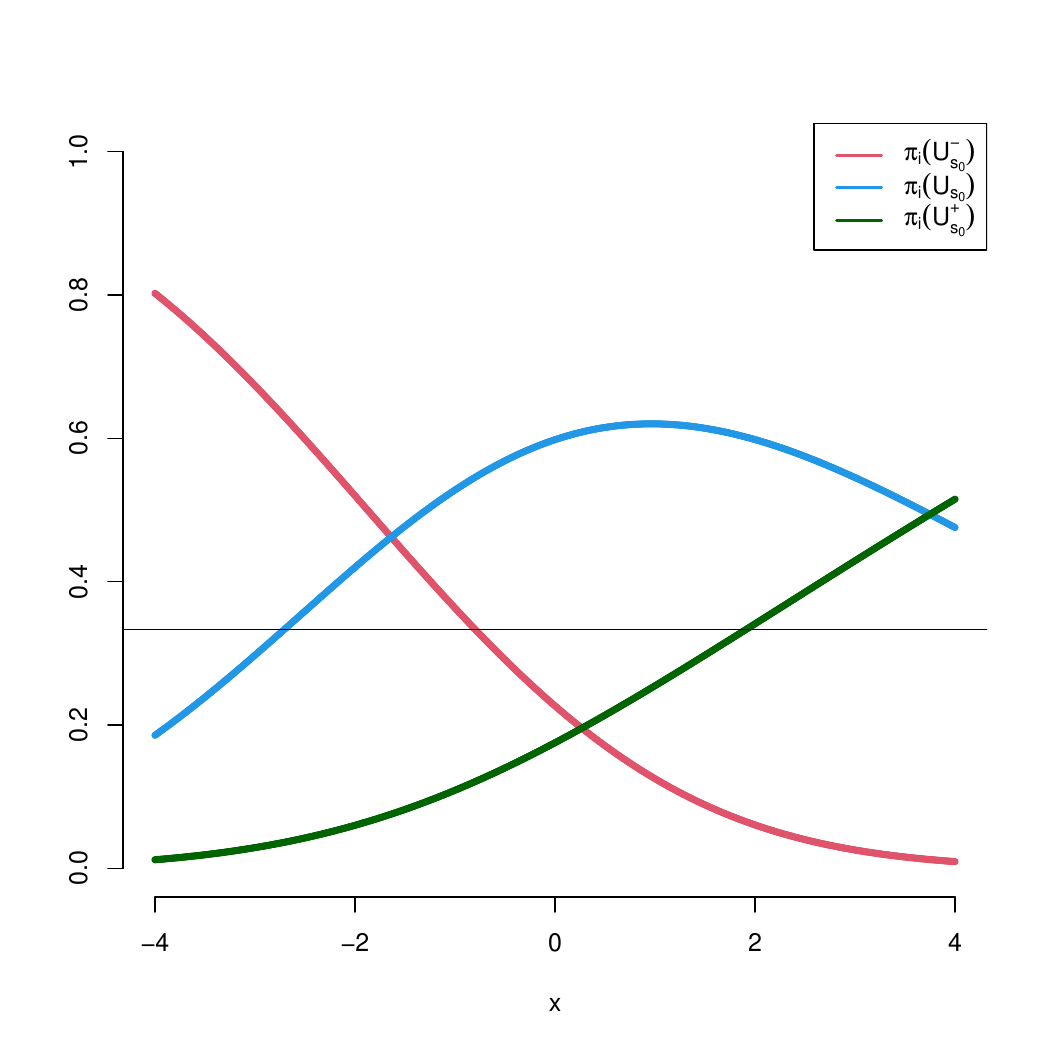}
	\end{tabular} 
\end{center}
\caption{Simulation setting 2. Subpopulation membership probabilities as a function of the covariate.}   \label{tab:sim2_pGx}
\end{figure}

In scenarios 1-4,  the dependence of the subpopulation membership on the covariate is relatively strong, whereas in scenarios 5-8, it is weaker (see Figure \ref{tab:sim2_pGx} showing the subpopulation membership probabilities as a function of the covariate).

The association between the forcing variable and the covariate in each subpopulation, $\Usetm$, $\Usetp$ and $\Uset$ ($\beta_X^{-}$, $\beta_X^{+}$, and $\beta_X$) is relatively strong in scenarios 1-2 and 5-6, and it is weaker in scenarios 3-4 and 7-8.
Table~\ref{tab:rho.s.y} shows the correlation coefficients between the covariate and the forcing variable in the $2 \times 2 \times 2$ scenarios.

\begin{table}\caption{Simulation setting 2. Correlation coefficients between the forcing variable and the covariate in $\Usetm$, $\Usetp$ and $\Uset$: $\rho_{X,S}^{-}$, $\rho_{X,S}^{+}$, and $\rho_{X,S}$.} \label{tab:rho.x.s}
$$
\begin{array}{lc|ccc}	
	\hbox{Scenario}&	{(\beta_X^{-},\beta_X^{+}, \beta_X)}&\rho_{X,S}^{-}& \rho_{X,S}^{+}& \rho_{X,S}\\
	\hline
	1,2,5,6&	(1.50, 1.25, 1.75) &  0.83 &0.78& 0.87\\
	3,4,7,8&	(0.75, 0.70, 0.80) & 0.60 & 0.57& 0.63\\
\end{array}  
$$
\end{table}

Finally, a smaller conditional outcome variability is used in scenarios 1, 3, 5, and 7, where we set 
$\tau^2_{-} =\tau^2_{+}=\tau^{2, z=0}=\tau^{2, z=1}=0.5^2$, and a larger conditional outcome variability is used in scenarios 2,4, 6 and 8 where we set $\tau^2_{-} =\tau^2_{+}=\tau^{2}_{z=0}=\tau^{2}_{z=1}=0.75^2$. 

\subsubsection*{Data Generating Process: Setting 3}
\begin{figure}
\begin{center}
	\begin{tabular}{ccc}
		Scenarios 1 and 3 &$\qquad$&	Scenarios 2 and 4\\
		Strong dependence between&$\qquad$& 
		Weak dependence  between \\
		subpopulation membership and $S_i$ &$\qquad$& 	subpopulation membership and $S_i$\\
		$\beta_0^{-}=-2.00; \beta_0^{+}=2.25; \beta_0=-1.00 $ &$\qquad$&	$\beta_0^{-}=-0.50; \beta_0^{+}=0.75; \beta_0=-0.25 $ \\
		\includegraphics[width=6.5cm]{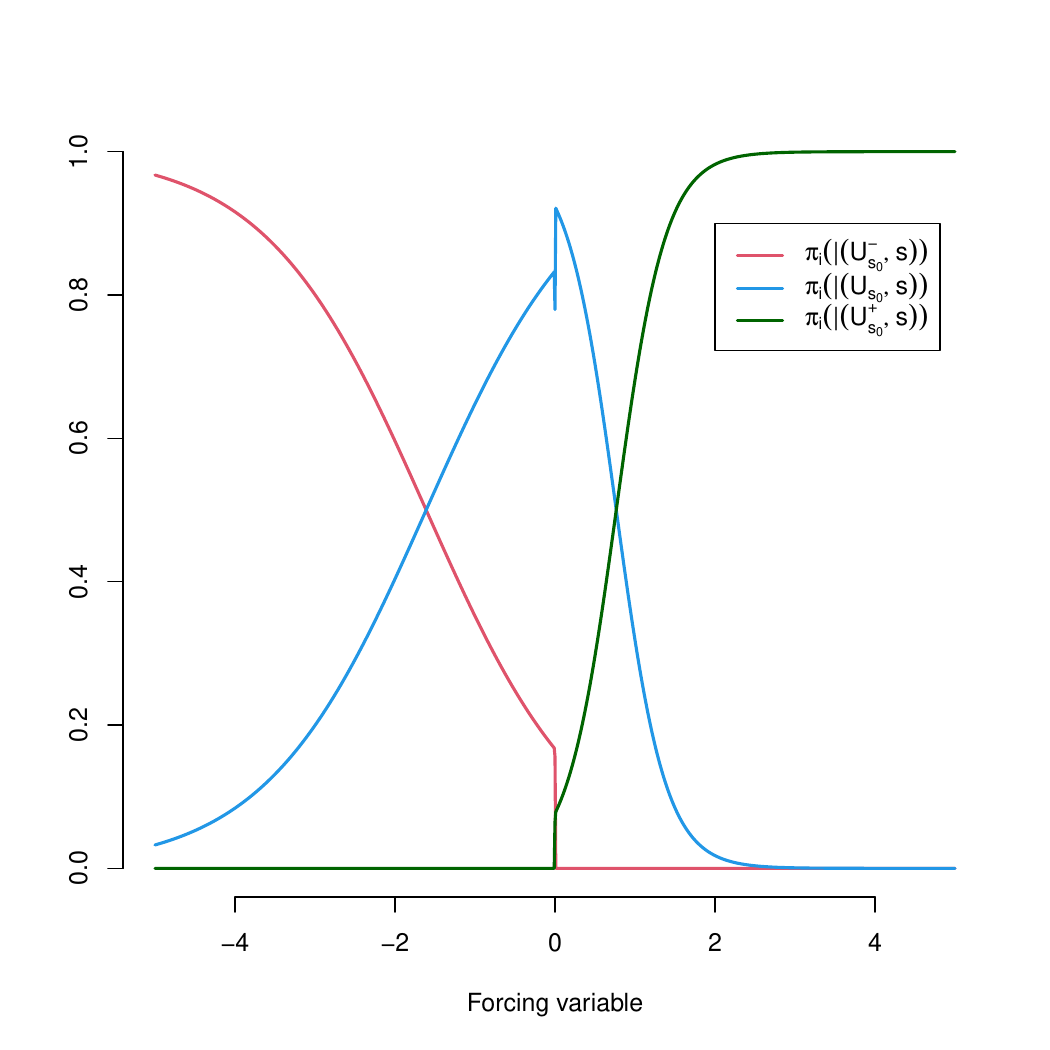} &$\qquad$& \includegraphics[width=6.5cm]{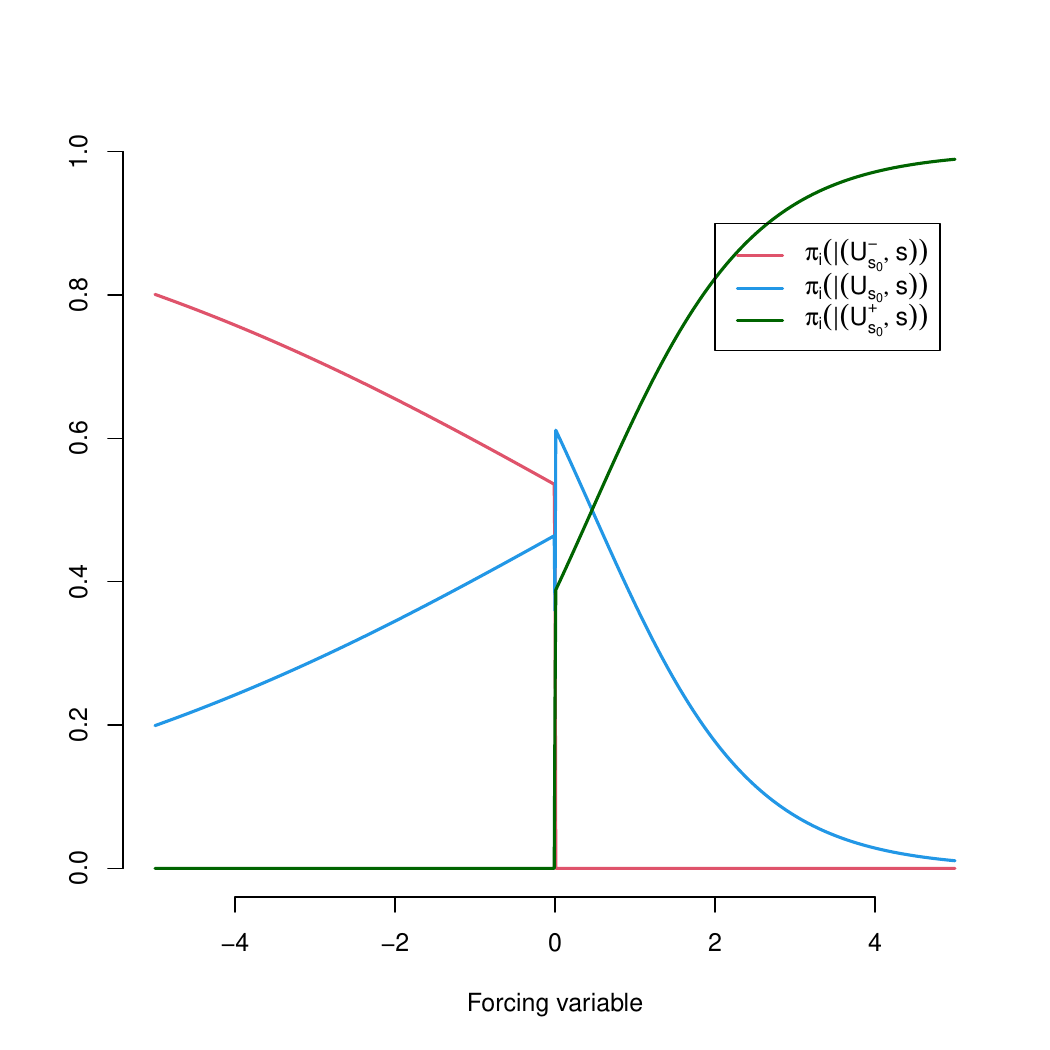}
	\end{tabular} 
\end{center}
\caption{Simulation setting 3. Subpopulation membership probabilities as a function of the forcing variable.}   \label{tab:sim3_pGs}
\end{figure}

Under the third setting, data are generated using the following process:

$$
pi_i(\Usetm) = 0.35, \qquad 
	\pi_i(\Usetp) = 0.25, \qquad	\pi_i(\Uset) =0.40;
 $$
$$
\begin{array}{c}
S_i \mid   i \in \Usetm \sim TN\left(\beta_0^{-}; \sigma^2_{-}; -\infty, s_0 \right),\\
\vspace{-0.3cm}\\
S_i \mid   i \in \Usetp \sim TN\left(\beta_0^{+}; \sigma^2_{+}; s_0, \infty \right), \\
\vspace{-0.3cm}\\
S_i \mid   i \in \Uset \sim N\left(\beta_0; \sigma^2  \right);
\end{array}
$$
and 
$$
\begin{array}{c}
Y_i(s) \mid    i \in \Usetm \sim  N\left(\gamma_{0}^{-} + (s-s_0) \gamma_{1}^{-};    \tau^2_{-} \right),\\
\vspace{-0.3cm}\\
Y_i(s) \mid   i \in \Usetp  \sim  N\left(\gamma_{0}^{+} + (s-s_0) \gamma_{1}^{+}; \tau^2_{+} \right),\\
\vspace{-0.3cm}\\
Y_i(0) \mid    i \in \Uset \sim  N\left(\gamma_{0,z=0}, \tau^2_{z=0} \right),\\
\vspace{-0.3cm}\\
Y_i(1) \mid    i \in \Uset \sim  N\left(\gamma_{0,z=1}, \tau^2_{z=1} \right).
\end{array}
$$
We set  $\sigma^2_{-}= \sigma^2_{+}= \sigma^2=1$, $(\gamma_{0}^{-}, \gamma_{1}^{-})=(1.75,1.00)$
$(\gamma_{0}^{+}, \gamma_{1}^{+})=(0.75,0.60)$, $\gamma_{0,z=0}=0.5$ and $\gamma_{0,z=1}=1.5$, 
and we vary $\beta_0^{-}$, $\beta_0^{+}$, and $\beta_0$, which describe the strength of association between the forcing variable and subpopulation membership, and outcome variances, $\tau^2_{-}$, $\tau^2_{+}$ and  $\tau^2_{z=0}$ and  $\tau^2_{z=1}$. Specifically, we consider $2 \times 2$ scenarios by setting
$$
\beta_0^{-}=-2.00; \beta_0^{+}=2.25; \beta_0=-1.00 
\qquad \hbox{ and }\qquad   \beta_0^{-}=-0.50; \beta_0^{+}=0.75; \beta_0=-0.25;
$$
and 
$$
\tau^2_{-}=\tau^2_{+}=\tau^2_{z=0}=\tau^2_{z=1}=0.50^2 \qquad \hbox{ and }\qquad  \tau^2_{-}=\tau^2_{+}=\tau^2_{z=0}=\tau^2_{z=1}=0.75^2.
$$
Figure \ref{tab:sim3_pGs} shows the conditional probabilities of belonging to each membership  given the forcing variable, which we derive using the Bayes Theorem as follows:
\begin{eqnarray*}
\pi_i(\Usetm\mid S_i=s) &=&\dfrac{\pi_i(\Usetm) f_{S\mid \Usetm}(s)}{\pi_i(\Usetm) f_{S\mid \Usetm}(s)+\pi_i(\Usetp) f_{S\mid \Usetp}(s)+\pi_i(\Uset) f_{S\mid \Uset}(s)},\\
\pi_i(\Usetp\mid S_i=s) &=&\dfrac{\pi_i(\Usetp) f_{S\mid \Usetp}(s)}{\pi_i(\Usetm) f_{S\mid \Usetm}(s)+\pi_i(\Usetp) f_{S\mid \Usetp}(s)+\pi_i(\Uset) f_{S\mid \Uset}(s)},
\\
\pi_i(\Uset \mid S_i=s) &=&\dfrac{\pi(\Uset ) f_{S\mid \Uset }(s)}{\pi_i(\Usetm) f_{S\mid \Usetm}(s)+\pi_i(\Usetp) f_{S\mid \Usetp}(s)+\pi_i(\Uset) f_{S\mid \Uset}(s)},
\end{eqnarray*}
with $f_{S\mid \Usetm}(s)=0$ for $s > s_0$ and $f_{S\mid \Usetp}(s)=0$ for $s \leq s_0$.

\subsection*{Simulation Results} \label{sec:simres}
For each scenario in each simulation setting, we apply our Bayesian model-based mixture approach to clustering to $50$ simulated datasets of size $N=10\,000$. 
We evaluate the performance of our procedure using the average of the posterior means over the 50 datasets, the bias, the mean squared error (MSE), and the coverage of the 95\% and 99\% highest density interval of each causal estimand of interest, the proportion of units classified as member of the subpopulation $\Uset$:
\begin{equation}\label{puset}
\pi(\Uset)= \dfrac{N_{\Uset}}{N},
\end{equation}
where $N_{\Uset}$ is the number of units belonging to $\Uset$, and the finite sample average causal effect in $\Uset$:
\begin{equation}\label{aceuset}
ACE_{\Uset}= \dfrac{1}{N_{\Uset}} \sum_{i \in \Uset} Y_i(1) - \dfrac{1}{N_{\Uset}} \sum_{i \in \Uset} Y_i(0).
\end{equation}

We also implement the traditional approach based on the continuity assumption, using uniform and triangular kernel functions to construct local-polynomial estimators with order of the local-polynomial equal to 1 and 2 and two-sided Mean Square Error-optimal bandwidth selectors  (below and above the cutoff).
For each simulation setting, we show the average of the point estimates and of the 95\% and 99\% confidence intervals for the average causal effect at the threshold:
\begin{equation}\label{aces0}
ACE_{s_0} = \mathbb{E}[Y_i(1) \mid S_i=s_0] - \mathbb{E}[Y_i(0) \mid S_i=s_0] 
\end{equation}
Except that in simulation setting 1, we do not report the bias and the mean squared error (MSE) of the local-polynomial estimators and the coverage of the 95\% and 99\% confidence intervals because the causal estimands of interest are generally not causal effects at the threshold in our simulation study.

\subsubsection*{Simulation Results: Setting 1} 
Simulation setting 1 focuses on studies where there is no unit in $\Uset$, aiming to investigate if our procedure successfully detects that the subpopulation $\Uset$ is empty rather than create a ``fake'' subpopulation $\Uset$ by selecting observations such that the dependence between the outcome and the forcing variable is broken. Table~\ref{tab:sim1resUset} shows that our Bayesian model-based mixture approach provides extremely small posterior means for the proportion of units in $\Uset$ with a very high precision (small bias, small MSE, and high coverage) under each scenario, successfully suggesting that the subpopulation $\Uset$ is empty. As we expected, the stronger the association between the forcing variable and the outcome, the stronger the evidence that $\Uset$ is empty is.

In this setting, looking at causal effects for units in the subpopulation $\Uset$ does not make sense. It might be sensible to look at causal effects at the threshold, such as the average causal effect at $s_0$ in Equation~\eqref{aces0}.
Under our data generating process, $ACE_{s_0} = \gamma_{0}^{-}-  \gamma_{0}^{+}=1.75-0.75=1$. 
Table~\ref{tab:sim1resUset} shows the results, which suggest that our Bayesian model-based mixture approach leads to a posterior distribution of the average causal effect at $s_0$ centered around the true value with good frequentistic properties.

\begin{table}\caption{Simulation setting 1: Simulation results on $\pi(\Uset)$. Average of the posterior means, bias, MSE, and 95\% and 99\% coverage for the proportion of units in $\Uset$.  True value: $\pi(\Uset)=0$.} \label{tab:sim1resUset}
$$
\begin{array}{ll ccccc}
	\multicolumn{7}{c}{\hbox{Conditional outcome variance given the forcing variable: } \tau^2_{-}=\tau^2_{+}=0.50^2}\\
	\hline
	&&\hbox{Average of the} &  &  & \multicolumn{2}{c}{\hbox{Coverage}}
	\\
	\cline{6-7}
	\multicolumn{2}{l}{\hbox{$S-Y$ association} }& \hbox{Posterior Means} & \hbox{Bias}& \hbox{MSE}& \hbox{95\% HDI } &
	\hbox{99\% HDI}\\
	\hline
	\hbox{Strong: } & \gamma_{1}^{-}=1.5; \gamma_{1}^{+}=1.2& 0.000006 & 0.000006 & 0.0000 & 1.00 & 1.00 \\
	\vspace{-0.35cm}\\
	\hbox{Medium: }& \gamma_{1}^{-}=1.0; \gamma_{1}^{+}=0.6& 0.000035 & 0.000035 & 1.0/10^7 & 1.00 & 1.00 \\ 
	\vspace{-0.35cm}\\
	\hbox{Weak: } & \gamma_{1}^{-}=0.5; \gamma_{1}^{+}=0.3	& 0.000346 & 0.000346 & 2.2/10^6 & 0.98 & 0.98 \\ 
	\hline
\end{array}
$$

$$
\begin{array}{ll ccccc}
	\multicolumn{7}{c}{\hbox{Conditional outcome variance given the forcing variable: } \tau^2_{-}=\tau^2_{+}=0.75^2}\\
	\hline
	&&\hbox{Average of the} &  &  & \multicolumn{2}{c}{\hbox{Coverage}}
	\\
	\cline{6-7}
	\multicolumn{2}{l}{\hbox{$S-Y$ association} }& \hbox{Posterior Means} & \hbox{Bias}& \hbox{MSE}& \hbox{95\% HDI } &
	\hbox{99\% HDI}\\
	\hline
	\hbox{Strong: } & \gamma_{1}^{-}=1.5; \gamma_{1}^{+}=1.2& 0.000070 & 0.000070 & 1.0/10^7 & 0.98 & 0.98 \\
	\vspace{-0.35cm}\\
	\hbox{Medium: }& \gamma_{1}^{-}=1.0; \gamma_{1}^{+}=0.6& 0.000077 & 0.000077 & 3.0/10^7 & 0.98 & 0.98 \\ 
	\vspace{-0.35cm}\\
	\hbox{Weak: } & \gamma_{1}^{-}=0.5; \gamma_{1}^{+}=0.3	& 0.000003 & 0.000003 & 0.0000 & 1.00 & 1.00 \\ 
	\hline
\end{array}
$$
\end{table}

\begin{table}\caption{Simulation setting 1: Simulation results on $ACE_{s_0}$. Average of the posterior means, bias, MSE, and 95\% and 99\% coverage for  $ACE_{s_0}$.  True value: $ACE_{s_0}=1$.} \label{tab:sim1restau}
$$
\begin{array}{ll ccccc}
	\multicolumn{7}{c}{\hbox{Conditional outcome variance given the forcing variable: } \tau^2_{-}=\tau^2_{+}=0.50^2}\\
	\hline
	&&\hbox{Average of the} &  &  & \multicolumn{2}{c}{\hbox{Coverage}}
	\\
	\cline{6-7}
	\multicolumn{2}{l}{\hbox{$S-Y$ association} }& \hbox{Posterior Means} & \hbox{Bias}& \hbox{MSE}& \hbox{95\% HDI } &
	\hbox{99\% HDI}\\
	\hline
	\hbox{Strong: } & \gamma_{1}^{-}=1.5; \gamma_{1}^{+}=1.2& 0.9940 & -0.0060 & 0.0007 & 0.96 & 0.98 \\
	\vspace{-0.35cm}\\
	\hbox{Medium: }& \gamma_{1}^{-}=1.0; \gamma_{1}^{+}=0.6& 0.9942 & -0.0058 & 0.0008 & 0.96 & 0.98 \\ 
	\vspace{-0.35cm}\\
	\hbox{Weak: } & \gamma_{1}^{-}=0.5; \gamma_{1}^{+}=0.3	& 0.9939 & -0.0061 & 0.0008 & 0.96 & 1.00 \\ 
	\hline
\end{array}
$$

$$
\begin{array}{ll ccccc}
	\multicolumn{7}{c}{\hbox{Conditional outcome variance given the forcing variable: } \tau^2_{-}=\tau^2_{+}=0.75^2}\\
	\hline
	&&\hbox{Average of the} &  &  & \multicolumn{2}{c}{\hbox{Coverage}}
	\\
	\cline{6-7}
	\multicolumn{2}{l}{\hbox{$S-Y$ association} }& \hbox{Posterior Means} & \hbox{Bias}& \hbox{MSE}& \hbox{95\% HDI } &
	\hbox{99\% HDI}\\
	\hline
	\hbox{Strong: } & \gamma_{1}^{-}=1.5; \gamma_{1}^{+}=1.2& 0.9911 & -0.0089 & 0.0017 & 0.96 & 0.96 \\
	\vspace{-0.35cm}\\
	\hbox{Medium: }& \gamma_{1}^{-}=1.0; \gamma_{1}^{+}=0.6& 0.9900 & -0.0100 & 0.0018 & 0.96 & 0.98 \\ 
	\vspace{-0.35cm}\\
	\hbox{Weak: } & \gamma_{1}^{-}=0.5; \gamma_{1}^{+}=0.3	& 0.9905 & -0.0095 & 0.0018 & 0.96 & 1.00 \\ 
	\hline
\end{array}
$$
\end{table}

Table~\ref{tab:sim1resSA} shows the simulation results for $ACE_{s_0}$ derived using  local polynomial estimators  under the continuity assumption. 
Under simulation setting 1,  $ACE_{s_0}$ is the causal estimand we focus on and the target estimand of local polynomial estimators under the continuity assumption. Therefore, above and beyond the average of the point estimates and of the 95\% and 99\% confidence interval, we show the bias, the MSE of the local polynomial estimators, and the coverage of the 95\% and 99\% confidence intervals. The local polynomial estimators we consider perform relatively well in terms of bias and MSE, although our Bayesian model-based mixture approach outperforms the local polynomial estimators in terms of MSE. 
Also, 95\% confidence intervals derived using local polynomial estimators seem to uncover the true value of the average causal effect under some scenarios, whereas the 95\%   highest posterior density intervals always reach the nominal level.

\begin{table}
\caption{Simulation setting 1: Simulation results on $ACE_{s_0}$ under the continuity assumption. Average of the point estimates, and of 95\% and 99\% confidence intervals, bias, MSE, and 95\% and 99\% coverage for   $ACE_{s_0}$.  True value: $ACE_{s_0}=1$.} \label{tab:sim1resSA}
$$
 \hspace*{-2cm}
\begin{array}{ll cccr c cc}
	\multicolumn{9}{c}{\hbox{Conditional outcome variance given the forcing variable: } \tau^2_{-}=\tau^2_{+}=0.50^2}\\
	\hline
	&&\multicolumn{3}{c}{\hbox{Average of}} & & & \multicolumn{2}{c}{\hbox{Coverage}}
	\\
\cline{3-5}	\cline{8-9}
	\multicolumn{2}{l}{\hbox{$S-Y$ association} }& \hbox{Point estimates} & 
  \hbox{95\% CI } &  \hbox{99\% CI } &
 \hbox{Bias}& \hbox{MSE}& \hbox{95\% CI } &	\hbox{99\% CI}\\
	\hline
	\hbox{Strong: } & \gamma_{1}^{-}=1.5; \gamma_{1}^{+}=1.2&   \\
 \multicolumn{2}{r}{\hbox{Uniform Kernel ($p=1$)}}& 1.0040 & (0.8835;  1.1244) & (0.8456;  1.1623) & 0.0040 & 0.0041 & 0.98 & 1.00 \\ 
 \multicolumn{2}{r}{\hbox{Triangular Kernel ($p=1$)}}&  1.0020 & (0.8747; 1.1293) & (0.8347;1.1693) & 0.0020 & 0.0070 & 0.88 & 1.00 \\ 
 \multicolumn{2}{r}{\hbox{Uniform Kernel ($p=2$)}}&  1.0055 & (0.8370; 1.1740) & (0.7840; 1.2269) & 0.0055 & 0.0103 & 0.90 & 1.00 \\ 
 \multicolumn{2}{r}{\hbox{Triangular Kernel ($p=2$)}} & 1.0066 & (0.8280; 1.1852) & (0.7718; 1.2413) & 0.0066 & 0.0126 & 0.88 & 0.98 \\ 
	\vspace{-0.35cm}\\
	\hbox{Medium: }& \gamma_{1}^{-}=1.0; \gamma_{1}^{+}=0.6&   \\ 
 \multicolumn{2}{r}{\hbox{Uniform Kernel ($p=1$)}}& 1.0035 & (0.8825; 1.1245) & (0.8445; 1.1626)& 0.0035 & 0.0039 & 0.98 & 1.00 \\ 
\multicolumn{2}{r}{\hbox{Triangular Kernel ($p=1$)}} & 0.9999 & (0.8721; 1.1276) & (0.8320; 1.1678) & -0.0001 & 0.0070 & 0.88 & 1.00 \\
 \multicolumn{2}{r}{\hbox{Uniform Kernel ($p=2$)}}& 1.0035 & (0.8350; 1.1719) & (0.7821; 1.2249) & 0.0035 & 0.0106 & 0.90 & 1.00 \\ 
\multicolumn{2}{r}{\hbox{Triangular Kernel ($p=2$)}} & 1.0053 & (0.8271; 1.1834) & (0.7712; 1.2394) & 0.0053 & 0.0126 & 0.88 & 0.98 \\
	\vspace{-0.35cm}\\
	\hbox{Weak: } & \gamma_{1}^{-}=0.5; \gamma_{1}^{+}=0.3	&   \\ 
 \multicolumn{2}{r}{\hbox{Uniform Kernel ($p=1$)}} & 1.0017 & (0.8804; 1.1229) & (0.8423; 1.1610) & 0.0017 & 0.0039 & 0.98 & 1.00 \\ 
\multicolumn{2}{r}{\hbox{Triangular Kernel ($p=1$)}} & 0.9990 & (0.8703; 1.1277) & (0.8298; 1.1681) & -0.0010 & 0.0070 & 0.88 & 1.00 \\ 
 \multicolumn{2}{r}{\hbox{Uniform Kernel ($p=2$)}} & 1.0004 & (0.8305; 1.1702 )& (0.7771; 1.2236) & 0.0004 & 0.0104 & 0.90 & 1.00 \\ 
\multicolumn{2}{r}{\hbox{Triangular Kernel ($p=2$)}} & 1.0063 & (0.8255; 1.1870) & (0.7688; 1.2438) & 0.0063 & 0.0126 & 0.88 & 0.98 \\ 
	\hline
\end{array}
 \hspace*{-2cm}
$$
$$
 \hspace*{-2cm}
\begin{array}{ll cccr c cc}
	\multicolumn{9}{c}{\hbox{Conditional outcome variance given the forcing variable: } \tau^2_{-}=\tau^2_{+}=0.75^2}\\
	\hline
	&&\multicolumn{3}{c}{\hbox{Average of}} & & & \multicolumn{2}{c}{\hbox{Coverage}}
	\\
\cline{3-5}	\cline{8-9}
	\multicolumn{2}{l}{\hbox{$S-Y$ association} }& \hbox{Point estimates} & 
  \hbox{95\% CI } &  \hbox{99\% CI } &
 \hbox{Bias}& \hbox{MSE}& \hbox{95\% CI } &	\hbox{99\% CI}\\
	\hline
	\hbox{Strong: } & \gamma_{1}^{-}=1.5; \gamma_{1}^{+}=1.2&   \\
  \multicolumn{2}{r}{\hbox{Uniform kernel ($p=1$)}} & 1.0047 & (0.8232; 1.1862) & (0.7662; 1.2433) & 0.0047 & 0.0088 & 0.98 & 1.00 \\ 
 \multicolumn{2}{r}{\hbox{Triangular kernel ($p=1$)}} & 0.9998 & (0.8082; 1.1914) & (0.7479; 1.2516) & -0.0002 & 0.0157 & 0.88 & 1.00 \\ 
 \multicolumn{2}{r}{\hbox{Uniform kernel ($p=2$)}} & 1.0046 & (0.7519; 1.2573) & (0.6725; 1.3367) & 0.0046 & 0.0237 & 0.90 & 1.00 \\ 
 \multicolumn{2}{r}{\hbox{Triangular kernel ($p=2$)}} & 1.0079 & (0.7407; 1.2751) & (0.6567; 1.3591) & 0.0079 & 0.0284 & 0.88 & 0.98 \\
	\vspace{-0.35cm}\\
	\hbox{Medium: }& \gamma_{1}^{-}=1.0; \gamma_{1}^{+}=0.6&  \\ 
  \multicolumn{2}{r}{\hbox{Uniform kernel ($p=1$)}} & 1.0060 & (0.8239;1.1880) & (0.7668; 1.2452) & 0.0060 & 0.0087 & 0.98 & 1.00 \\ 
 \multicolumn{2}{r}{\hbox{Triangular kernel ($p=1$)}} & 1.0013 & (0.8083; 1.1944) & (0.7476; 1.2551) & 0.0013 & 0.0156 & 0.88 & 1.00 \\ 
 \multicolumn{2}{r}{\hbox{Uniform kernel ($p=2$)}} & 1.0059 & (0.7512; 1.2606) & (0.6712; 1.3407) & 0.0059 & 0.0231 & 0.90 & 1.00 \\ 
 \multicolumn{2}{r}{\hbox{Triangular kernel ($p=2$)}} & 1.0117 & (0.7414; 1.2821) & (0.6565; 1.3670) & 0.0117 & 0.0285 & 0.88 & 0.98 \\ 
	\vspace{-0.35cm}\\
	\hbox{Weak: } & \gamma_{1}^{-}=0.5; \gamma_{1}^{+}=0.3	&\\ 
  \multicolumn{2}{r}{\hbox{Uniform kernel ($p=1$)}} & 1.0019 & (0.8196; 1.1842) & (0.7623; 1.2415) & 0.0019 & 0.0088 & 0.98 & 1.00 \\ 
 \multicolumn{2}{r}{\hbox{Triangular kernel ($p=1$)}} & 0.9983 & (0.8042; 1.1924) & (0.7432; 1.2534) & -0.0017 & 0.0157 & 0.88 & 1.00 \\ 
 \multicolumn{2}{r}{\hbox{Uniform kernel ($p=2$)}} & 0.9986 & (0.7421; 1.2550) & (0.6616; 1.3356) & -0.0014 & 0.0236 & 0.90 & 1.00 \\ 
 \multicolumn{2}{r}{\hbox{Triangular kernel ($p=2$)}} & 1.0102 & (0.7377; 1.2827) & (0.6520; 1.3683) & 0.0102 & 0.0283 & 0.88 & 0.98 \\ 
	\hline
\end{array}
 \hspace*{-2cm}
$$
\end{table}

\subsubsection*{Simulation Results: Setting 2} 
In setting 2, we focus on the proportion of units in $\Uset$ in Equation~\eqref{puset} and the average causal effect for units in $\Uset$ in Equation~\eqref{aceuset}.
Results shown in Table \ref{tab:sim2resUset} highlight the importance of observing covariates that are predictive of the subpopulation membership and the forcing variable. In particular, our simulation study reveals that the association between the covariate and forcing variable plays a special role. The stronger the association between the covariate and forcing variable, the better the performance of the Bayesian model-based approach to clustering is.

\begin{table}[t]\caption{Simulation setting 2: Simulation results. Average of the posterior means, bias, MSE, and 95\% and 99\% coverage for $\pi(\Uset)$ and $ACE_{\Uset}$.}  \label{tab:sim2resUset}
$$
\begin{array}{ccc rcrccc}
\multicolumn{9}{c}{\hbox{Conditional outcome variance given the forcing variable: } \tau^2_{-}=\tau^2_{+}=\tau^2= 0.50^2}\\
\multicolumn{9}{c}{\hbox{Dependence of the subpopulation membership on the covariate: } \alpha_X^{-}=0.8 \hbox{ and } \alpha_X^{+}=-0.40}\\
\vspace{-0.35cm}\\
\hline
\multicolumn{3}{l}{\hbox{$S-X$ association} }&  &\hbox{Average of the} &  &  & \multicolumn{2}{c}{\hbox{Coverage}}
\\
\cline{8-9}
\beta_X^{-}	& \beta_X^{+} & \beta_X & \hbox{True value}& \hbox{Posterior Means} & \hbox{Bias}& \hbox{MSE}& \hbox{95\% HDI } &
\hbox{99\% HDI}\\
\hline
1.50 & 1.25 & 1.75 &  \pi(\Uset)= 0.5211 & 0.5232 & 0.0020 & 0.0000 & 0.98 & 1.00 \\ 
&      &      & ACE_{\Uset}= 0.9995 & 0.9917 & -0.0077 & 0.0006 & 0.90 & 1.00 \\ 
\vspace{-0.25cm}\\
0.75 & 0.70 & 0.80	& \pi(\Uset)=0.5221 & 0.5096 & -0.0125 & 0.0039 & 0.64 & 0.84 \\ 		
& & &ACE_{\Uset}=  0.9996 & 1.0449 & 0.0454 & 0.0573 & 0.82 & 0.92 \\ 
\hline
\end{array}
$$

$$
\begin{array}{ccc rcrccc}
	\multicolumn{9}{c}{\hbox{Conditional outcome variance given the forcing variable: } \tau^2_{-}=\tau^2_{+}=\tau^2= 0.50^2}\\
	\multicolumn{9}{c}{\hbox{Dependence of the subpopulation membership on the covariate: } \alpha_X^{-}=0.4 \hbox{ and } \alpha_X^{+}=-0.20}\\
	\vspace{-0.35cm}\\
	\hline
	\multicolumn{3}{l}{\hbox{$S-X$ association} }&  &\hbox{Average of the} &  &  & \multicolumn{2}{c}{\hbox{Coverage}}
	\\
	\cline{8-9}
	\beta_X^{-}	& \beta_X^{+} & \beta_X & \hbox{True value}& \hbox{Posterior Means} & \hbox{Bias}& \hbox{MSE}& \hbox{95\% HDI } &
	\hbox{99\% HDI}\\
	\hline
	1.50 & 1.25 & 1.75 &  \pi(\Uset)= 0.5749 & 0.5777 & 0.0028 & 0.0000 & 0.82 & 0.94 \\  
	&      &      & ACE_{\Uset}=  1.0023 & 1.0018 & -0.0005 & 0.0005 & 0.92 & 1.00\\
	\vspace{-0.25cm}\\
	0.75 & 0.70 & 0.80	& \pi(\Uset)=  0.5746 & 0.5753 & 0.0008 & 0.0029 & 0.60 & 0.82 \\ 		
	& & &ACE_{\Uset}= 0.9996 & 1.0074 & 0.0078 & 0.0197 & 0.96 & 0.98   \\ 
	\hline
\end{array}
$$

$$
\begin{array}{ccc rcrccc}
	\multicolumn{9}{c}{\hbox{Conditional outcome variance given the forcing variable: } \tau^2_{-}=\tau^2_{+}=\tau^2= 0.75^2}\\
	\multicolumn{9}{c}{\hbox{Dependence of the subpopulation membership on the covariate: } \alpha_X^{-}=0.8 \hbox{ and } \alpha_X^{+}=-0.40}\\
	\vspace{-0.35cm}\\
	\hline
	\multicolumn{3}{l}{\hbox{$S-X$ association} }&   &\hbox{Average of the} &  &  & \multicolumn{2}{c}{\hbox{Coverage}}
	\\
	\cline{8-9}
	\beta_X^{-}	& \beta_X^{+} & \beta_X & \hbox{True value}& \hbox{Posterior Means} & \hbox{Bias}& \hbox{MSE}& \hbox{95\% HDI } &
	\hbox{99\% HDI}\\
	\hline
	1.50 & 1.25 & 1.75 &  \pi(\Uset)= 0.5214 & 0.5263 & 0.0049 & 0.0000 & 0.82 & 0.94 \\     
	&      &      & ACE_{\Uset}= 0.9990 & 0.9865 & -0.0125 & 0.0012 & 0.94 & 0.98 \\ 
	\vspace{-0.25cm}\\
	0.75 & 0.70 & 0.80	& \pi(\Uset)= 0.5224 & 0.5227 & 0.0004 & 0.0022 & 0.64 & 0.86  \\ 		
	& & &ACE_{\Uset}=   0.9988 & 1.0003 & 0.0015 & 0.0321 & 0.88 & 0.96 \\ 
	\hline
\end{array}
$$

$$
\begin{array}{ccc rcrccc}
	\multicolumn{9}{c}{\hbox{Conditional outcome variance given the forcing variable: } \tau^2_{-}=\tau^2_{+}=\tau^2= 0.75^2}\\
	\multicolumn{9}{c}{\hbox{Dependence of the subpopulation membership on the covariate: } \alpha_X^{-}=0.4 \hbox{ and } \alpha_X^{+}=-0.20}\\
	\vspace{-0.35cm}\\
	\hline
	\multicolumn{3}{l}{\hbox{$S-X$ association} }&   &\hbox{Average of the} &  &  & \multicolumn{2}{c}{\hbox{Coverage}}
	\\
	\cline{8-9}
	\beta_X^{-}	& \beta_X^{+} & \beta_X & \hbox{True value}& \hbox{Posterior Means} & \hbox{Bias}& \hbox{MSE}& \hbox{95\% HDI } &
	\hbox{99\% HDI}\\
	\hline
	1.50 & 1.25 & 1.75 &  \pi(\Uset)=  0.5745 & 0.5798 & 0.0053 & 0.0001 & 0.78 & 0.88 \\ 
	&      &      & ACE_{\Uset}= 1.0032 & 0.9988 & -0.0044 & 0.0011 & 0.94 & 1.00 \\ 
	\vspace{-0.25cm}\\
	0.75 & 0.70 & 0.80	& \pi(\Uset)=  0.5748 & 0.5768 & 0.0020 & 0.0051 & 0.34 & 0.54  \\ 		
	& & &ACE_{\Uset}=   1.0004 & 1.0011 & 0.0007 & 0.0205 & 0.86 & 0.94  \\ 
	\hline
\end{array}
$$
\end{table}

Table~\ref{tab:sim2resSA} shows the simulation results derived using local polynomial estimators under the continuity assumption. 

\begin{table}[h]
\caption{Simulation setting 2: Local polynomial estimators under the continuity assumption. Average of the point estimates and of  95\% and 99\% confidence intervals.}   \label{tab:sim2resSA}
\small
\begin{center}
Conditional outcome variance given the forcing variable: $\tau^2_{-}=\tau^2_{+}=\tau^2= 0.50^2$
\end{center}
\vspace{-0.1cm}
$$
 \hspace*{-2cm}
\begin{array}{cccc ccccc c ccccc}
\hline
& & & & \multicolumn{11}{c}{\hbox{Dependence of the subpopulation membership on the covariate: } }\\
\cline{5-15}
& & & & \multicolumn{5}{c}{\alpha_X^{-}=0.8 \hbox{ and } \alpha_X^{+}=-0.40} &&
\multicolumn{5}{c}{\alpha_X^{-}=0.4 \hbox{ and } \alpha_X^{+}=-0.20} \\
\cline{5-9}\cline{11-15}
\multicolumn{3}{l}{\hbox{$S-X$ association} }&&\multicolumn{5}{c}{\hbox{Average of the}}&&\multicolumn{5}{c}{\hbox{Average of the}}
\\
\cline{5-9}\cline{11-15}
\beta_X^{-}	& \beta_X^{+} & \beta_X & &\hbox{Estimates}& \multicolumn{2}{c}{\hbox{95\% CI} }&  \multicolumn{2}{c}{\hbox{99\% CI} } && \hbox{Estimates}& \multicolumn{2}{c}{\hbox{95\% CI} }&  \multicolumn{2}{c}{\hbox{99\% CI} }\\
\hline
1.50 & 1.25 & 1.75 & \\ 
\cline{1-3}
\multicolumn{4}{r}{\hbox{Uniform kernel ($p=1$)}} & 1.0923 & (0.9771; & 1.2075) & (0.9409; & 1.2437) & & 1.2461 & (1.1322; & 1.3599) & (1.0965; & 1.3957)\\ 
\multicolumn{4}{r}{\hbox{Triangular kernel ($p=1$)}} & 1.0920 & (0.9812; & 1.2028) & (0.9464 & 1.2377) &&1.2472 & (1.1372; & 1.3571) & (1.1026; & 1.3917) \\ 
\multicolumn{4}{r}{\hbox{Uniform kernel ($p=2$)}}& 1.0578 & (0.9329; & 1.1828) & (0.8936; & 1.2220) &&  1.2252 & (1.0979; & 1.3525) & (1.0579; & 1.3926)\\ 
\multicolumn{4}{r}{\hbox{Triangular kernel ($p=2$)}}& 1.0601 & (0.9393; & 1.1808) & (0.9014; & 1.2188) && 1.2257 &(1.1022; & 1.3491) & (1.0634; & 1.3879)  \\ 
\vspace{-0.25cm}\\
0.75 & 0.70 & 0.80	& \\ 		
\cline{1-3}
\multicolumn{4}{r}{\hbox{Uniform kernel ($p=1$)}} & 1.0107 & (0.8752; & 1.1462) & (0.8326; & 1.1888)&& 1.0597 & (0.9276; & 1.1917) & (0.8862; & 1.2332)  \\ 
\multicolumn{4}{r}{\hbox{Triangular kernel ($p=1$)}} & 1.0151 & (0.8844; & 1.1457) & (0.8434; & 1.1868) &&  1.0567 & (0.9281; & 1.1852) & (0.8877; & 1.2256) \\
\multicolumn{4}{r}{\hbox{Uniform kernel ($p=2$)}}& 0.9728 & (0.8274; & 1.1181) & (0.7818; & 1.1637) &&  1.0298 & (0.8860; & 1.1735) & (0.8408; & 1.2187) \\
\multicolumn{4}{r}{\hbox{Triangular kernel ($p=2$)}}&0.9858 & (0.8444; & 1.1271) & (0.7999; & 1.1716) && 1.0380 & (0.8948; & 1.1813) & (0.8497; & 1.2263) \\
\hline
\end{array}
 \hspace*{-2cm}
$$
\begin{center}
Conditional outcome variance given the forcing variable: $\tau^2_{-}=\tau^2_{+}=\tau^2= 0.75^2$
\end{center}
\vspace{-0.1cm}
$$
 \hspace*{-2cm}
\begin{array}{cccc ccccc c ccccc}
\hline
& & & & \multicolumn{11}{c}{\hbox{Dependence of the subpopulation membership on the covariate: } }\\
\cline{5-15}
& & & & \multicolumn{5}{c}{\alpha_X^{-}=0.8 \hbox{ and } \alpha_X^{+}=-0.40} &&
\multicolumn{5}{c}{\alpha_X^{-}=0.4 \hbox{ and } \alpha_X^{+}=-0.20} \\
\cline{5-9}\cline{11-15}
\multicolumn{3}{l}{\hbox{$S-X$ association} }&&\multicolumn{5}{c}{\hbox{Average of the}}&&\multicolumn{5}{c}{\hbox{Average of the}}
\\
\cline{5-9}\cline{11-15}
\beta_X^{-}	& \beta_X^{+} & \beta_X & &\hbox{Estimates}& \multicolumn{2}{c}{\hbox{95\% CI} }&  \multicolumn{2}{c}{\hbox{99\% CI} } && \hbox{Estimates}& \multicolumn{2}{c}{\hbox{95\% CI} }&  \multicolumn{2}{c}{\hbox{99\% CI} }\\
\hline
1.50 & 1.25 & 1.75 & \\ 
\multicolumn{4}{r}{\hbox{Uniform kernel ($p=1$)}} &  1.1019 & (0.9560; & 1.2478) & (0.9102; & 1.2937) &&  1.2550 & (1.1150; & 1.3949) & (1.0710; & 1.4389) \\ 
\multicolumn{4}{r}{\hbox{Triangular kernel ($p=1$)}} & 1.0959 & (0.9565; & 1.2353) & (0.9127; & 1.2791) && 1.2563 & (1.1210; & 1.3916) & (1.0785; & 1.4341)\\ 
\multicolumn{4}{r}{\hbox{Uniform kernel ($p=2$)}}&   1.0513 & (0.8923; & 1.2104) & (0.8423; & 1.2604) && 1.2264 & (1.0672; & 1.3856) & (1.0172; & 1.4357)\\ 
\multicolumn{4}{r}{\hbox{Triangular kernel ($p=2$)}}& 1.0599 & (0.9061; & 1.2137) & (0.8578; & 1.2620) &&1.2339 & (1.0783; & 1.3895) & (1.0294; & 1.4383)   \\ 
\vspace{-0.25cm}\\
0.75 & 0.70 & 0.80	& \\ 		
\cline{1-3}
\multicolumn{4}{r}{\hbox{Uniform kernel ($p=1$)}} & 1.0196 & (0.8563; & 1.1829) & (0.8050; & 1.2342) & & 1.0610 & (0.9043; & 1.2178) & (0.8551; & 1.2670) \\ 
\multicolumn{4}{r}{\hbox{Triangular kernel ($p=1$)}} & 1.0176 & (0.8606; & 1.1745) & (0.8113; & 1.2238) && 1.0568 & (0.9044; & 1.2091) & (0.8565; & 1.2570)  \\
\multicolumn{4}{r}{\hbox{Uniform kernel ($p=2$)}} & 0.9766 & (0.8008; & 1.1523) & (0.7456; & 1.2076) &&  1.0251 & (0.8531; & 1.1970) & (0.7991; & 1.2511) \\
\multicolumn{4}{r}{\hbox{Triangular kernel ($p=2$)}} & 0.9866 & (0.8140; & 1.1591) & (0.7597; & 1.2134) && 1.0369 & (0.8648; & 1.2090) & (0.8107; & 1.2631) \\
\hline
\end{array}
 \hspace*{-2cm}
$$
\end{table}

\subsubsection*{Simulation Results: Setting 3} 
In setting 3, we again focus on the proportion of units in $\Uset$, $\pi(\Uset)$ (Equation~\eqref{puset}) and the average causal effect for units in $\Uset$, $ ACE_{\Uset}$ (Equation~\eqref{aceuset}).
Table \ref{tab:sim3res}  shows the results, which suggest that 
the Bayesian model-mixture approach to clustering we propose leads to more accurate and more efficient estimates of the causal estimands  (that is, estimates with smaller bias, smaller MSE, and higher coverage) in studies where the dependence of subpopulation membership on the forcing variable
is stronger.

\begin{table}\caption{Simulation setting 3: Simulation results. Average of the posterior means, bias, MSE, and 95\% and 99\% coverage for $\pi(\Uset)$ and $ACE_{\Uset}$.}
	\label{tab:sim3res}
$$
\begin{array}{ccc rcrccc}
	\multicolumn{9}{c}{\hbox{Conditional outcome variance given the forcing variable: } \tau^2_{-}=\tau^2_{+}=\tau^2= 0.50^2}\\
	\vspace{-0.35cm}\\
	\hline
	&&& &  \hbox{Average of the} &  &  & \multicolumn{2}{c}{\hbox{Coverage}}
	\\
	\cline{8-9}
	\beta_0^{-}	& \beta_0^{+} & \beta_0 & \hbox{True value}&\hbox{Posterior Means} & \hbox{Bias}& \hbox{MSE}& \hbox{95\% HDI } &
	\hbox{99\% HDI}\\
	\hline
	-2.00 & 2.25 & -1.00 & \pi(\Uset)=0.4002 & 0.3999 & -0.0002 & 0.0001 & 0.94 & 1.00 \\     
	&&& ACE_{\Uset}=1.0012 & 1.0014 & 0.0003 & 0.0007 & 0.96 & 1.00 \\ 
	\vspace{-0.25cm}\\
	-0.50 & 0.75 & -0.25	& \pi(\Uset)=0.4006 & 0.3979 & -0.0026 & 0.0003 & 0.94 & 0.98 	\\
	&&& ACE_{\Uset}=1.0010 & 1.0036 & 0.0026 & 0.0014 & 0.94 & 0.96 \\      		
	\vspace{-0.25cm}\\
	\hline
\end{array}
$$

$$
\begin{array}{ccc rcrccc}
	\multicolumn{9}{c}{\hbox{Conditional outcome variance given the forcing variable: } \tau^2_{-}=\tau^2_{+}=\tau^2= 0.75^2}\\
	\vspace{-0.35cm}\\
	\hline
	&&& &  \hbox{Average of the} &  &  & \multicolumn{2}{c}{\hbox{Coverage}}
	\\
	\cline{8-9}
	\beta_0^{-}	& \beta_0^{+} & \beta_0 & \hbox{True value} &\hbox{Posterior Means} & \hbox{Bias}& \hbox{MSE}& \hbox{95\% HDI } &
	\hbox{99\% HDI}\\
	\hline
	-2.00 & 2.25 & -1.00 & \pi(\Uset)=   0.3999 & 0.3979 & -0.0020 & 0.0002 & 0.88 & 0.98  \\ 
	&      &      &  ACE_{\Uset}= 1.0020 & 1.0025 & 0.0004 & 0.0018 & 0.96 & 0.98 \\ 
	\vspace{-0.25cm}\\
	-0.50 & 0.75 & -0.25	&  \pi(\Uset)= 0.4002 & 0.3957 & -0.0045 & 0.0009 & 0.90 & 1.00\\ 		
	& & &  ACE_{\Uset}=  1.0008 & 1.0003 & -0.0005 & 0.0068 & 0.90 & 0.98\\ 
	\hline
\end{array}
$$
\end{table}

Table~\ref{tab:sim3resSA} shows the simulation results derived using local polynomial estimators under the continuity assumption. 

\begin{table}[h]
\caption{Simulation setting 3: Local polynomial estimators under the continuity assumption. Average of the point estimates and of  95\% and 99\% confidence intervals.}   \label{tab:sim3resSA}
$$
 \hspace*{-2.4cm}
\begin{array}{cccc ccccc c ccccc}
\hline
& & & & \multicolumn{11}{c}{\hbox{Conditional outcome variance given the forcing variable: } }\\
\cline{5-15}
& & & & \multicolumn{5}{c}{\tau^2_{-}=\tau^2_{+}=\tau^2= 0.50^2} &&
\multicolumn{5}{c}{ \tau^2_{-}=\tau^2_{+}=\tau^2= 0.75^2} \\
\cline{5-9}\cline{11-15}
\multicolumn{3}{l}{}&&\multicolumn{5}{c}{\hbox{Average of the}}&&\multicolumn{5}{c}{\hbox{Average of the}}
\\
\cline{5-9}\cline{11-15}
\beta_0^{-}	& \beta_0^{+} & \beta_0 & &\hbox{Estimates}& \multicolumn{2}{c}{\hbox{95\% CI} }&  \multicolumn{2}{c}{\hbox{99\% CI} } && \hbox{Estimates}& \multicolumn{2}{c}{\hbox{95\% CI} }&  \multicolumn{2}{c}{\hbox{99\% CI} }\\
\hline
-2.00 & 2.25 & -1.00 & \\ 
\cline{1-3}
\multicolumn{4}{r}{\hbox{Uniform kernel ($p=1$)}}&   1.0469 & (0.9326 & 1.1612) & (0.8967 & 1.1971) && 1.0674 & (0.9147 & 1.2202) & (0.8667 & 1.2682)\\
\multicolumn{4}{r}{\hbox{Triangular kernel ($p=1$)}}   & 1.0454 & (0.9339 & 1.1570) & (0.8989 & 1.1920)&&1.0484 & (0.8951 & 1.2016) & (0.8470 & 1.2498) \\
\multicolumn{4}{r}{\hbox{Uniform kernel ($p=2$)}}& 1.0104 & (0.8858 & 1.1349) &(0.8467 & 1.1741 )&& 1.0074 & (0.8361 & 1.1787) & (0.7823 & 1.2325)\\
\multicolumn{4}{r}{\hbox{Triangular kernel ($p=2$)}}&  1.0126 & (0.8946 & 1.1307) & (0.8575& 1.1677)&&1.0097& (0.8437 & 1.1757) & (0.7916 & 1.2278) \\
\vspace{-0.25cm}\\
-0.50 & 0.75 & -0.25 & \\ 		
\cline{1-3}
\multicolumn{4}{r}{\hbox{Uniform kernel ($p=1$)}}&  1.0440 & (0.9716 & 1.1163) & (0.9489 & 1.1390) && 1.0489 & (0.9470 & 1.1509) & (0.9150 & 1.1829)\\
\multicolumn{4}{r}{\hbox{Triangular kernel ($p=1$)}} &  1.0450 & (0.9766 & 1.1134) & (0.9551 & 1.1349) && 1.0461 & (0.9488 & 1.1434) & (0.9182 & 1.1740)\\
\multicolumn{4}{r}{\hbox{Uniform kernel ($p=2$)}}& 1.0304 &( 0.9430 & 1.1177) & 0.9156 & 1.1452)  && 1.0311 & (0.9075 & 1.1547) & (0.8686 & 1.1935) \\ 
\multicolumn{4}{r}{\hbox{Triangular kernel ($p=2$)}}&  1.0322 &( 0.9479 & 1.1166) & (0.9214 & 1.1431)&&   1.0300 & (0.9081 & 1.1519) & (0.8698 & 1.1902)\\
\hline
\end{array}
 \hspace*{-1cm}
$$
\end{table}
\clearpage

\section{Mixture-model Bayesian Analysis: Prior versus Posterior Distributions of Individual Mixing Probabilities}  \label{sec:BA_PriorPost}

\begin{figure}[t]
	\begin{center}
		\begin{subfigure}[b]{1\textwidth}
			\centering
			\begin{tabular}{ccc}
				\includegraphics[width=4.0cm]{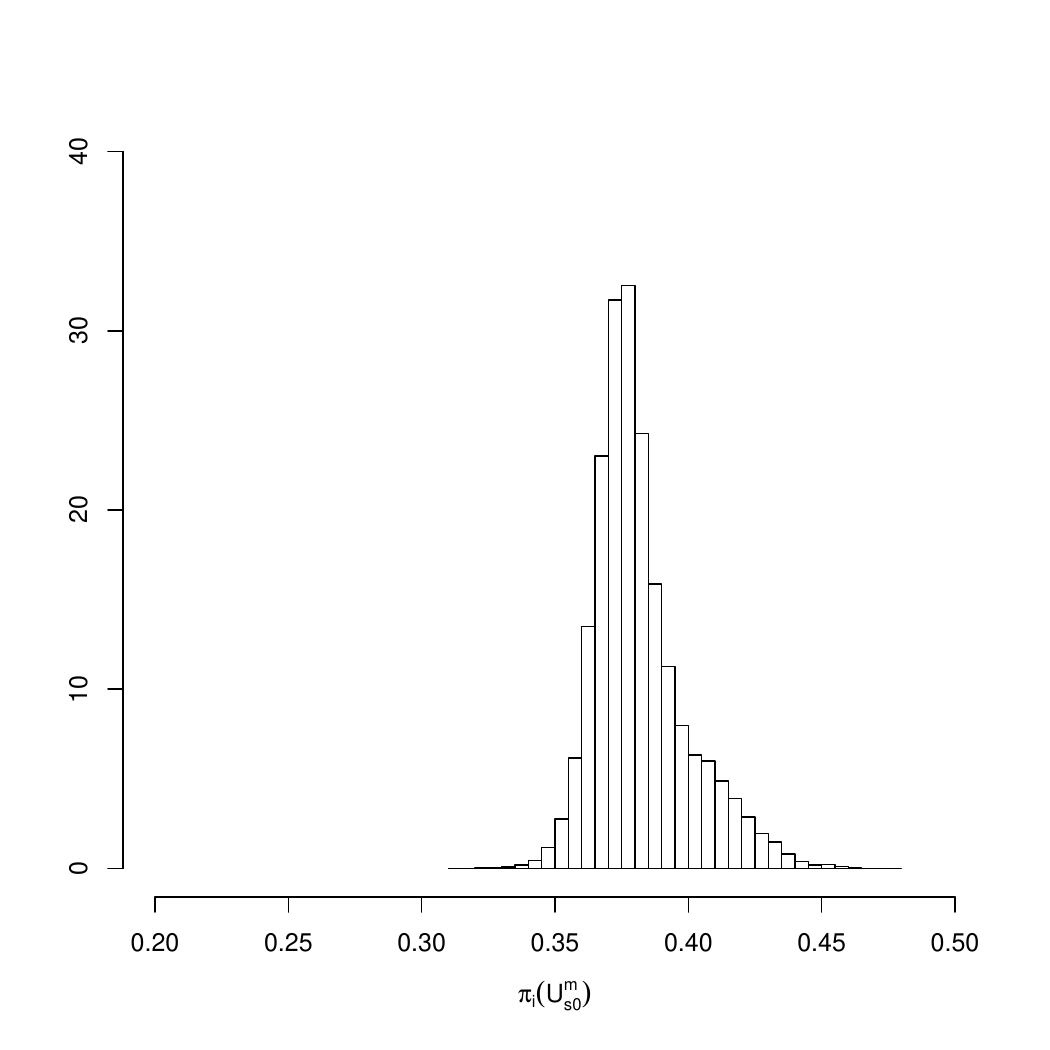} &
				\includegraphics[width=4.0cm]{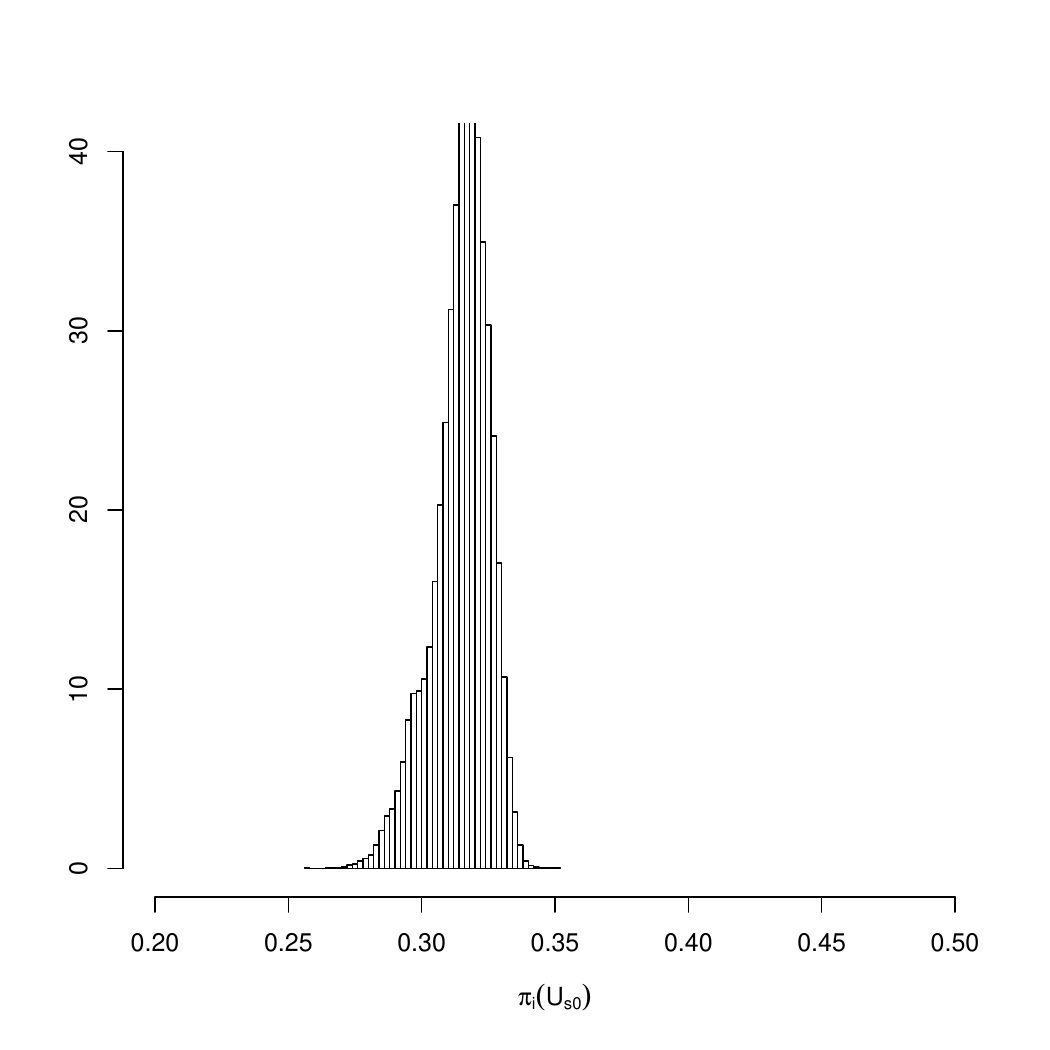}&
				\includegraphics[width=4.0cm]{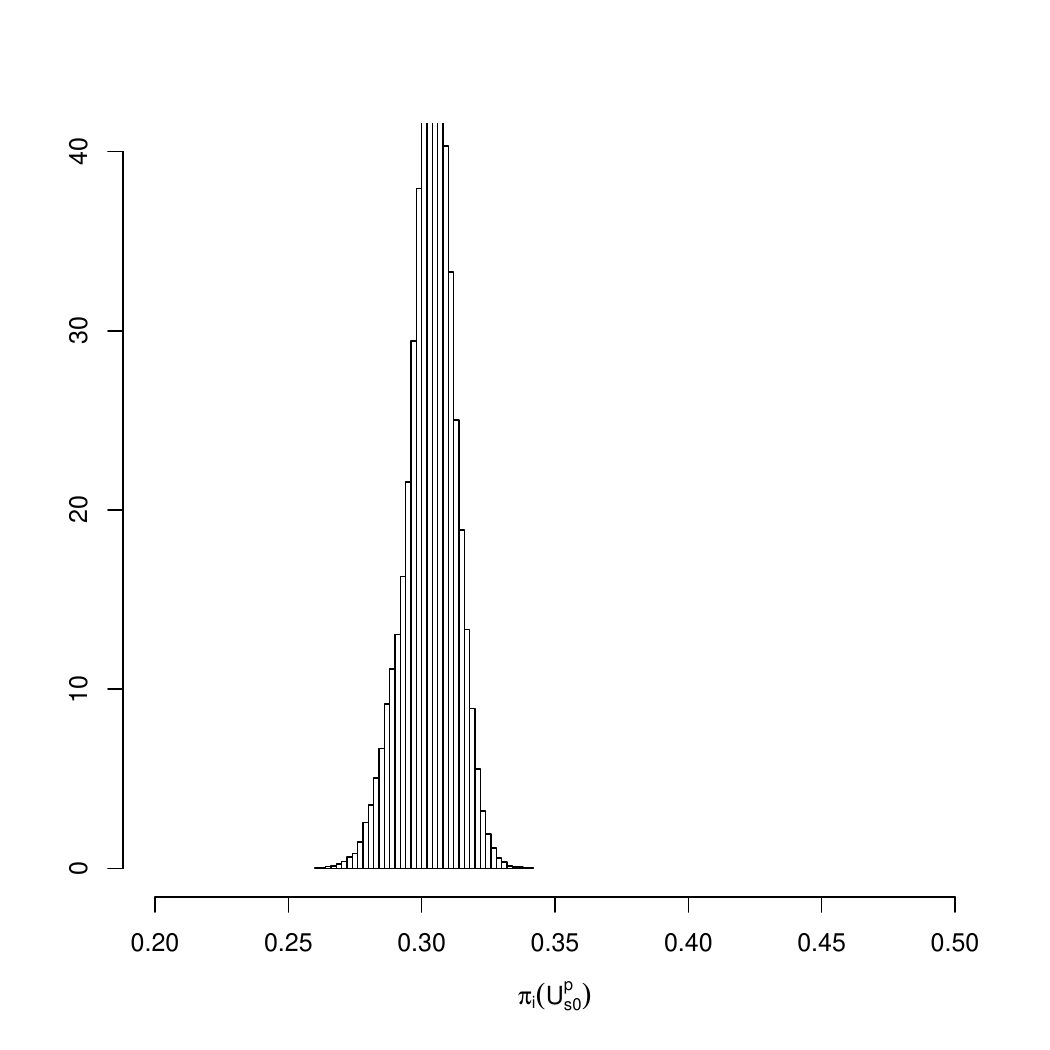}
			\end{tabular}
			\caption{Prior mean of individual  mixing probabilities.} 
				\label{fig:PriorG}
		\end{subfigure}
		\begin{subfigure}[b]{1\textwidth}
			\centering
			\begin{tabular}{ccc}
				\includegraphics[width=4.0cm]{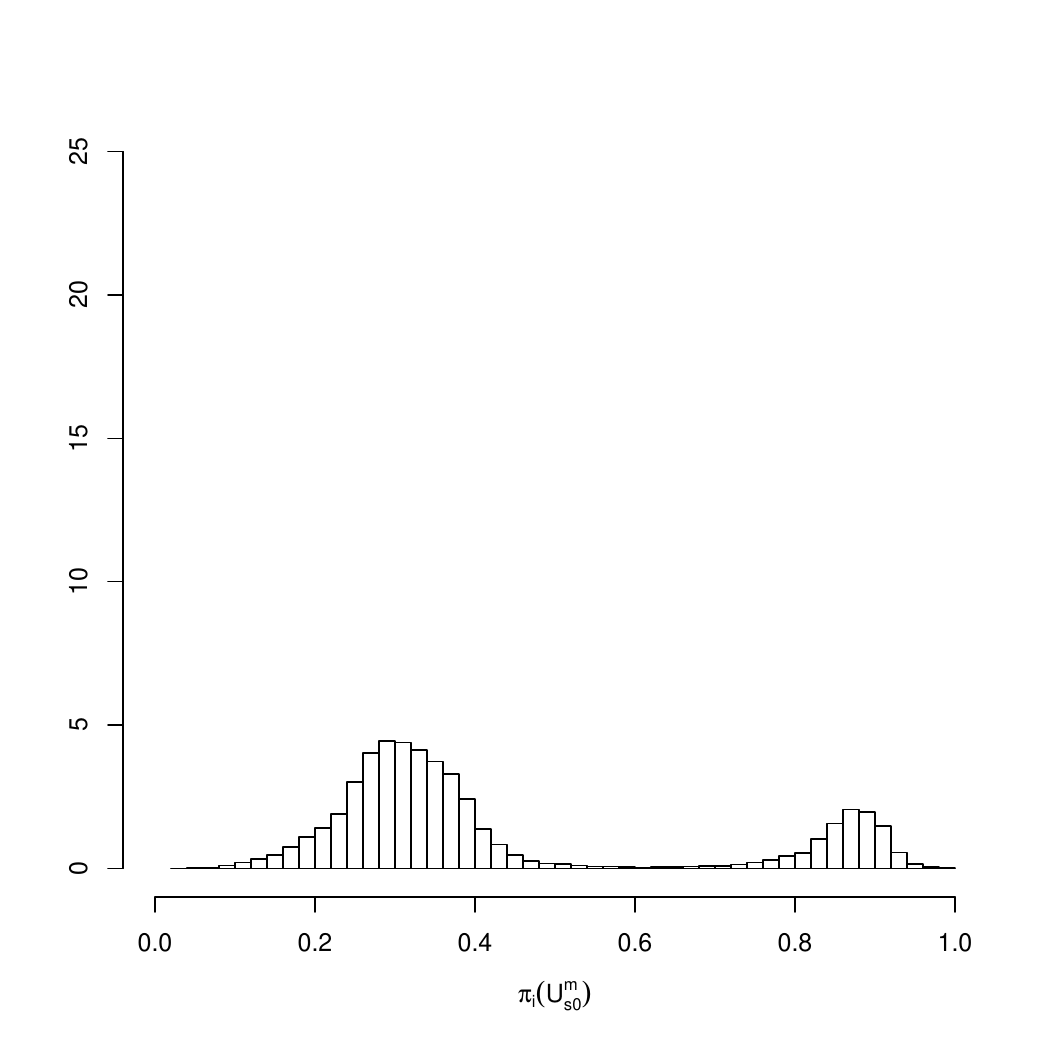} &
				\includegraphics[width=4.0cm]{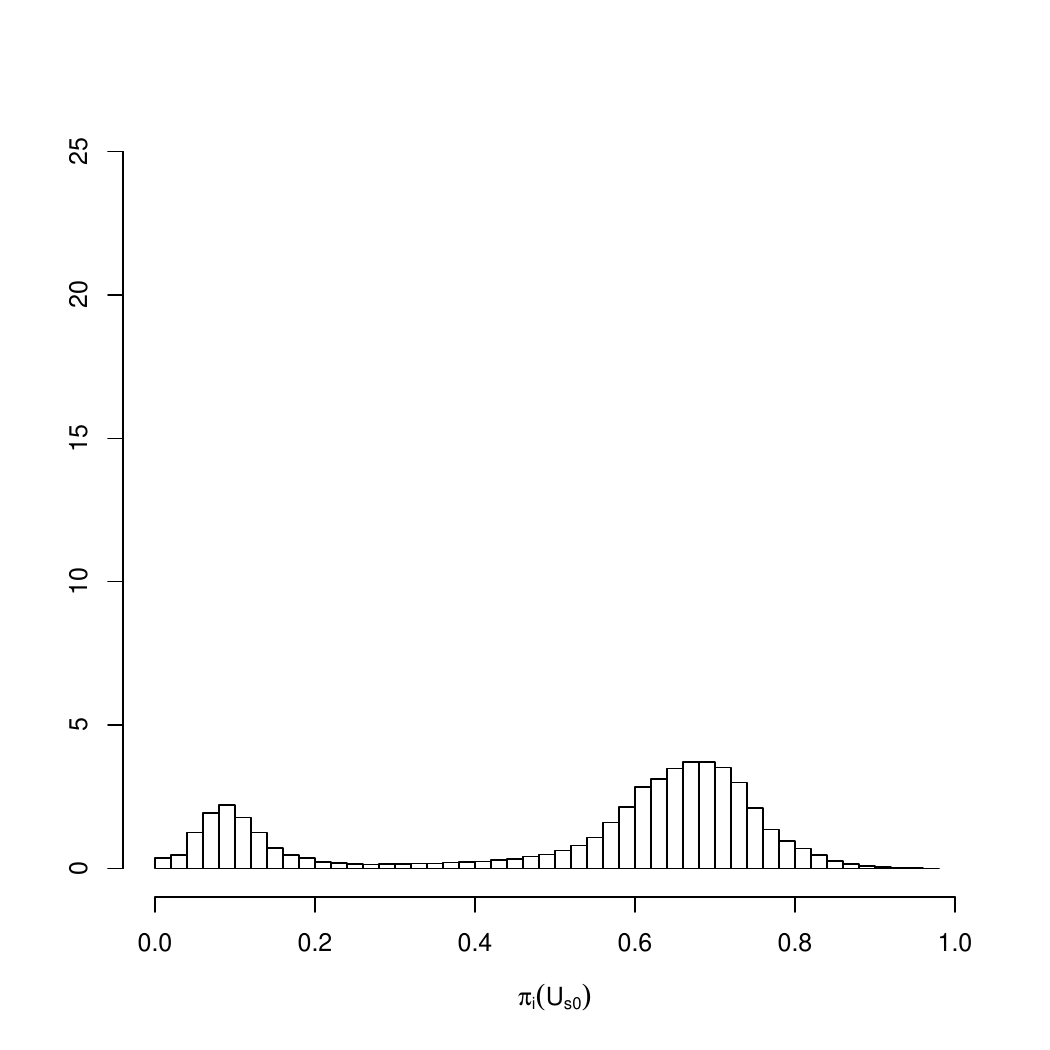}&
				\includegraphics[width=4.0cm]{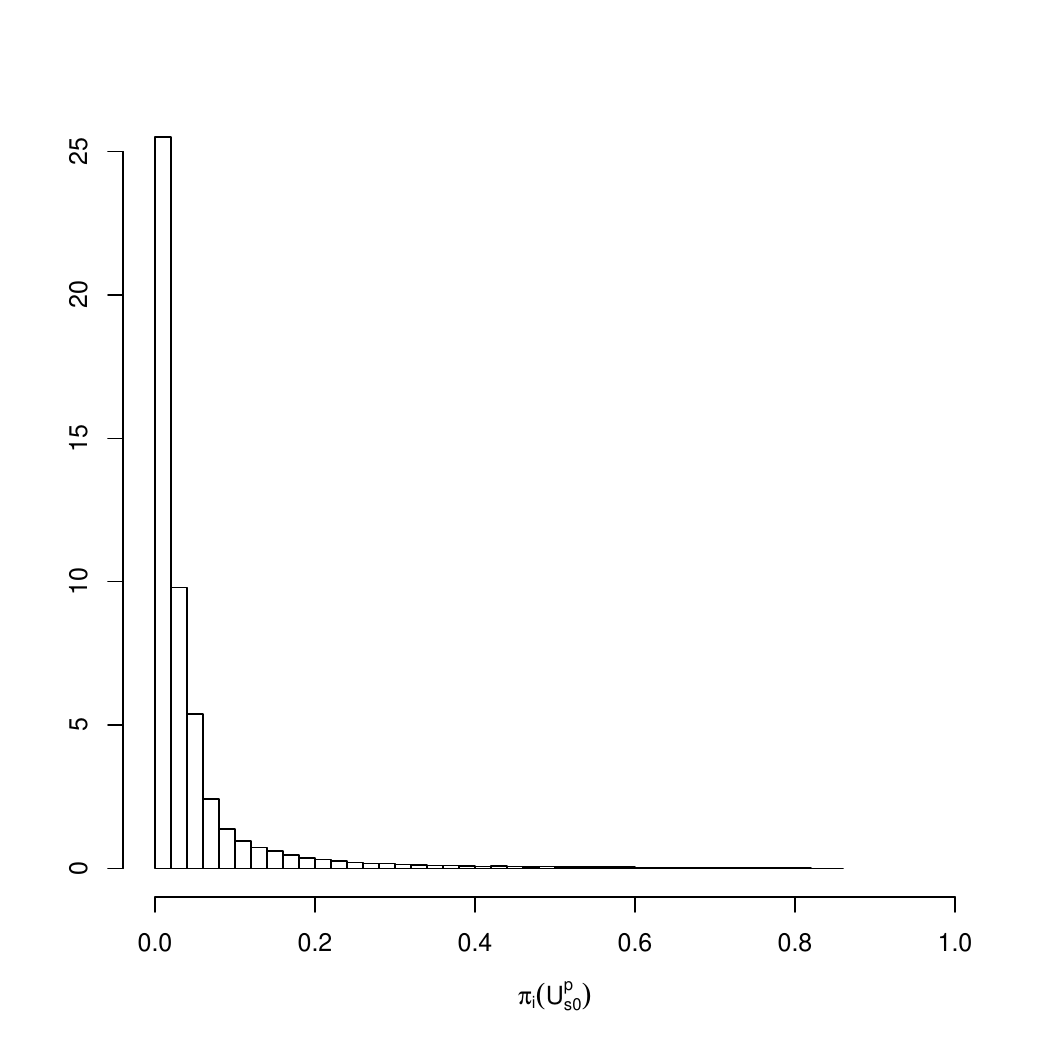}
			\end{tabular}
			\caption{Posterior mean of individual  mixing probabilities.} \label{fig:PostG}
		\end{subfigure}
	\end{center}
	\caption{Bolsa Fam\'ilia study: Mixture-model Bayesian analysis. 	Prior and posterior mean of individual  mixing probabilities.}
		\label{fig:mixprobs}
\end{figure}

Figure~\ref{fig:mixprobs} shows the distribution of the prior mean and the posterior mean of the individual mixing probabilities $\pi_i(\Usetm)$, $\pi_i(\Usetp)$ and $\pi_i(\Uset)$, across individuals. 
As we can see in Figure~\ref{fig:mixprobs},  data are informative about subpopulation membership: the information in the data is able to update the prior and shift the posterior distributions of the mixing probabilities.

\section{Mixture-model Bayesian analysis: Covariate Distributions by Subpopulation Membership}  \label{sec:BA_balance}

Table \ref{A-tab:DistrX} shows the posterior median of the sample means and standard deviations of the covariates by sub-population membership.

\begin{figure}[t]
\caption{Bolsa Fam\'ilia study: Mixture-model Bayesian analysis.  Posterior median of the Normalized mean differences for families in $\Uset$.} \label{A-fig:bal} 
\begin{center}
\begin{tabular}{ccc}
\includegraphics[width=5cm]{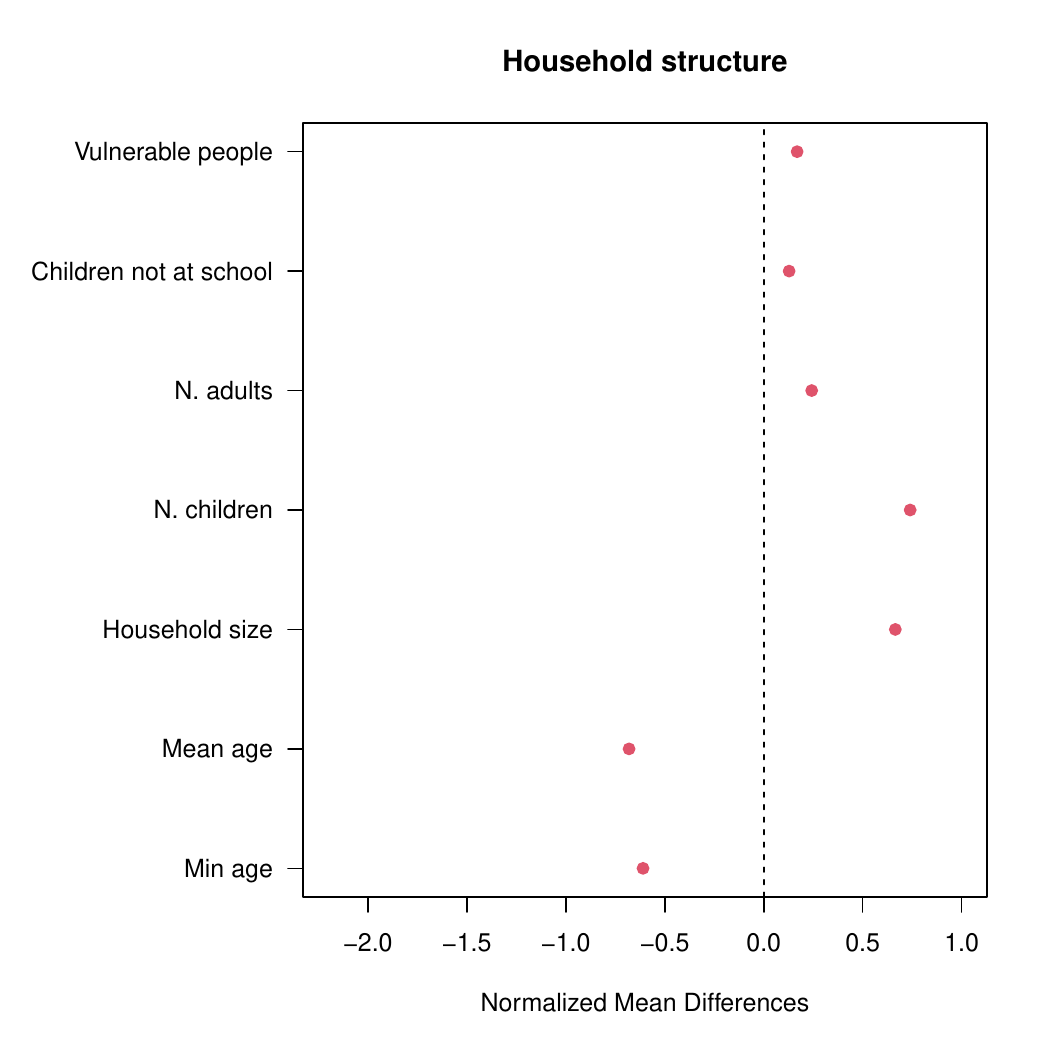} & 	\includegraphics[width=5cm]{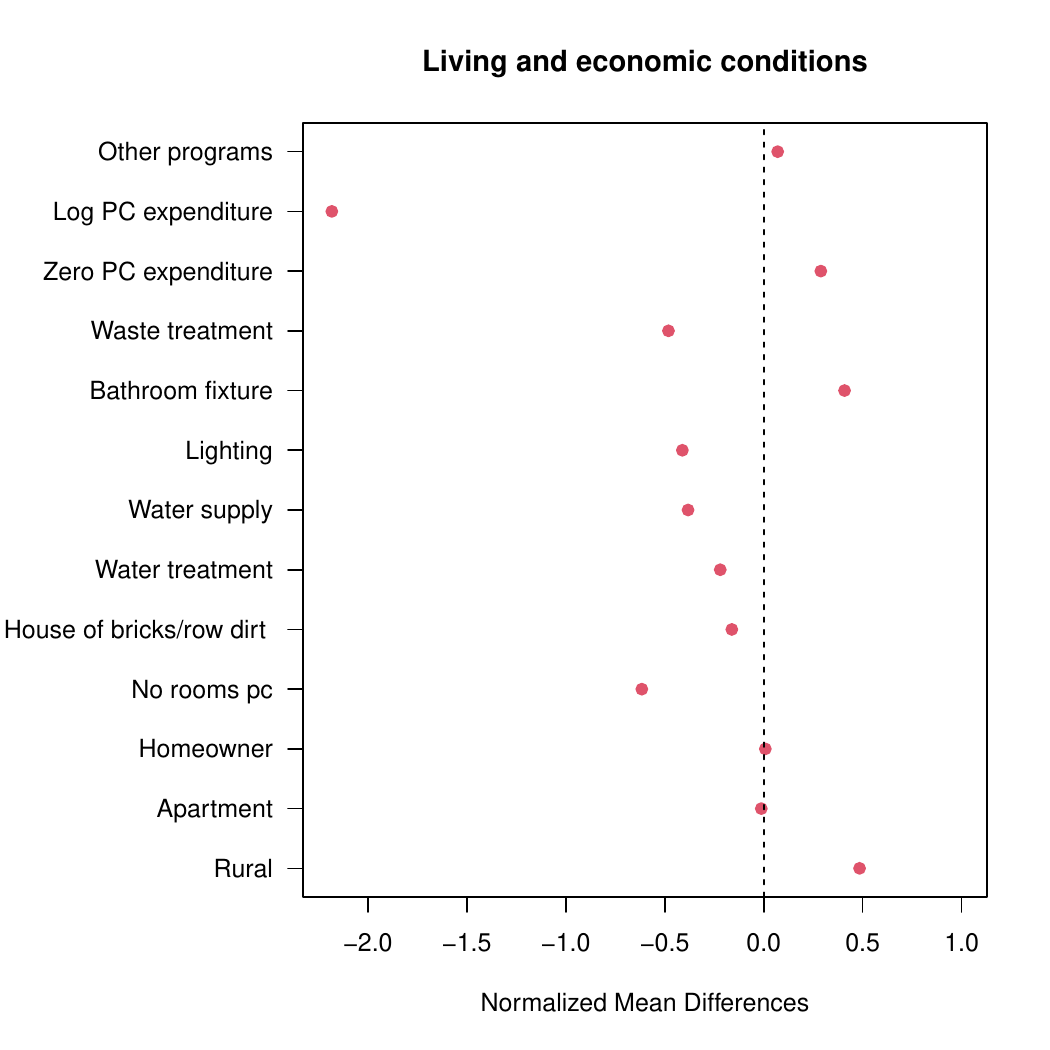} &
\includegraphics[width=5cm]{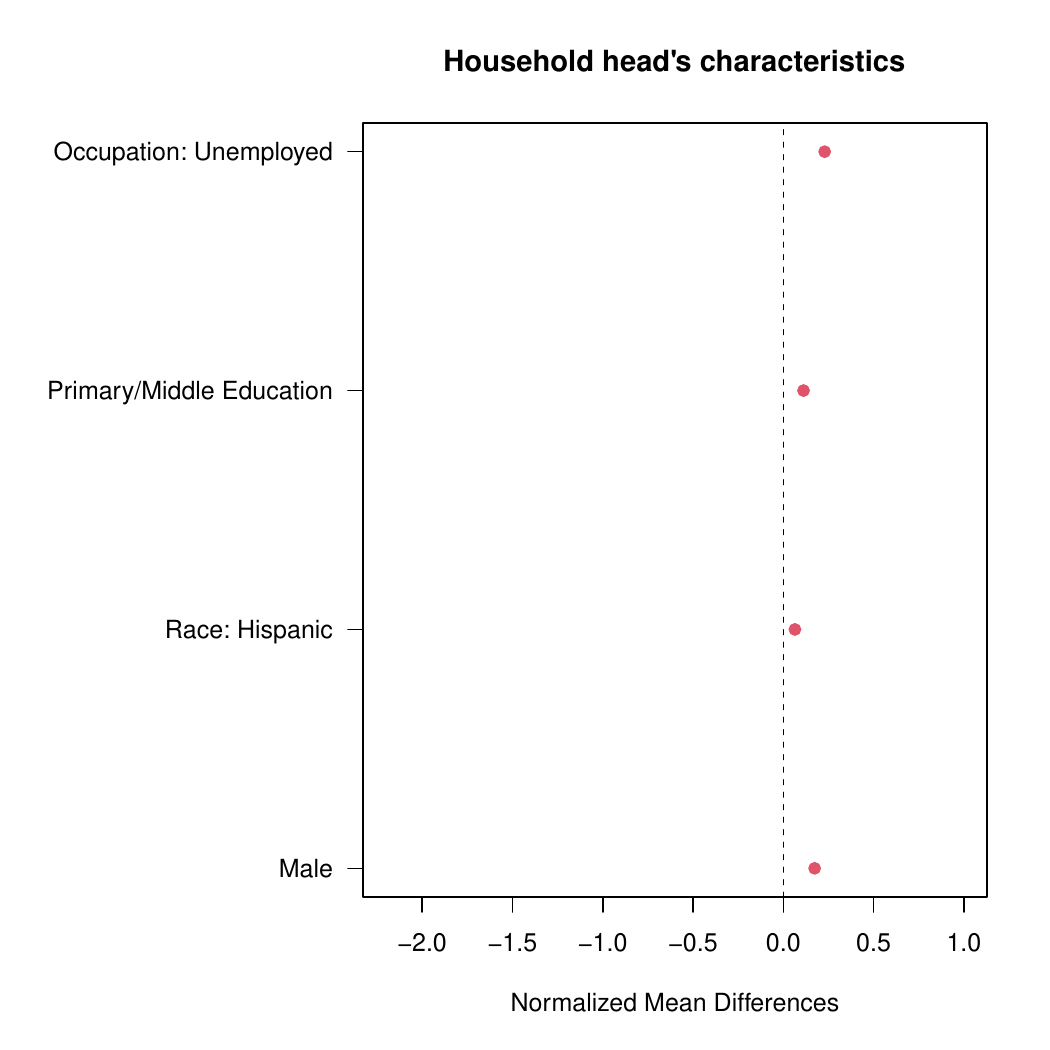}\\
\end{tabular}
\end{center}
\end{figure}

Table \ref{A-tab:bal} shows the posterior median of some measure of covariate balance between eligible and ineligible families in the subpopulation $\Uset$. Specifically for families in $\Uset$, Table \ref{A-tab:bal} presents:
\begin{enumerate}
	\item The posterior median of the sample averages and  standard deviations by eligibility status:
	$$
	\overline{X}_{j,0}= \dfrac{1}{N_{\Uset,0}} \sum_{i \in \Uset} (1-Z_i) X_{ij}  \qquad \hbox{and} \qquad \overline{X}_{j,1}= \dfrac{1}{N_{\Uset,1}} \sum_{i \in \Uset}  Z_i X_{ij},
	$$
	and
	$$
	s^2_{j,0}= \dfrac{\sum_{i \in \Uset} (1-Z_i) \left(X_{ij} - \overline{X}_{j,0}\right)^2 }{N_{\Uset,0}-1} \qquad  \hbox{and} \quad s^2_{j,1}= \dfrac{\sum_{i \in \Uset}  Z_i \left(X_{ij} - \overline{X}_{j,1}\right)^2}{N_{\Uset,1}-1},
	$$
	where $N_{\Uset,0}= \sum_{i \in \Uset} (1-Z_i) $ and $N_{\Uset,1}=\sum_{i \in \Uset}Z_i$;
	\item The posterior median of the difference in means by treatment group, normalized by the square root of the average within-group squared standard deviation:
	$$
	\Delta_{j}=\dfrac{\overline{X}_{j,1}-\overline{X}_{j,0}}{\sqrt{\left(s^2_{j,0}+s^2_{j,1}\right)/2}};
	$$
	\item The posterior median of the log of the ratio of the sample standard deviations:
	$$
	\Gamma_{j}=\log \left(\dfrac{\sqrt{s^2_{j,1}}}{\sqrt{s^2_{j,0}}}\right) = \dfrac{1}{2}\left[\log(s^2_{j,1})- \log(s^2_{j,0})\right];
	$$
	and 
	\item The posterior median of the Mahalanobis distance between the means with   respect to the $\left[\left(\Sigma_0 + \Sigma_1\right)/2\right]^{-1}$ inner product:
	$$
	\Delta^{\mathrm{mv}} = \sqrt{\left(\overline{\bX}_{1}- \overline{\bX}_{0}\right)' \left(\dfrac{\Sigma_0 + \Sigma_1}{2}\right)^{-1}\left(\overline{\bX}_{1}- \overline{\bX}_{0}\right)},
	$$
	where $\overline{\bX}_{z} =  \left[\overline{X}_{1,z}, \ldots, \overline{X}_{p,z}\right]'$, $z=0,1$, and 
	$$
	\Sigma_0 = \dfrac{1}{N_{\Uset,0}-1} \sum_{i \in \Uset} (1-Z_i) \left(\bX_{i} - \overline{\bX}_{0}\right)\left(\bX_{i} - \overline{\bX}_{0}\right)'
	$$
	and
	$$
	\Sigma_1 = \dfrac{1}{N_{\Uset,1}-1} \sum_{i \in \Uset} Z_i \left(\bX_{i} - \overline{\bX}_{1}\right)\left(\bX_{i} - \overline{\bX}_{1}\right)'.
	$$
\end{enumerate}
Figure \ref{A-fig:bal} shows graphically the posterior median of the normalized differences, $\Delta_j$.

\begin{table}[ht]\caption{Bolsa Fam\'ilia study:  Mixture-model Bayesian analysis.  Posterior median of the sample means and standard deviations of the covariates by subpopulation membership.} \label{A-tab:DistrX} 

{\small
$$
\begin{array}{l rr c rr c rr } 
\hline 
\vspace{-0.3cm}\\
&\multicolumn{2}{c}{ \Usetm}& &\multicolumn{2}{c}{\Uset} && \multicolumn{2}{c}{\Usetp}   \\
\cline{2-3} \cline{5-6} \cline{8-9}
\vspace{-0.3cm}\\
\hbox{Covariate } & \hbox{Mean} & \hbox{SD}  && \hbox{Mean} & \hbox{SD} & & \hbox{Mean} & \hbox{SD} \\ 
\hline
\vspace{-0.3cm}\\	
\multicolumn{7}{l}{\hbox{\textit{Household structure}}}\\															
\vspace{-0.3cm}\\															
\hbox{Min age}  	&	9.87	&	13.12	&	&	11.01	&	14.69	&	&	19.61	&	19.11	\\
\hbox{Mean age}  	&	21.97	&	10.92	&	&	22.09	&	11.78	&	&	34.05	&	15.88	\\
\hbox{Household size}  	&	3.12	&	1.34	&	&	2.92	&	1.30	&	&	2.71	&	1.05	\\
\hbox{N. Children}  	&	1.40	&	1.11	&	&	1.34	&	1.10	&	&	0.73	&	0.75	\\
\hbox{N. Adults}  	&	1.67	&	0.78	&	&	1.54	&	0.69	&	&	1.61	&	0.88	\\
\hbox{Children not at school}  	&	0.05	&	0.21	&	&	0.03	&	0.18	&	&	0.02	&	0.14	\\
\hbox{Presence of vulnerable people}  	&	0.22	&	0.42	&	&	0.23	&	0.42	&	&	0.17	&	0.37	\\
\vspace{-0.3cm}\\															
\multicolumn{7}{l}{\hbox{\textit{Living and economic conditions}}}\\															
\vspace{-0.3cm}\\															
\hbox{Rural}  	&	0.41	&	0.49	&	&	0.39	&	0.49	&	&	0.23	&	0.42	\\
\hbox{Apartment}  	&	0.95	&	0.21	&	&	0.95	&	0.23	&	&	0.96	&	0.20	\\
\hbox{Home ownership: Homeowner}  	&	0.62	&	0.49	&	&	0.56	&	0.50	&	&	0.71	&	0.45	\\
\hbox{No rooms pc}  	&	1.54	&	0.99	&	&	1.62	&	1.07	&	&	1.94	&	1.10	\\
\hbox{House of bricks/row dirt }  	&	0.93	&	0.26	&	&	0.90	&	0.30	&	&	0.96	&	0.21	\\
\hbox{Water treatment}  	&	0.78	&	0.41	&	&	0.79	&	0.41	&	&	0.85	&	0.36	\\
\hbox{Water supply}  	&	0.63	&	0.48	&	&	0.62	&	0.48	&	&	0.76	&	0.43	\\
\hbox{Lighting}  	&	0.80	&	0.40	&	&	0.78	&	0.41	&	&	0.91	&	0.29	\\
\hbox{Bathroom fixture}  	&	0.61	&	0.49	&	&	0.62	&	0.49	&	&	0.48	&	0.50	\\
\hbox{Waste treatment}  	&	0.62	&	0.49	&	&	0.64	&	0.48	&	&	0.80	&	0.40	\\
\hbox{Zero PC expenditure}  	&	0.43	&	0.50	&	&	0.04	&	0.19	&	&	0.21	&	0.41	\\
\hbox{Log PC expenditure}  	&	2.15	&	2.05	&	&	3.55	&	0.91	&	&	3.57	&	2.00	\\
\hbox{Other programs}  	&	0.06	&	0.23	&	&	0.06	&	0.24	&	&	0.06	&	0.25	\\
\vspace{-0.3cm}\\															
\multicolumn{7}{l}{\hbox{\textit{Household head's characteristics}}}\\															
\vspace{-0.3cm}\\															
\hbox{Male}  	&	0.88	&	0.33	&	&	0.85	&	0.35	&	&	0.84	&	0.36	\\
\hbox{Race: White}  	&	0.88	&	0.32	&	&	0.89	&	0.32	&	&	0.85	&	0.35	\\
\hbox{Primary/Middle Education}  	&	0.47	&	0.50	&	&	0.47	&	0.50	&	&	0.46	&	0.50	\\
\hbox{Occupation: Unemployed}  	&	0.49	&	0.50	&	&	0.49	&	0.50	&	&	0.36	&	0.48	\\
\hline
\end{array}		$$		
}
\end{table}

\begin{table}[h]\caption{Bolsa Fam\'ilia study: Mixture-model Bayesian analysis.  Posterior median of sample averages and standard deviations by eligibility status and some measure of covariance balance for families in $\Uset$.}\label{A-tab:bal} 
$$
\begin{array}{lrrrrrr}
\hline 
\vspace{-0.3cm}\\
&\multicolumn{2}{c}{\hbox{Ineligible}}& \multicolumn{2}{c}{\hbox{Eligible}} &\hbox{Norm.} &  \\
\hbox{Covariate }& \hbox{Mean} & \hbox{SD}  & \hbox{Mean} & \hbox{SD} & \hbox{Diff.} & \log(s_1/s_0) \\
\vspace{-0.3cm}\\
\hline
\vspace{-0.3cm}\\
\multicolumn{7}{l}{\hbox{\textit{Household structure}}}\\													
\vspace{-0.3cm}\\													
\hbox{Min age}  	&	21.08	&	19.27	&	10.70	&	14.42	&	-0.61	&	-0.29	\\
\hbox{Mean age}  	&	30.86	&	14.76	&	21.83	&	11.57	&	-0.68	&	-0.24	\\
\hbox{Household size}  	&	2.19	&	0.95	&	2.94	&	1.30	&	0.66	&	0.32	\\
\hbox{N. Children}  	&	0.67	&	0.72	&	1.36	&	1.11	&	0.74	&	0.44	\\
\hbox{N. Adults}  	&	1.38	&	0.66	&	1.54	&	0.69	&	0.24	&	0.04	\\
\hbox{Children not at school}  	&	0.01	&	0.12	&	0.03	&	0.18	&	0.13	&	0.42	\\
\hbox{Presence of vulnerable people}  	&	0.16	&	0.37	&	0.23	&	0.42	&	0.17	&	0.13	\\
\vspace{-0.3cm}\\													
\multicolumn{7}{l}{\hbox{\textit{Living and economic conditions}}}\\													
\vspace{-0.3cm}\\													
\hbox{Rural}  	&	0.19	&	0.39	&	0.40	&	0.49	&	0.48	&	0.23	\\
\hbox{Apartment}  	&	0.95	&	0.22	&	0.95	&	0.23	&	-0.01	&	0.02	\\
\hbox{Home ownership: Homeowner}  	&	0.55	&	0.50	&	0.56	&	0.50	&	0.01	&	0.00	\\
\hbox{No rooms pc}  	&	2.36	&	1.39	&	1.60	&	1.05	&	-0.62	&	-0.28	\\
\hbox{House of bricks/row dirt }  	&	0.94	&	0.24	&	0.90	&	0.30	&	-0.16	&	0.25	\\
\hbox{Water treatment}  	&	0.87	&	0.34	&	0.78	&	0.41	&	-0.22	&	0.19	\\
\hbox{Water supply}  	&	0.79	&	0.41	&	0.62	&	0.49	&	-0.38	&	0.18	\\
\hbox{Lighting}  	&	0.92	&	0.27	&	0.78	&	0.42	&	-0.41	&	0.43	\\
\hbox{Bathroom fixture}  	&	0.43	&	0.49	&	0.63	&	0.48	&	0.41	&	-0.02	\\
\hbox{Waste treatment}  	&	0.84	&	0.37	&	0.63	&	0.48	&	-0.48	&	0.27	\\
\hbox{Zero PC expenditure}  	&	0.00	&	0.00	&	0.04	&	0.20	&	0.29	&	\infty	\\
\hbox{Log PC expenditure}  	&	4.93	&	0.26	&	3.51	&	0.89	&	-2.18	&	1.22	\\
\hbox{Other programs}  	&	0.04	&	0.21	&	0.06	&	0.24	&	0.07	&	0.14	\\
\vspace{-0.3cm}\\													
\multicolumn{7}{l}{\hbox{\textit{Household head's characteristics}}}\\												
\vspace{-0.3cm}\\													
\hbox{Male}  	&	0.79	&	0.41	&	0.86	&	0.35	&	0.17	&	-0.15	\\
\hbox{Race: White}  	&	0.87	&	0.34	&	0.89	&	0.32	&	0.06	&	-0.07	\\
\hbox{Primary/Middle Education}  	&	0.41	&	0.49	&	0.47	&	0.50	&	0.11	&	0.01	\\
\hbox{Occupation: Unemployed}  	&	0.38	&	0.49	&	0.49	&	0.50	&	0.23	&	0.03	\\
\vspace{-0.2cm}\\
\hbox{Multivariate measure} & & & & &3.15\\
\vspace{-0.3cm}\\
\hline
\end{array}		$$		
\end{table}
\newpage

\section{Mixture-model Bayesian analysis: Results on the causal risk difference}
 \label{sec:BA_riskD}
In the Bolsa Fam\'ilia study, we focus on the finite sample causal relative risk in the subpopulation $\Uset$, because leprosy is a rare outcome, and the causal relative risk is generally more informative than the causal risk difference for rare outcomes.
Nevertheless, it might be worthwhile to show results for the causal risk difference. We focus on the finite sample causal risk difference in the subpopulation $\Uset$ defined as follows:
$$
ACE_{\Uset} = \dfrac{1}{N_{\Uset}}\sum_{i: i \in \Uset} Y_i(1)-\dfrac{1}{N_{\Uset}}\sum_{i: i \in \Uset} Y_i(0)
$$

\begin{figure}[t]
	\begin{center}
	\includegraphics[width=9cm]{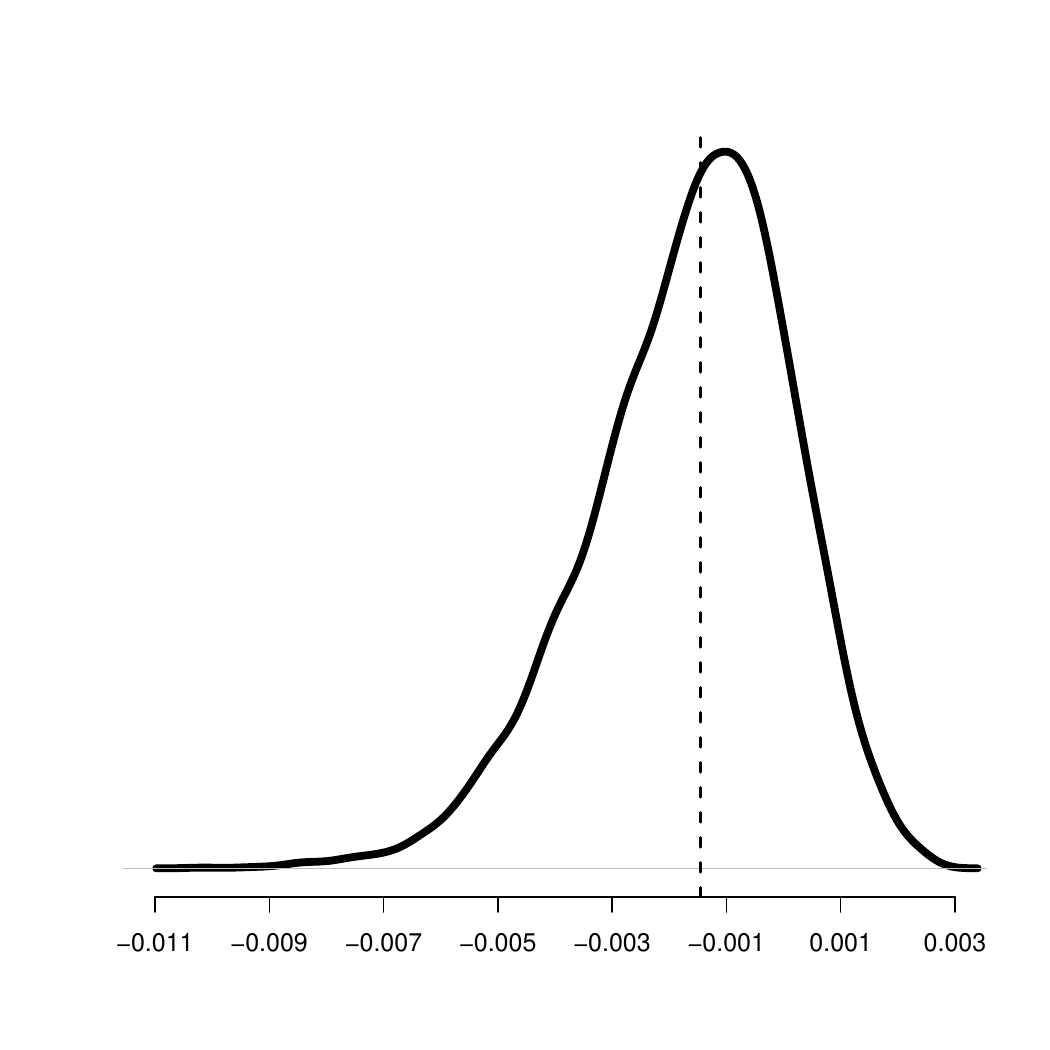} 
	\end{center}
	\caption{Bolsa Fam\'ilia study: Mixture-model Bayesian analysis. Posterior distributions of the finite sample causal risk difference (Dashed line = Posterior median).}  \label{fig:mbard}
\end{figure}

Figure~\ref{fig:mbard} and  Table~\ref{tab:mbard} show the posterior distribution of the causal risk difference and some summary statistics of it. 
The posterior median of the causal relative difference is equal to $-0.0014$,  and the $95\%$  highest density interval covers zero, including values from $-0.0053$ to $0.0015$. The posterior probability that the causal relative difference is negative is approximately $83.0\%$, but the posterior distribution of $ACE_{\Uset}$ is skewed to the left. 
Therefore, results based on the causal relative difference are in line with those based on the causal relative risk, providing some evidence that being eligible for the Bolsa Fam\'ilia program based on per capita income reduces the risk of leprosy.  

\begin{table*}	
\caption{Bolsa Fam\'ilia study: Mixture-model Bayesian approach.		Summary statistics of the posterior distributions of the finite sample causal risk difference, $ACE_{\Uset}$.}\label{tab:mbard}
	$$
	\begin{array}{ccccc}
		\hline
		\vspace{-0.35cm}\\
		&\multicolumn{3}{c}{95\% \hbox{ HDI}}  \\ 
		\vspace{-0.35cm}\\
		\cline{2-4}
		\vspace{-0.35cm}\\
		\hbox{Median}  & \hbox{Lower bound}& \hbox{Upper bound}& \hbox{(width)} &Pr\left(ACE_{\Uset}  < 0\right)\\
		\vspace{-0.35cm}\\	\hline
		\vspace{-0.35cm}\\
		-0.0014   & -0.0053  & 0.0015 &(0.0068)&  0.830\\ 
		\vspace{-0.35cm}\\
		\hline
	\end{array}
	$$
 \end{table*}

\section{Mixture-model Bayesian analysis: Model checks}
\label{sec:BA_modelchecks}

We evaluate the influence of the model assumptions through model checking using posterior predictive checks.  We compute a posterior predictive $p-$value (PPPV) for various discrepancy measures, defined as functions of the observed data and possibly of missing data (the subpopulation membership) and the model parameter vector.

In our study, we consider  seven posterior predictive discrepancy measures: the proportion of units in the subpopulation $\Uset$   
$$
p^d(\Uset) = \dfrac{\sum_{i}  \mathbb{I}\{i \in \Uset^{d}\}}{N},
$$
and six  posterior predictive discrepancy measures defined as a function of
$$
\mathcal{A}^{d}_{z}(\Uset)=\{Y_{i}^d:  Z^d_i=z, i \in \Uset\} \qquad z=0,1,
$$
where  $d=rep$ for the measures that are functions of data from a replicated study and $d=obs$ for the measures that are functions of observed data from the Bolsa Fam\'ilia study.
Specifically, we use the following discrepancy measures based on $\mathcal{A}^{d}_{z}(\Uset)$, $z=0,1$; $d=obs, rep$: 
\begin{itemize}
    \item[$(i)$] the absolute value of
the difference between the mean of $\mathcal{A}^{d}_{1}(\Uset)$  and the mean of   $\mathcal{A}^{d}_{0}(\Uset)$:
$$
\left|\Delta^d \right| = \left| \dfrac{\sum_{i: i \in \Uset^d} Y_{i}^d Z_{i}^d}{\sum_{i: i \in \Uset^d}  Z_{i}^d}-\dfrac{\sum_{i: i \in \Uset^d} Y_{i}^d (1-Z_{i}^d)}{\sum_{i: i \in \Uset^d}  (1-Z_{i}^d)}\right| \equiv  \left|  p_1^d-p_0^d\right|;
$$
\item[$(ii)$] the standard error of the difference in means based on a simple two-sample comparison for this difference:
$$
SE(\Delta^d) =  \sqrt{\dfrac{p_1^d (1-p_1^d)}{\sum_{i: i \in \Uset^d}  Z_{i}^d} +\dfrac{p_0^d (1-p_0^d)}{\sum_{i: i \in \Uset^d}  (1-Z_{i}^d)} };
$$
\item[$(iii)$] the ratio of $\left|\Delta^d \right|$ to $SE(\Delta^d)$; 
\item[$(iv)$] the absolute value of the logarithm of the ratio between the mean of $\mathcal{A}^{d}_{1}(\Uset)$  and the mean of   $\mathcal{A}^{d}_{0}(\Uset)$:
$$
\left|\log R^d\right| = \left|  \log\left(\dfrac{p_1^d}{p_0^d}\right)\right| =  \left|  \log(p_1^d)-\log(p_0^d)\right|;
$$
\item[$(v)$] the standard error of the logarithm of the ratio calculated using the delta method:
$$
SE(\log R ^d) = \sqrt{\dfrac{1-p_1^d}{p_1^d \sum_{i: i \in \Uset^d}  Z_{i}^d} + \dfrac{1-p_0^d}{p_0^d \sum_{i: i \in \Uset^d}  (1-Z_{i}^d)} }
;  
$$
and 
\item[$(vi)$] the ratio of $\left|\log R^d\right|$ to $SE(\log R^d)$.  
\end{itemize}
It is worth noting that the measures $(i)-(iii)$ and $(iv)-(vi)$ are not causal effects. Nevertheless, they allow us to assess whether the model
can preserve broad features of signal-discrepancy measures $(i)$ and $(iv)$, noise-discrepancy measures $(ii)$ and $(v)$, and signal-to-noise ratio-discrepancy measures $(iii)$ and $(vi)$ of the outcome distribution in the subpopulation $\Uset$.
We think these measures can be influential in estimating the treatment effects of interest.
We calculate a  PPPV for each discrepancy measure as the proportion of draws in which the replicated discrepancy measure exceeded the value of the study's discrepancy measure.

As Table~\ref{tab:pppvs} shows, the PPPVs range between 0.360 and 0.692, suggesting that the model - 	defined by the prior distributions and the likelihood - can successfully replicate the classification of the units in the subpopulation $\Uset$ and adequately preserve features of signal, noise, and signal-to-noise ratio of the outcome distribution in the subpopulation $\Uset$.


\begin{table*}\caption{Bolsa Fam\'ilia study: Mixture-model Bayesian approach. Posterior predictive $p-$values} \label{tab:pppvs} 
$$\begin{array}{c c ccc c ccc cccc}
\hline
\vspace{-0.25cm}\\
p(\Uset) & \qquad & |\Delta| & &SE(\Delta) & &|\Delta| / SE(\Delta) &\qquad  & |\log R| & &SE(\log R) & &|\log R| / SE(\log R)  \\
\vspace{-0.25cm}\\
\hline 
\vspace{-0.25cm}\\
0.602 &\qquad  & 0.646 & & 0.398 & & 0.692 &\qquad & 0.668 & & 0.360 & & 0.674\\
\vspace{-0.25cm}\\
\hline
\end{array}
$$
\end{table*}

\begin{figure}
	\begin{center}
		\includegraphics[width=8cm]{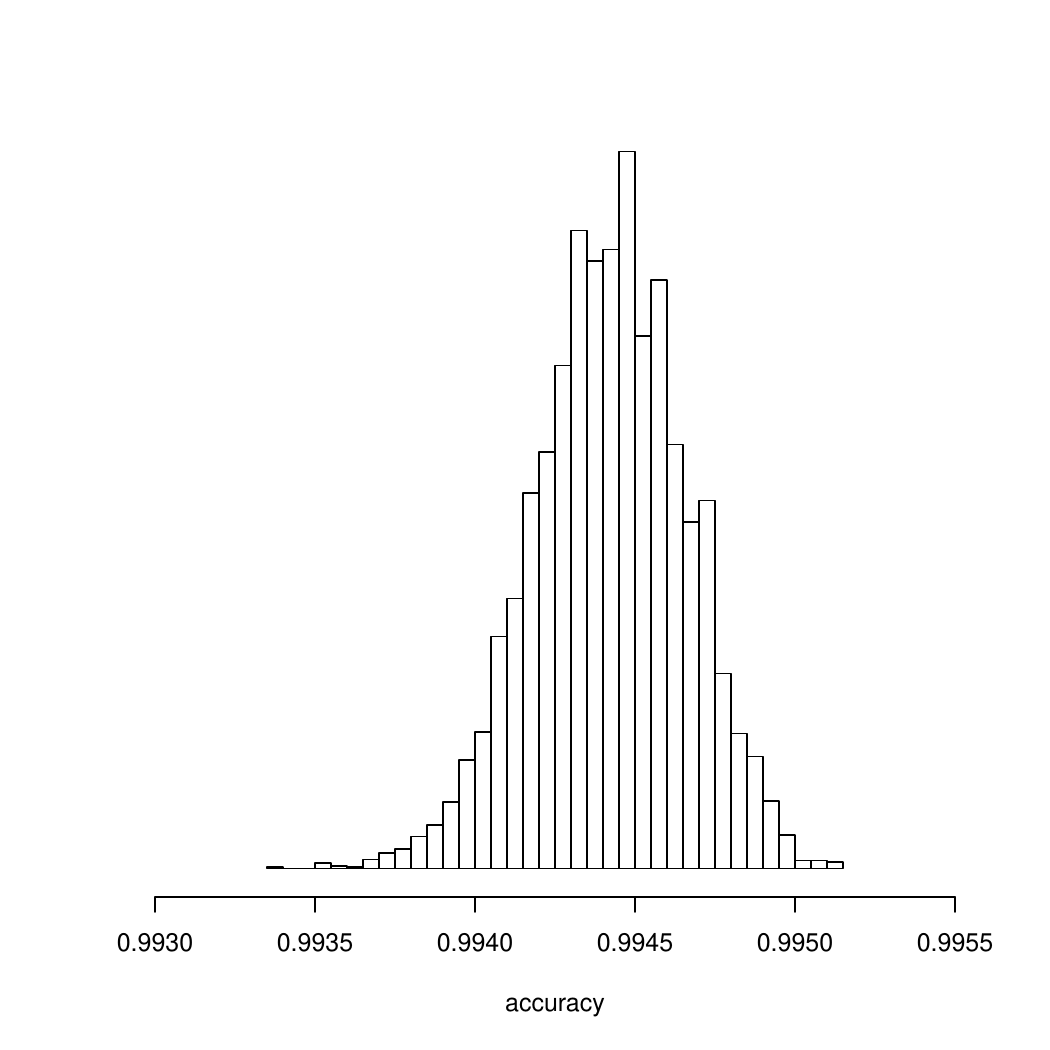} 
	\end{center}
	\caption{Posterior Predictive Checks: Proportion of correct outcome prediction in each replicated dataset.}  \label{fig-r1} 
\end{figure}

We also consider an accuracy measure for the outcome submodel, defined as the proportion of correct outcome prediction in each replicated dataset. 
Figure~\ref{fig-r1}   shows that the proportion of correct outcome prediction in each replicated dataset is very high, with a wrong outcome prediction for less than 1\,000 families over $147\,399$.

\section{Multiple imputation: Combining $M$ complete-membership inferences}\label{sec:MI}
Let $\tau$ be the causal estimand of interest, for instance, the finite sample causal relative risk  for the subpopulation $\Uset$, $\tau=RR_{\Uset}$ in the local randomization framework, or the average causal effect at the threshold, $s_0$, $\tau=\mathbb{E}[Y_i(1)-Y_i(0) \mid S_i=s_0]$, in the traditional framework.	
Let $\widehat{\tau}_m$ and $\widehat{V}_m$, $m=1, \ldots, M$, be $M$ estimates  of $\tau$ and their associated sampling variance, calculated from the $M$ completed membership datasets. The final point estimate of $\tau$ can be obtained by averaging together the $M$ completed-membership estimates: $\widehat{\tau}_{\mathrm{MI}} = M^{-1} \sum_{\ell=1}^{M} \widehat{\tau}_m$. The variability associated with this estimate is $\widehat{\theta}_{\mathrm{MI}}$ is $ W_M + \left(1+M^{-1}\right)B_M$, where 	$W_M = M^{-1} \sum_{m=1}^{M} \widehat{V}_m$ is the average within-imputation variance,  $B_M = (M-1)^{-1}\sum_{m=1}^{M} \left(\widehat{\tau}_m - \widehat{\tau}_{\mathrm{MI}}\right)^2$  is the between-imputation variance, and the factor $ \left(1+M^{-1}\right)$ reflects the fact that only a finite number of completed-membership estimates $\widehat{\tau}_m$, $m=1, \ldots, M$, are averaged together to obtain the final point estimate \cite[][]{Rubin1987}.

\section{Covariate Balance in Specific MSE-optimal Sub-populations}
\label{sec:MSEUs0_balance}

Four MSE-optimal subpopulations have been selected using  the data-driven bandwidth selectors based on MSE-optimal criteria proposed by \cite{CalonicoCattaneoTitiunik2014}. We use uniform and triangular kernel functions to construct local-polynomial estimators of order one and two and two different MSE-optimal bandwidth selectors: one for below and one for above the threshold.	The four selected  MSE-optimal subpopulations are shown in Table~\ref{tab:Usets0}. 	The sample size of the selected subpopulations, and especially the number of eligible families falling in these subpopulations (namely the left bandwidth), is strongly affected by both the kernel function and the order of the local-polynomial estimator. The smallest subpopulation is obtained using the uniform kernel function and the local linear regression estimator ($p=1$); it comprises $69\, 796$ families ($47.3\%$) of which $64\, 944$ are eligible families. The largest subpopulation is obtained using  the triangular kernel function and the local-polynomial estimator of order $p=2$; it  comprises all the $138\, 220$ eligible families and $4\,879$ ineligible families for a total of $143\, 099$ families  ($97.1\%$).

\begin{table}[t]\caption{Bolsa Fam\'ilia study: Subpopulations defined using MSE-optimal bandwidth selectors based on uniform and triangular kernel functions to construct local-polynomial estimators of order $p$.} \label{tab:Usets0}
	$$
  \hspace*{-1cm}
	\begin{array}{l rr c rcr}
		\hline
		\vspace{-0.35cm}\\
		&\multicolumn{2}{c}{\hbox{Bandwidths}} &&\\
		\hbox{Kernel function } (p)  &\hbox{(Left)} &  \hbox{(Right)} & \hbox{Suppopulation: } \Uset &  N_{\Uset} & \sum_{i \in \Uset}(1-Z_i)
		& \sum_{i \in \Uset}Z_i \\ 
		\vspace{-0.35cm}\\
		\hline
		\vspace{-0.35cm}\\
		\hbox{Uniform kernel } (p=1) & 76.5 & 43.9 & \{i \in \mathcal{U}:  43.5 \leq S_i\leq 163.9\}& 69\,796 &4\,852 &  64\,944 \\ 
		\vspace{-0.15cm}\\
		\hbox{Triangular kernel } (p=1) &  81.0 & 54.5 &
		\{i \in \mathcal{U}:  39.0 \leq S_i\leq 174.5\}& 83\,326&5\,159 &78\,167 \\  
		\vspace{-0.15cm}\\
		\hbox{Uniform kernel } (p=2) & 120.0 &46.0&\{i \in \mathcal{U}: \ 0.0 \leq S_i\leq 166.0\} & 143\,102 &4\,882& 138\,220\\
		\vspace{-0.15cm}\\
		\hbox{Triangular kernel  } (p=2) & 120.0 &45.6 & \{i \in \mathcal{U}:  \ 0.0 \leq S_i\leq 165.6\} & 143\,099 &4\,879& 138\,220\\
		\vspace{-0.35cm}\\
		\hline
	\end{array}
  \hspace*{-1cm}
	$$
\end{table}

\begin{figure}[h]
\caption{Bolsa Fam\'ilia study: (Weighted) normalized mean differences within specific sub-populations selected using the MSE-optimal bandwidth approach based on uniform and triangular kernel functions and local-polynomial estimator of order $p=1,2$  (Weights are exploited when the triangular kernel is used).}\label{A-fig:bal1}
\vspace{-0.00cm}
\begin{center}
	\begin{tabular}{ccc}
		\multicolumn{3}{c}{Uniform Kernel - $p=1$: $\Uset=\{i: 43.5 \leq S_i \leq  163.9\}$}\\
		\includegraphics[width=4.5cm]{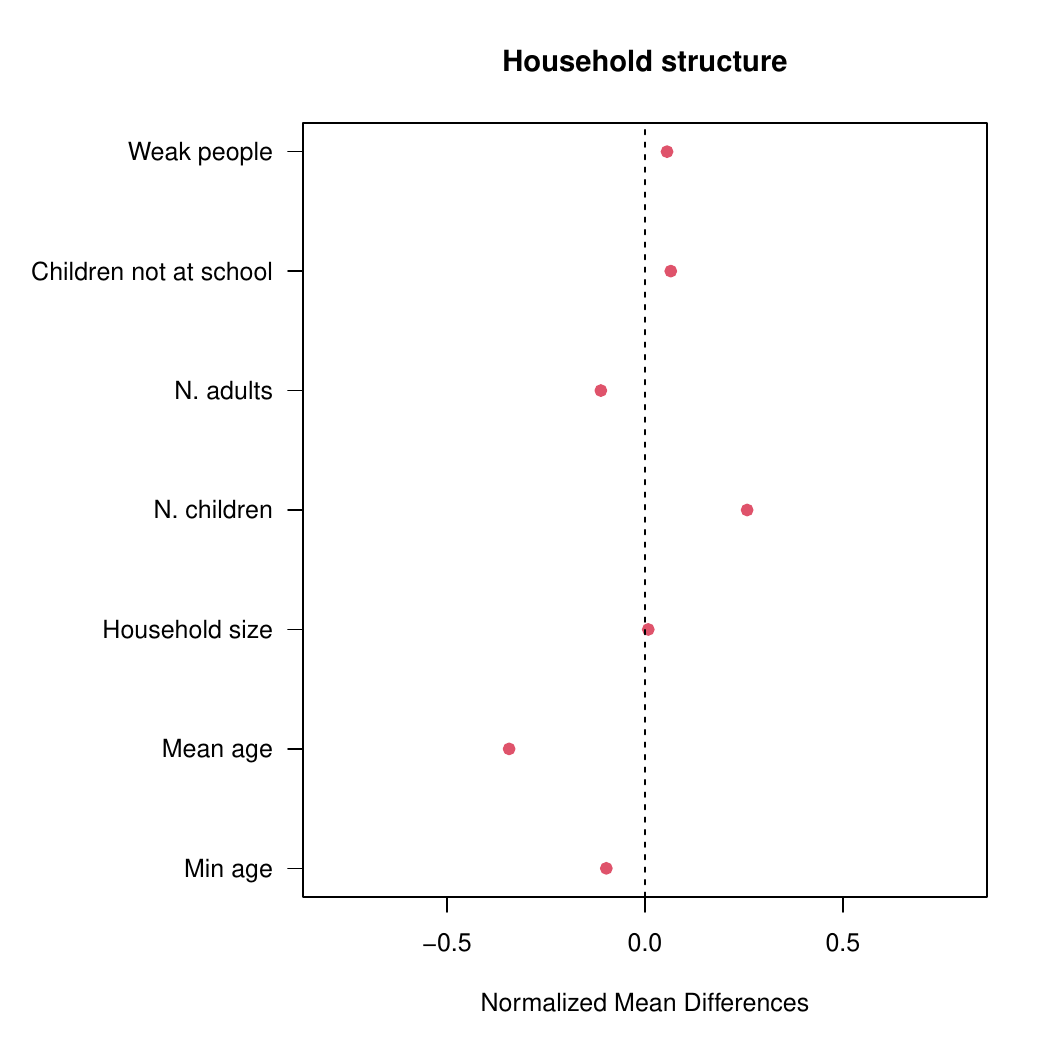} & 	\includegraphics[width=4.5cm]{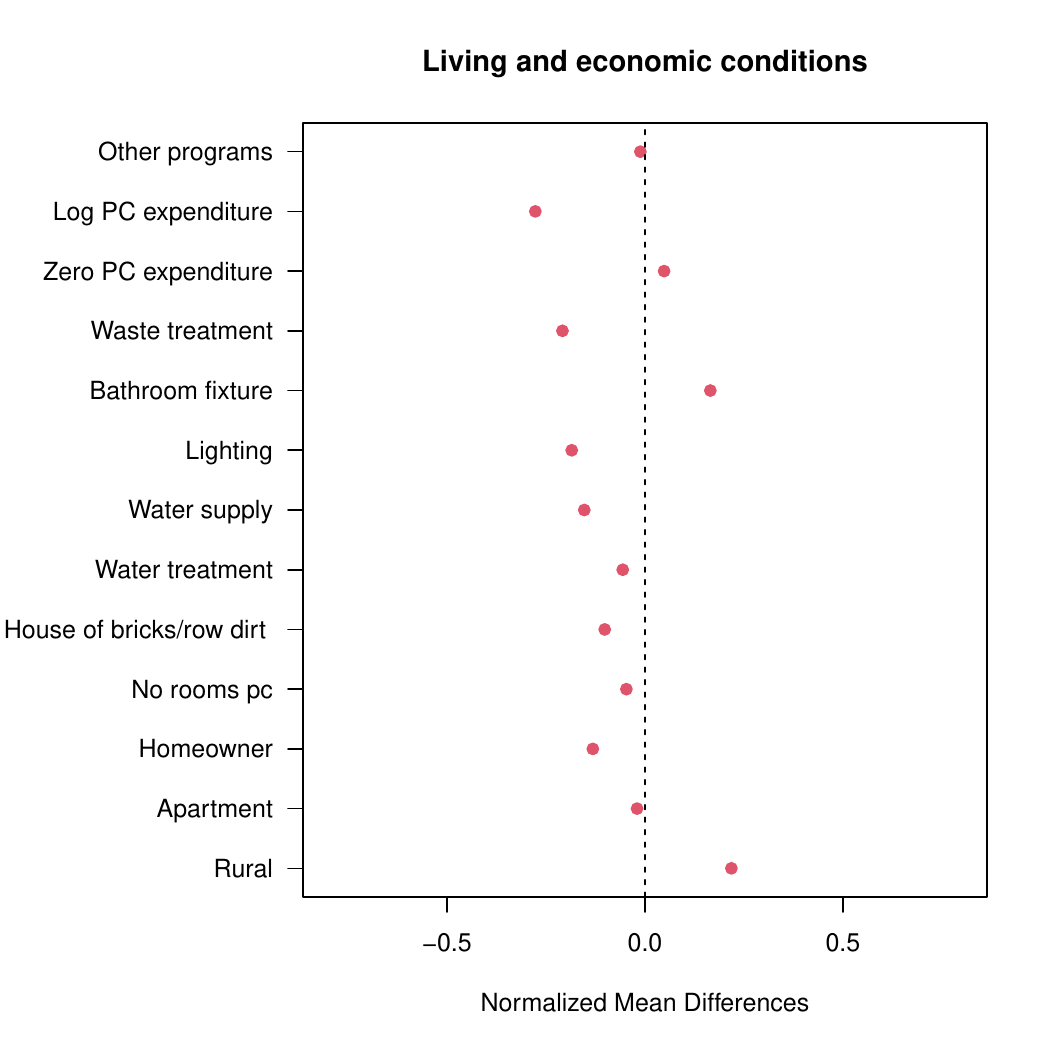} &
		\includegraphics[width=4.5cm]{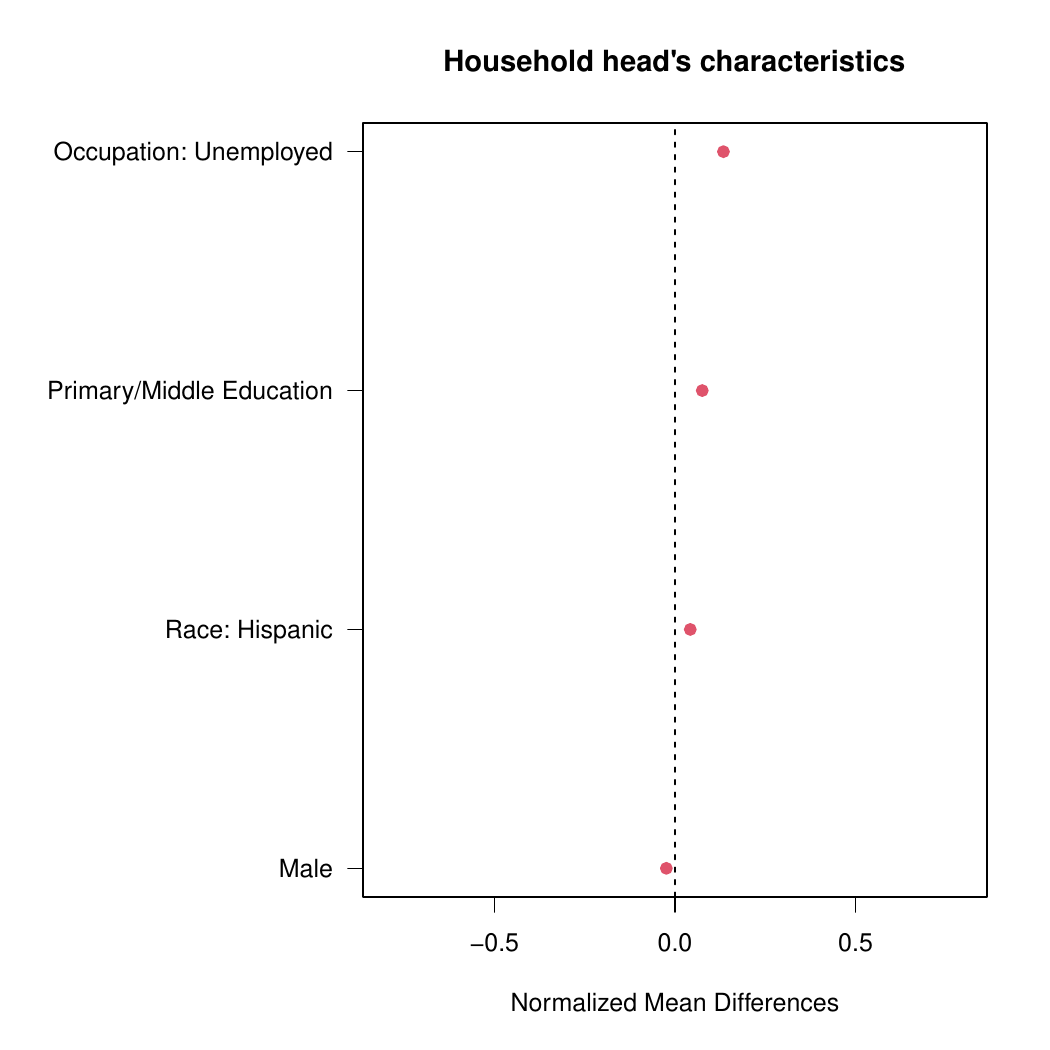}\\
		\vspace{-0.25cm}\\	 
		\multicolumn{3}{c}{Triangular Kernel - $p=1$: $\Uset=\{i:  39.0  \leq S_i \leq 174.5\}$}\\
		\includegraphics[width=4.5cm]{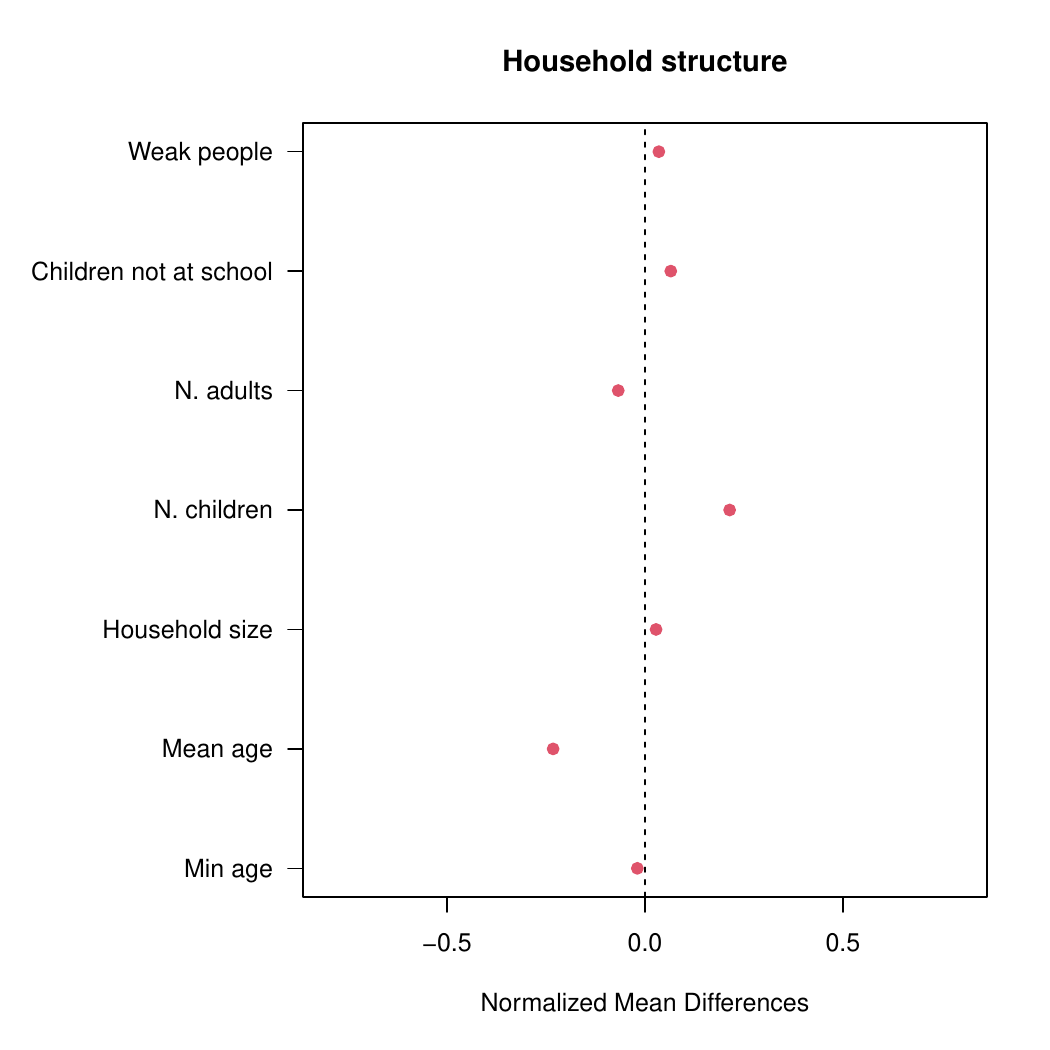} & 	\includegraphics[width=4.5cm]{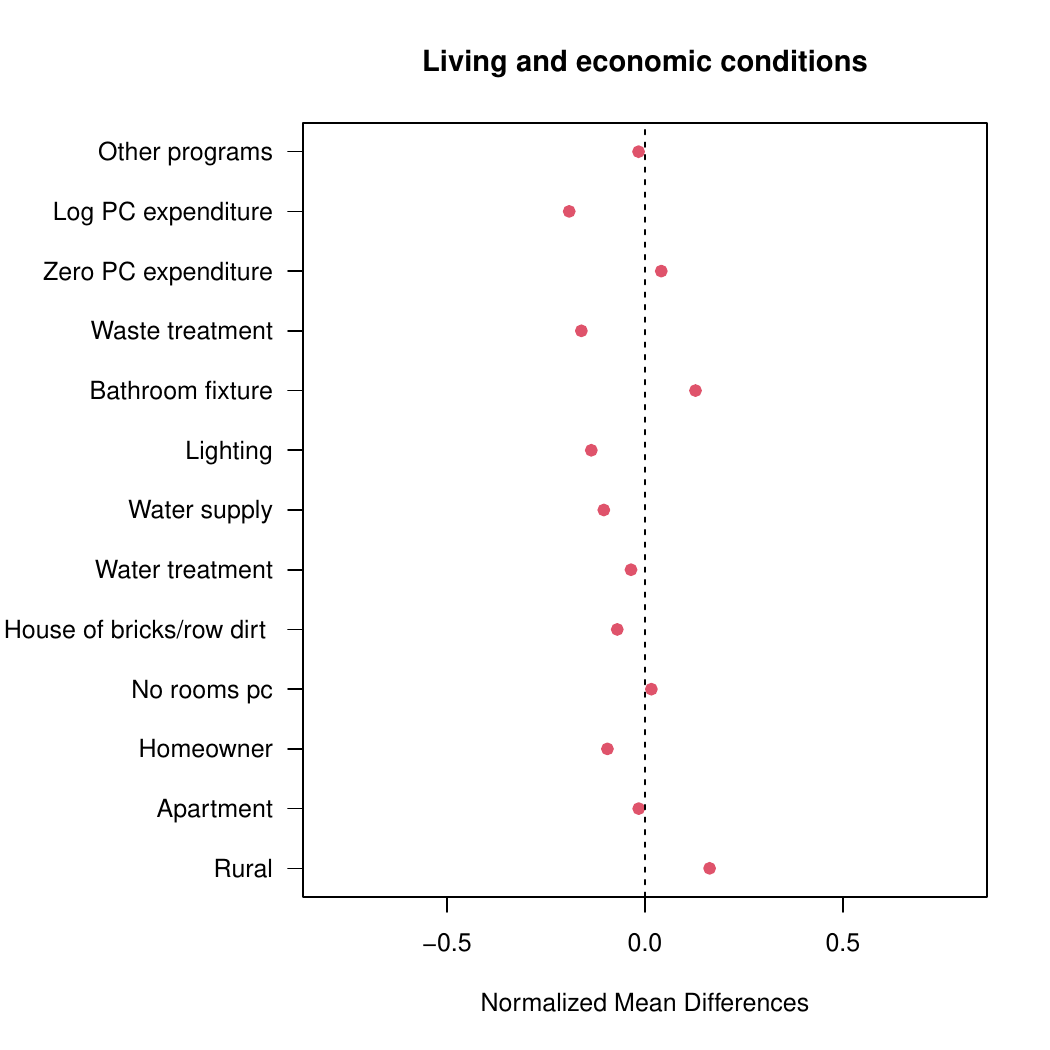} &
		\includegraphics[width=4.5cm]{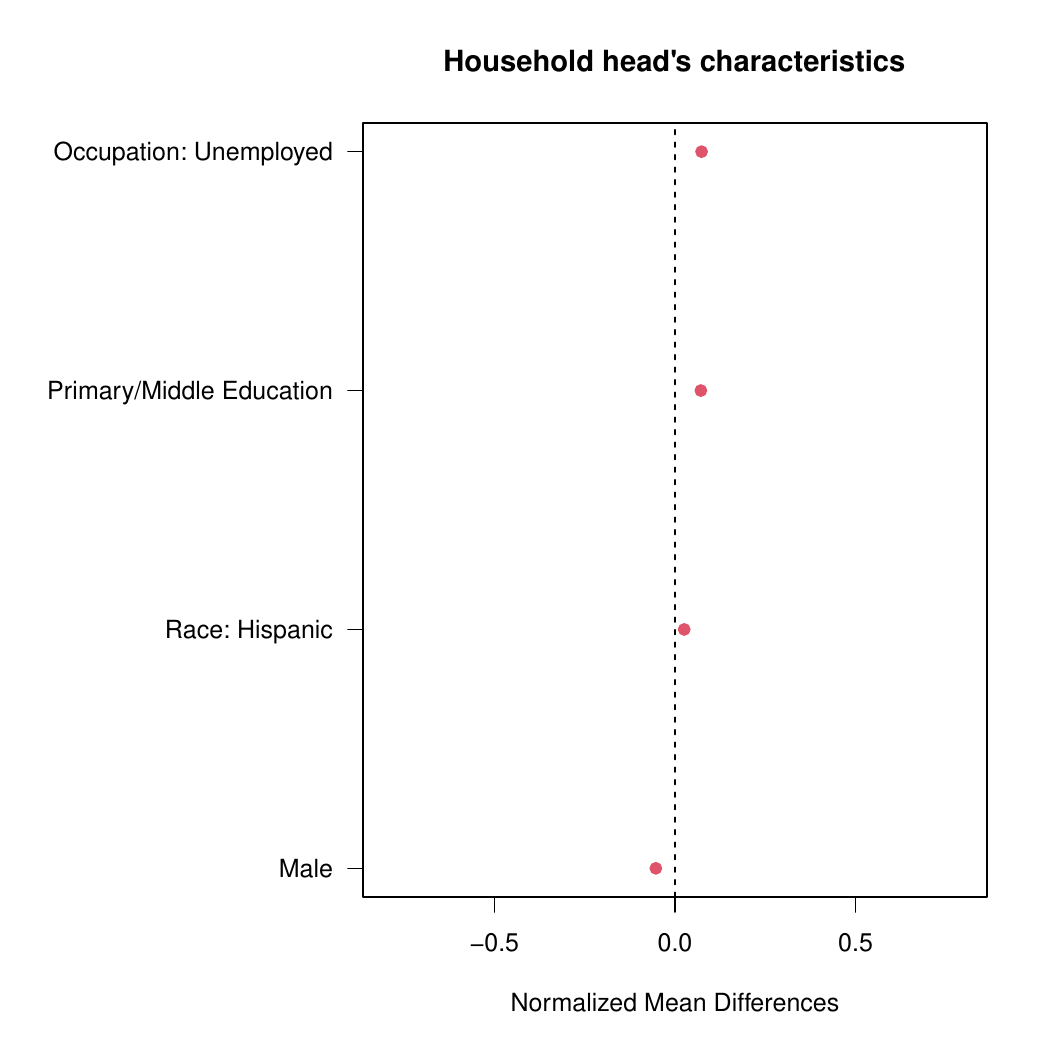}\\
		\vspace{-0.25cm}\\		 
		\multicolumn{3}{c}{Uniform Kernel - $p=2$: $\Uset=\{i: 0.0  \leq S_i \leq  166.0\}$}\\
		\includegraphics[width=4.5cm]{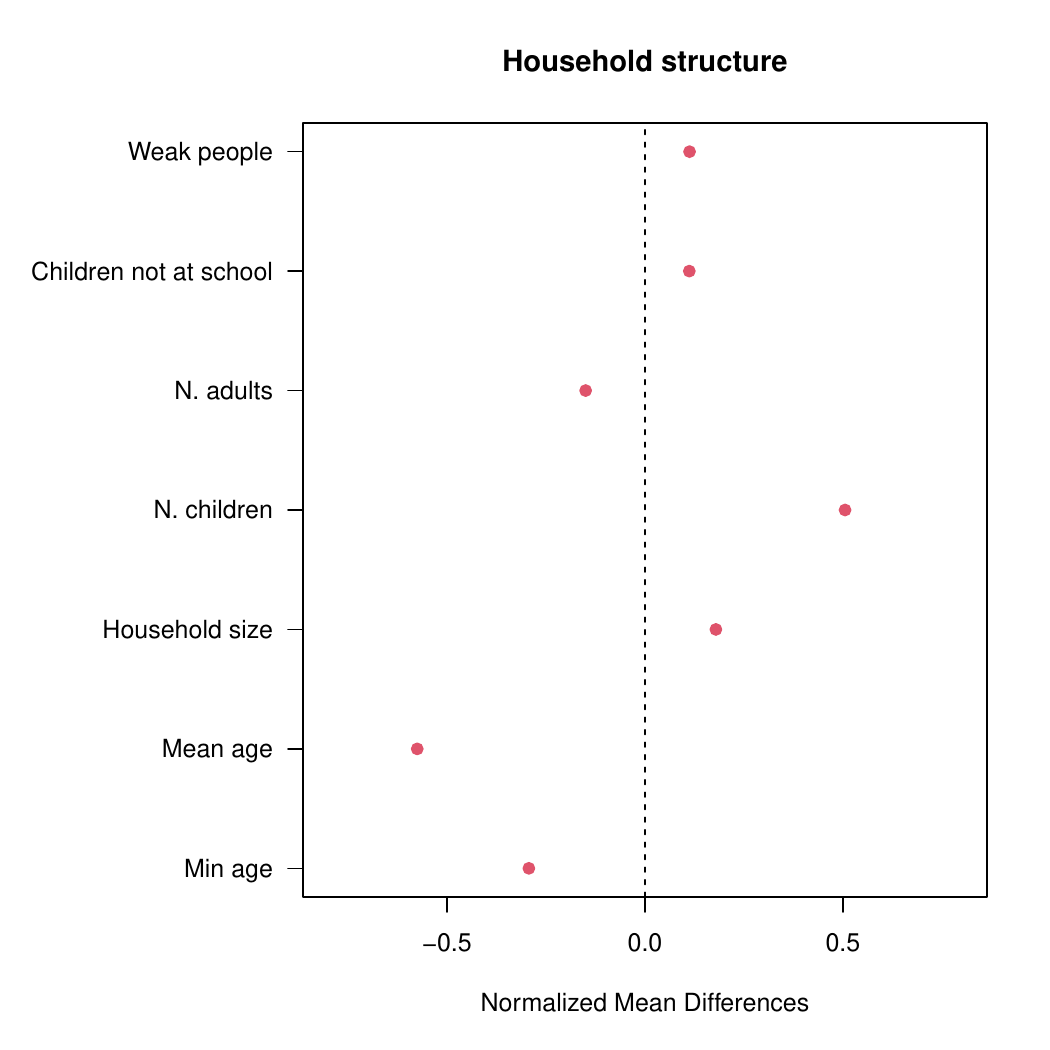} & 	\includegraphics[width=4.5cm]{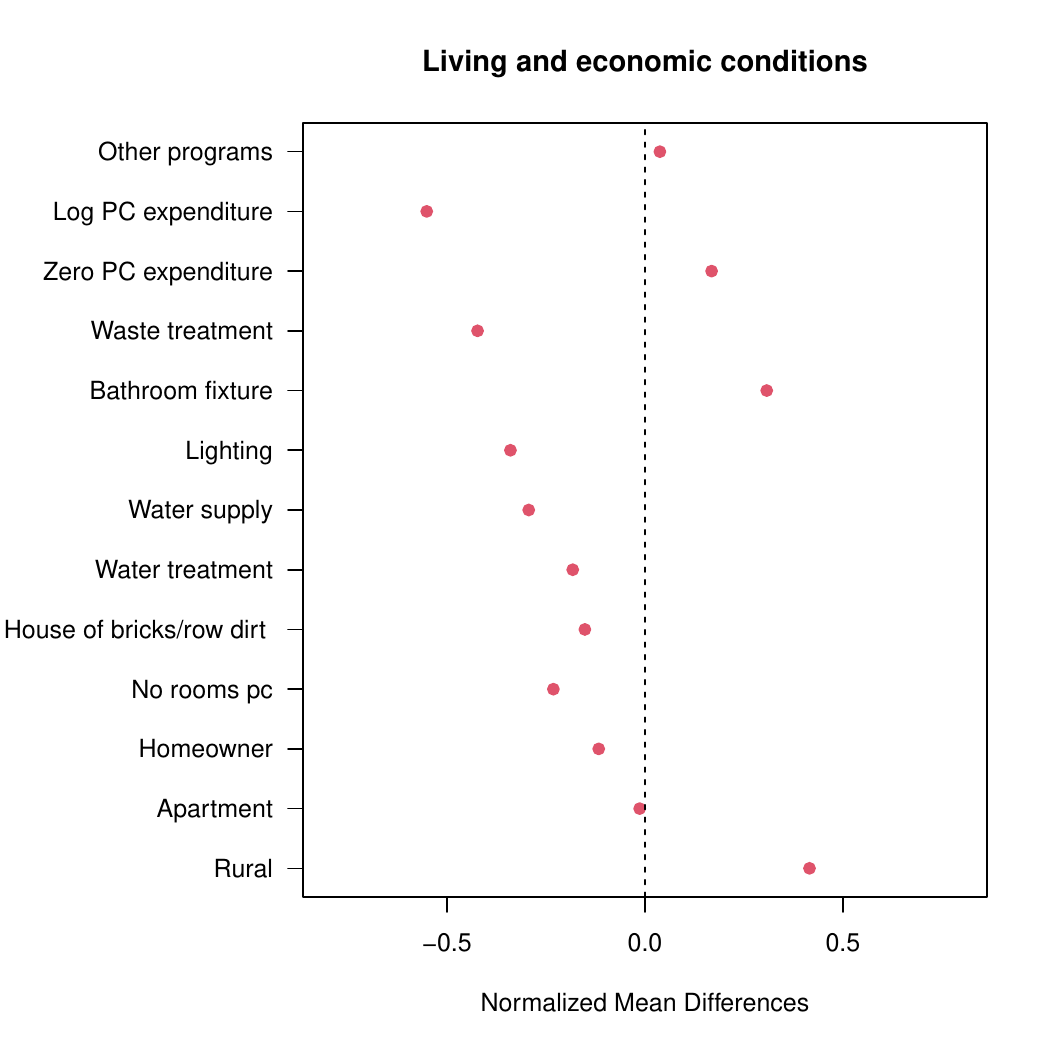} &
		\includegraphics[width=4.5cm]{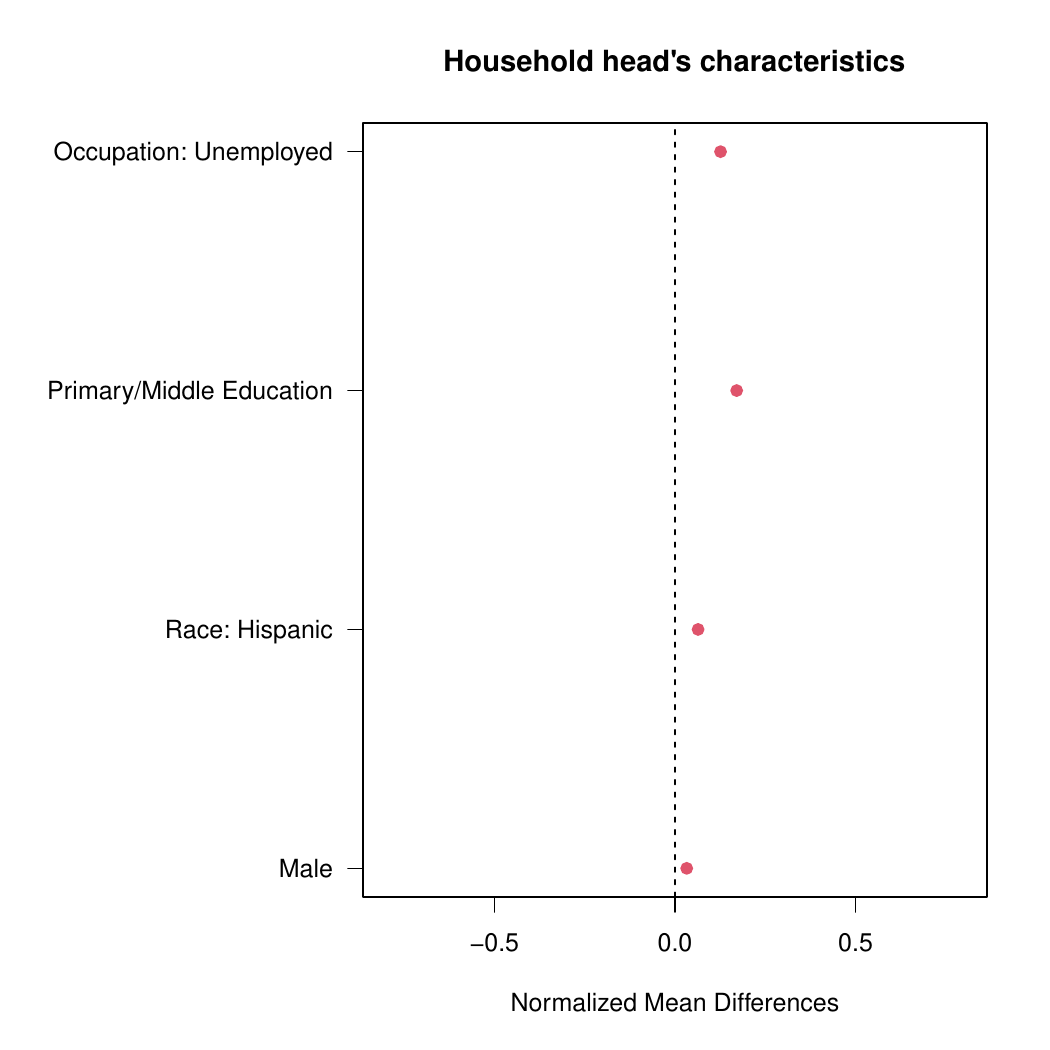}\\
		\vspace{-0.25cm}\\		 
		\multicolumn{3}{c}{Triangular Kernel - $p=2$: $\Uset=\{i: 0.0 \leq S_i \leq  165.6\}$}\\
		\includegraphics[width=4.5cm]{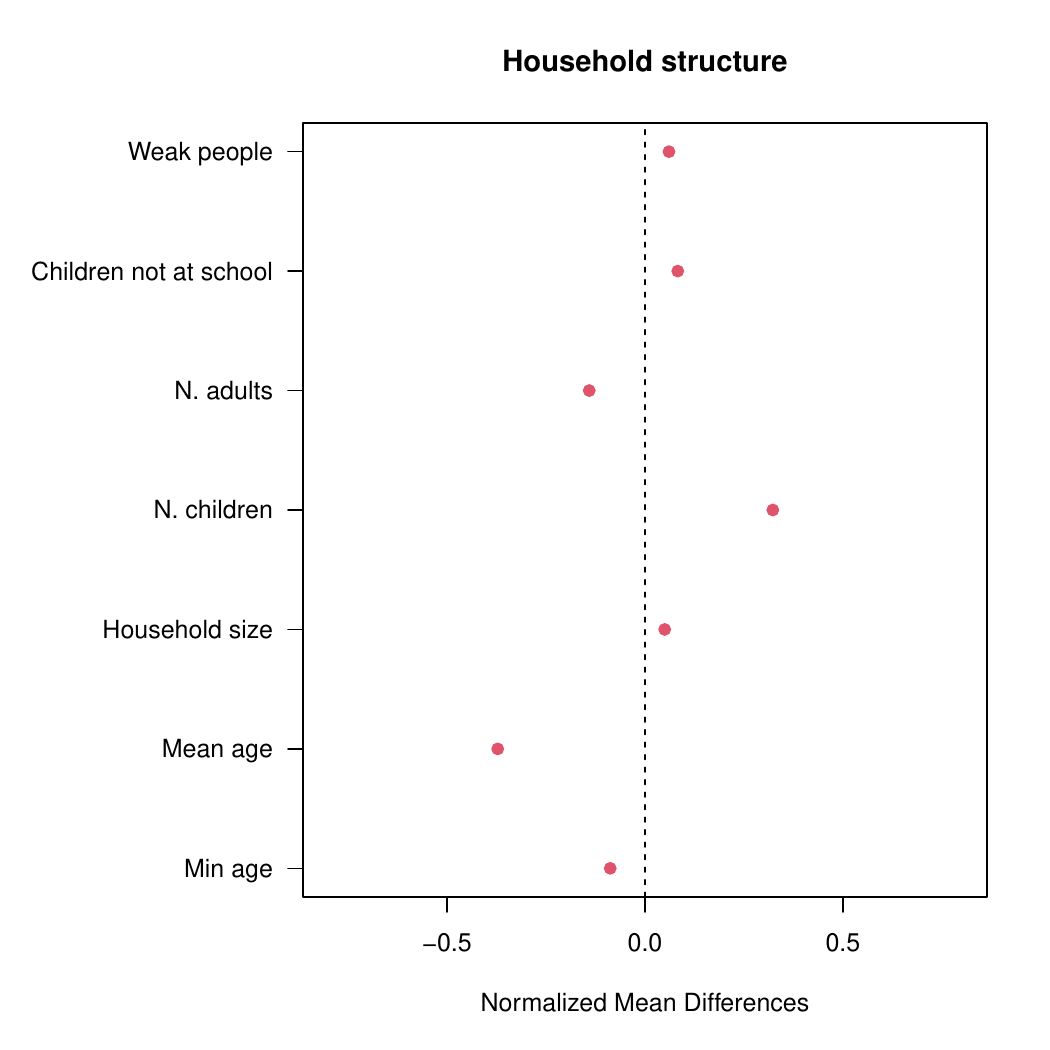} & 	\includegraphics[width=4.5cm]{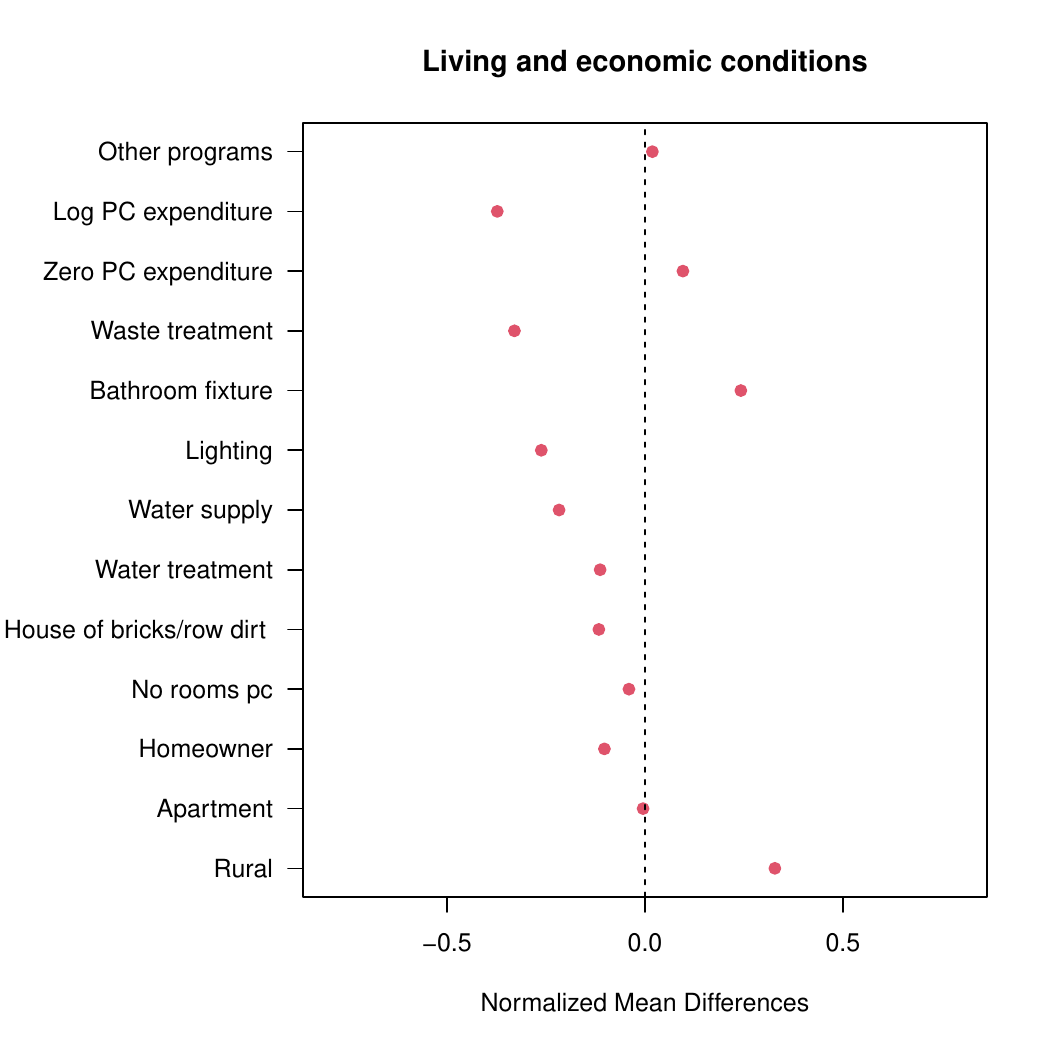} &
		\includegraphics[width=4.5cm]{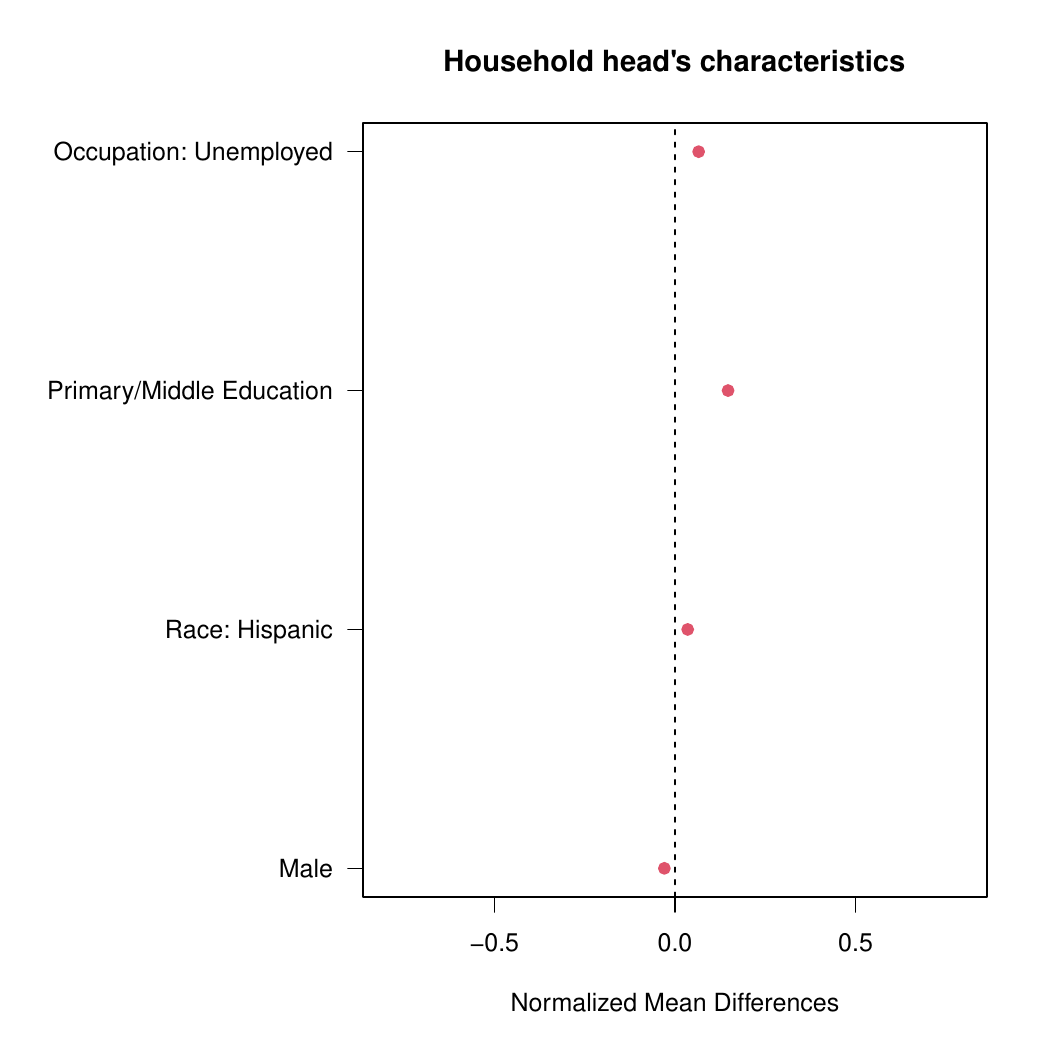}\\
	\end{tabular}
\end{center}
\end{figure}

Tables \ref{A-tab:bal1a}-\ref{A-tab:bal4a} show 
the sample averages and standard deviations by eligibility status, and the three measures of balancing--the normalized differences;  the log of the ratio of the sample standard deviations; and the Mahalanobis distance between the means--within the four  MSE-optimal subpopulations. 
Figure \ref{A-fig:bal1} shows graphically the normalized differences within the   four  MSE-optimal subpopulations.

\begin{table}[ht]\caption{Bolsa Fam\'ilia study:  Covariate balance within the  sub-population $\Uset = \{i: 43.5\leq S_i \leq  163.9\}$, selected using the MSE-optimal bandwidth approach  based on the uniform   kernel function  and the local-polynomial estimator  of order $p=1$.}  \label{A-tab:bal1a}
$$
\begin{array}{lrrrrrr}
	\hline 
	\vspace{-0.35cm}\\
	&\multicolumn{2}{c}{\hbox{Ineligible}}& \multicolumn{2}{c}{\hbox{Eligible}}\\
	N_{\Uset}= 69\,796 	 &\multicolumn{2}{c}{(4\,852)}& \multicolumn{2}{c}{(64\,944 )} &\hbox{Norm.} &  \\
	\vspace{-0.4cm}\\
	\hbox{Covariate}		 & \hbox{Mean} & \hbox{SD}  & \hbox{Mean} & \hbox{SD} & \hbox{Diff.} & \log(s_1/s_0)  \\ 
	\vspace{-0.35cm}\\
	\hline
	\vspace{-0.35cm}\\
	\multicolumn{7}{l}{\hbox{\textit{Household structure}}}\\
	\vspace{-0.4cm}\\
	\hbox{Min age}	&	14.68	&	15.74	&	13.15	&	15.56	&	-0.10	&	-0.01	\\
	\hbox{Mean age}	&	28.84	&	12.84	&	24.56	&	12.10	&	-0.34	&	-0.06	\\
	\hbox{Household size}	&	2.80	&	1.05	&	2.81	&	1.26	&	0.01	&	0.19	\\
	\hbox{N. Children}	&	0.90	&	0.76	&	1.12	&	0.97	&	0.26	&	0.25	\\
	\hbox{N. Adults}	&	1.71	&	0.78	&	1.63	&	0.76	&	-0.11	&	-0.03		\\
	\hbox{Children not at school}	&	0.02	&	0.14	&	0.03	&	0.17	&	0.07	&	0.20	\\
	\hbox{Presence of weak people}	&	0.18	&	0.38	&	0.20	&	0.40	&	0.06	&	0.04	\\
	\vspace{-0.3cm}\\
	\multicolumn{7}{l}{\hbox{\textit{Living and economic conditions}}}\\
	\vspace{-0.4cm}\\
	\hbox{Rural}	&	0.22	&	0.41	&	0.31	&	0.46	&	0.22	&	0.12	\\
	\hbox{Apartment}	&	0.95	&	0.21	&	0.95	&	0.22	&	-0.02	&	0.04		\\
	\hbox{Home ownership: Homeowner}	&	0.64	&	0.48	&	0.58	&	0.49	&	-0.13	&	0.03	\\
	\hbox{No rooms pc}	&	1.82	&	1.10	&	1.77	&	1.14	&	-0.05	&	0.04	\\
	\hbox{House of bricks/row dirt }	&	0.95	&	0.22	&	0.92	&	0.26	&	-0.10	&	0.18	\\
	\hbox{Water treatment}	&	0.85	&	0.35	&	0.83	&	0.37	&	-0.06	&	0.05	\\
	\hbox{Water supply}	&	0.76	&	0.43	&	0.69	&	0.46	&	-0.15	&	0.08 	\\
	\hbox{Lighting}	&	0.91	&	0.29	&	0.84	&	0.36	&	-0.19	&	0.21	 	\\
	\hbox{Bathroom fixture}	&	0.47	&	0.50	&	0.55	&	0.50	&	0.17	&	0.00 	\\
	\hbox{Waste treatment}	&	0.81	&	0.39	&	0.72	&	0.45	&	-0.21	&	0.13	 	\\
	\hbox{Zero PC expenditure}	&	0.15	&	0.36	&	0.17	&	0.38	&	0.05	&	0.04	 	\\
	\hbox{Log PC expenditure}	&	3.83	&	1.76	&	3.36	&	1.65	&	-0.28	&	-0.06 	\\
	\hbox{Other programs}	&	0.05	&	0.22	&	0.05	&	0.21	&	-0.01	&	-0.02 	\\
	\vspace{-0.3cm}\\
	\multicolumn{7}{l}{\hbox{\textit{Household head's characteristics}}}\\
	\vspace{-0.4cm}\\
	\hbox{Male}	&	0.85	&	0.35	&	0.85	&	0.36	&	-0.02	&	0.02 	\\
	\hbox{Race: White}	&	0.86	&	0.34	&	0.88	&	0.33	&	0.04	&	-0.05	 	\\
	\hbox{Primary/Middle Education}	&	0.39	&	0.49	&	0.42	&	0.49	&	0.08	&	0.01	 	\\
	\hbox{Occupation: Unemployed}	&	0.43	&	0.50	&	0.50	&	0.50	&	0.13	&	0.01	 	\\
	\vspace{-0.2cm}\\
	\hbox{Multivariate measure} & & & & & 0.91\\
	\vspace{-0.35cm}\\
	\hline 
\end{array}
$$
\end{table}
\newpage
\begin{table}[ht]\caption{Bolsa Fam\'ilia study: 
	Weighted covariate  balance within the  sub-population $\Uset=\{i: 39.0 \leq S_i \leq 174.5\}$, selected using the MSE-optimal bandwidth approach  based on the triangular kernel function  and the local-polynomial estimator  of order $p=1$  (Weights based on the triangular Kernel).}
	\label{A-tab:bal2a}
$$
\begin{array}{lrrrrrr}
	\hline 
	\vspace{-0.35cm}\\
	&\multicolumn{2}{c}{\hbox{Ineligible}}& \multicolumn{2}{c}{\hbox{Eligible}}\\
	N_{\Uset}=   83\,326 		 & \multicolumn{2}{c}{(=5\,159)}& \multicolumn{2}{c}{( 78\,167)} &\hbox{Norm.} &   \\
	\vspace{-0.4cm}\\
	\hbox{Covariate}	& \hbox{Mean} & \hbox{SD}  & \hbox{Mean} & \hbox{SD} & \hbox{Diff.} & \log(s_1/s_0) \\  
	\vspace{-0.35cm}\\
	\hline
	\vspace{-0.35cm}\\
	\multicolumn{7}{l}{\hbox{\textit{Household structure}}}\\
	\vspace{-0.4cm}\\
	\hbox{Min age}	 & 	13.25	 & 	14.87	 & 	12.96	 & 	15.17	 & 	-0.02	 & 	0.02 	\\
	\hbox{Mean age}	 & 	28.02	 & 	12.52	 & 	25.19	 & 	11.90	 & 	-0.23	 & 	-0.05	 	\\
	\hbox{Household size}	 & 	2.86	 & 	0.99	 & 	2.90	 & 	1.26	 & 	0.03	 & 	0.24	  	\\
	\hbox{N. Children}	 & 	0.95	 & 	0.75	 & 	1.13	 & 	0.95	 & 	0.21	 & 	0.24	  \\
	\hbox{N. Adults}	 & 	1.73	 & 	0.77	 & 	1.68	 & 	0.78	 & 	-0.07	 & 	0.01	  	\\
	\hbox{Children not at school}	 & 	0.02	 & 	0.15	 & 	0.03	 & 	0.18	 & 	0.07	 & 	0.19	  	\\
	\hbox{Presence of weak people}	 & 	0.19	 & 	0.39	 & 	0.20	 & 	0.40	 & 	0.04	 & 	0.03	 	\\
	\vspace{-0.3cm}\\
	\multicolumn{7}{l}{\hbox{\textit{Living and economic conditions}}}\\
	\vspace{-0.4cm}\\
	\hbox{Rural}	 & 	0.21	 & 	0.41	 & 	0.28	 & 	0.45	 & 	0.16	 & 	0.10	  	\\
	\hbox{Apartment}	 & 	0.95	 & 	0.22	 & 	0.95	 & 	0.22	 & 	-0.02	 & 	0.03	  	\\
	\hbox{Home ownership: Homeowner}	 & 	0.64	 & 	0.48	 & 	0.59	 & 	0.49	 & 	-0.09	 & 	0.02 	\\
	\hbox{No rooms pc}	 & 	1.73	 & 	1.01	 & 	1.75	 & 	1.14	 & 	0.02	 & 	0.13	  	\\
	\hbox{House of bricks/row dirt }	 & 	0.95	 & 	0.22	 & 	0.93	 & 	0.25	 & 	-0.07	 & 	0.13	 	\\
	\hbox{Water treatment}	 & 	0.85	 & 	0.35	 & 	0.84	 & 	0.37	 & 	-0.04	 & 	0.03	 	\\
	\hbox{Water supply}	 & 	0.76	 & 	0.43	 & 	0.72	 & 	0.45	 & 	-0.10	 & 	0.06	 	\\
	\hbox{Lighting}	 & 	0.91	 & 	0.29	 & 	0.87	 & 	0.34	 & 	-0.14	 & 	0.17	  	\\
	\hbox{Bathroom fixture}	 & 	0.46	 & 	0.50	 & 	0.53	 & 	0.50	 & 	0.13	 & 	0.00	 	\\
	\hbox{Waste treatment}	 & 	0.82	 & 	0.39	 & 	0.75	 & 	0.43	 & 	-0.16	 & 	0.11	  	\\
	\hbox{Zero PC expenditure}	 & 	0.15	 & 	0.36	 & 	0.17	 & 	0.37	 & 	0.04	 & 	0.04	 	\\
	\hbox{Log PC expenditure}	 & 	3.82	 & 	1.76	 & 	3.48	 & 	1.70	 & 	-0.19	 & 	-0.03 	\\
	\hbox{Other programs}	 & 	0.05	 & 	0.22	 & 	0.05	 & 	0.21	 & 	-0.02	 & 	-0.03	  	\\
	\vspace{-0.3cm}\\
	\multicolumn{7}{l}{\hbox{\textit{Household head's characteristics}}}\\
	\vspace{-0.4cm}\\
	\hbox{Male}	 & 	0.86	 & 	0.34	 & 	0.85	 & 	0.36	 & 	-0.05	 & 	0.05	  	\\
	\hbox{Race: White}	 & 	0.87	 & 	0.34	 & 	0.88	 & 	0.33	 & 	0.03	 & 	-0.03 	\\
	\hbox{Primary/Middle Education}	 & 	0.37	 & 	0.48	 & 	0.41	 & 	0.49	 & 	0.07	 & 	0.02	 	\\
	\hbox{Occupation: Unemployed}	 & 	0.45	 & 	0.50	 & 	0.49	 & 	0.50	 & 	0.07	 & 	0.00 	\\
	\vspace{-0.2cm}\\
	\hbox{Multivariate measure} & & & & &0.67\\
	\vspace{-0.3cm}\\
	\hline
\end{array}
$$
\end{table}
\newpage
\begin{table}[h]\caption{Bolsa Fam\'ilia study: Covariate balance within the  sub-population $\Uset=\{i: 0.0 \leq S_i \leq 166.0\}$, selected using the MSE-optimal bandwidth approach  based on the uniform   kernel function  and the local-polynomial estimator  of order $p=2$.}
	\label{A-tab:bal3a}
$$
\begin{array}{lrrrrrr}
	\hline 		\vspace{-0.35cm}\\
	&\multicolumn{2}{c}{\hbox{Ineligible}}& \multicolumn{2}{c}{\hbox{Eligible}}\\
	N_{\Uset}= 143\,102  &\multicolumn{2}{c}{(4\,882)}& \multicolumn{2}{c}{(138\,220)} &\hbox{Norm.} &  \\
	\vspace{-0.4cm}\\
	\hbox{Covariate}&  \hbox{Mean} & \hbox{SD}  & \hbox{Mean} & \hbox{SD} & \hbox{Diff.} & \log(s_1/s_0) \\  
	\vspace{-0.35cm}\\
	\hline
	\vspace{-0.35cm}\\
	\multicolumn{7}{l}{\hbox{\textit{Household structure}}}\\
	\vspace{-0.4cm}\\
	\hbox{Min age}	 & 	14.67	 & 	15.72	 & 	10.32	 & 	13.85	 & 	-0.29	 & 	-0.13	  	\\
	\hbox{Mean age}	 & 	28.84	 & 	12.82	 & 	21.89	 & 	11.28	 & 	-0.58	 & 	-0.13	 	\\
	\hbox{Household size}	 & 	2.81	 & 	1.05	 & 	3.02	 & 	1.32	 & 	0.18	 & 	0.23	 	\\
	\hbox{N. Children}	 & 	0.90	 & 	0.76	 & 	1.38	 & 	1.11	 & 	0.51	 & 	0.38	  \\
	\hbox{N. Adults}	 & 	1.72	 & 	0.78	 & 	1.60	 & 	0.74	 & 	-0.15	 & 	-0.07	 	\\
	\hbox{Children not at school}	 & 	0.02	 & 	0.14	 & 	0.04	 & 	0.20	 & 	0.11	 & 	0.32	 	\\
	\hbox{Presence of weak people}	 & 	0.18	 & 	0.38	 & 	0.23	 & 	0.42	 & 	0.11	 & 	0.08	 	\\
	\vspace{-0.3cm}\\
	\multicolumn{7}{l}{\hbox{\textit{Living and economic conditions}}}\\
	\vspace{-0.4cm}\\
	\hbox{Rural}	 & 	0.22	 & 	0.41	 & 	0.40	 & 	0.49	 & 	0.42	 & 	0.18	 	\\
	\hbox{Apartment}	 & 	0.95	 & 	0.21	 & 	0.95	 & 	0.22	 & 	-0.01	 & 	0.03	 	\\
	\hbox{Home ownership: Homeowner}	 & 	0.64	 & 	0.48	 & 	0.58	 & 	0.49	 & 	-0.12	 & 	0.03	  	\\
	\hbox{No rooms pc}	 & 	1.82	 & 	1.10	 & 	1.57	 & 	1.02	 & 	-0.23	 & 	-0.07	  	\\
	\hbox{House of bricks/row dirt }	 & 	0.95	 & 	0.22	 & 	0.91	 & 	0.29	 & 	-0.15	 & 	0.26	 	\\
	\hbox{Water treatment}	 & 	0.85	 & 	0.35	 & 	0.78	 & 	0.41	 & 	-0.18	 & 	0.15	  	\\
	\hbox{Water supply}	 & 	0.76	 & 	0.43	 & 	0.63	 & 	0.48	 & 	-0.29	 & 	0.12	 	\\
	\hbox{Lighting}	 & 	0.91	 & 	0.29	 & 	0.79	 & 	0.41	 & 	-0.34	 & 	0.34	 	\\
	\hbox{Bathroom fixture}	 & 	0.47	 & 	0.50	 & 	0.62	 & 	0.49	 & 	0.31	 & 	-0.03	  	\\
	\hbox{Waste treatment}	 & 	0.81	 & 	0.39	 & 	0.63	 & 	0.48	 & 	-0.42	 & 	0.21	  	\\
	\hbox{Zero PC expenditure}	 & 	0.15	 & 	0.36	 & 	0.22	 & 	0.41	 & 	0.17	 & 	0.14	  	\\
	\hbox{Log PC expenditure}	 & 	3.83	 & 	1.75	 & 	2.89	 & 	1.67	 & 	-0.55	 & 	-0.05	  	\\
	\hbox{Other programs}	 & 	0.05	 & 	0.22	 & 	0.06	 & 	0.23	 & 	0.04	 & 	0.07 	\\
	\vspace{-0.3cm}\\
	\multicolumn{7}{l}{\hbox{\textit{Household head's characteristics}}}\\
	\vspace{-0.4cm}\\
	\hbox{Male}	 & 	0.85	 & 	0.35	 & 	0.87	 & 	0.34	 & 	0.03	 & 	-0.03	 	\\
	\hbox{Race: White}	 & 	0.86	 & 	0.34	 & 	0.88	 & 	0.32	 & 	0.06	 & 	-0.07 	\\
	\hbox{Primary/Middle Education}	 & 	0.39	 & 	0.49	 & 	0.47	 & 	0.50	 & 	0.17	 & 	0.02	 	\\
	\hbox{Occupation: Unemployed}	 & 	0.43	 & 	0.50	 & 	0.49	 & 	0.50	 & 	0.13	 & 	0.01	 	\\	 
	\vspace{-0.2cm}\\
	\hbox{Multivariate measure} & & & & &1.33\\
	\vspace{-0.3cm}\\
	\hline
\end{array}
$$
\end{table}
\newpage
\begin{table}[h]\caption{Bolsa Fam\'ilia study:	Weighted covariate  balance within the  sub-population $\Uset=\{i: 0.0 \leq S_i \leq 165.6\}$, selected using the MSE-optimal bandwidth approach  based on the triangular   kernel function  and the local-polynomial estimator  of order $p=2$  (Weights based on the triangular Kernel).} \label{A-tab:bal4a}
$$
\begin{array}{lrrrrrr}
	\hline 		\vspace{-0.35cm}\\
	&\multicolumn{2}{c}{\hbox{Ineligible}}& \multicolumn{2}{c}{\hbox{Eligible}}\\
	N_{\Uset}= 143\,099 &\multicolumn{2}{c}{(4\,879)}& \multicolumn{2}{c}{(138\,220)} &\hbox{Norm.} &  \\
	\vspace{-0.4cm}\\
	\hbox{Covariate}& \hbox{Mean} & \hbox{SD}  & \hbox{Mean} & \hbox{SD} & \hbox{Diff.} & \log(s_1/s_0) \\
	\vspace{-0.35cm}\\
	\hline
	\vspace{-0.35cm}\\
	\multicolumn{7}{l}{\hbox{\textit{Household structure}}}\\
	\vspace{-0.4cm}\\
	\hbox{Min age}	 & 	12.80	 & 	14.56	 & 	11.52	 & 	14.56	 & 	-0.09	 & 	0.00	  	\\
	\hbox{Mean age}	 & 	27.75	 & 	12.43	 & 	23.26	 & 	11.67	 & 	-0.37	 & 	-0.06 	\\
	\hbox{Household size}	 & 	2.88	 & 	0.96	 & 	2.94	 & 	1.28	 & 	0.05	 & 	0.28	 	\\
	\hbox{N. Children}	 & 	0.96	 & 	0.74	 & 	1.25	 & 	1.03	 & 	0.32	 & 	0.33	 	\\
	\hbox{N. Adults}	 & 	1.74	 & 	0.77	 & 	1.63	 & 	0.75	 & 	-0.14	 & 	-0.02 	\\
	\hbox{Children not at school}	 & 	0.02	 & 	0.15	 & 	0.04	 & 	0.19	 & 	0.08	 & 	0.24 	\\
	\hbox{Presence of weak people}	 & 	0.19	 & 	0.39	 & 	0.21	 & 	0.41	 & 	0.06	 & 	0.04	 \\
	\vspace{-0.3cm}\\
	\multicolumn{7}{l}{\hbox{\textit{Living and economic conditions}}}\\
	\vspace{-0.4cm}\\
	\hbox{Rural}	 & 	0.21	 & 	0.41	 & 	0.35	 & 	0.48	 & 	0.33	 & 	0.16 	\\
	\hbox{Apartment}	 & 	0.95	 & 	0.22	 & 	0.95	 & 	0.22	 & 	0.00	 & 	0.01	  	\\
	\hbox{Home ownership: Homeowner}	 & 	0.63	 & 	0.48	 & 	0.58	 & 	0.49	 & 	-0.10	 & 	0.02 	\\
	\hbox{No rooms pc}	 & 	1.70	 & 	0.97	 & 	1.66	 & 	1.08	 & 	-0.04	 & 	0.11 	\\
	\hbox{House of bricks/row dirt }	 & 	0.95	 & 	0.22	 & 	0.92	 & 	0.27	 & 	-0.12	 & 	0.21 	\\
	\hbox{Water treatment}	 & 	0.85	 & 	0.35	 & 	0.81	 & 	0.39	 & 	-0.11	 & 	0.10 	\\
	\hbox{Water supply}	 & 	0.76	 & 	0.43	 & 	0.66	 & 	0.47	 & 	-0.22	 & 	0.10 	\\
	\hbox{Lighting}	 & 	0.91	 & 	0.29	 & 	0.82	 & 	0.38	 & 	-0.26	 & 	0.29	 	\\
	\hbox{Bathroom fixture}	 & 	0.46	 & 	0.50	 & 	0.58	 & 	0.49	 & 	0.24	 & 	-0.01 	\\
	\hbox{Waste treatment}	 & 	0.82	 & 	0.38	 & 	0.68	 & 	0.47	 & 	-0.33	 & 	0.20	  	\\
	\hbox{Zero PC expenditure}	 & 	0.15	 & 	0.36	 & 	0.19	 & 	0.39	 & 	0.10	 & 	0.08	 	\\
	\hbox{Log PC expenditure}	 & 	3.81	 & 	1.76	 & 	3.17	 & 	1.67	 & 	-0.37	 & 	-0.05	  	\\
	\hbox{Other programs}	 & 	0.05	 & 	0.22	 & 	0.05	 & 	0.22	 & 	0.02	 & 	0.04	 	\\
	\vspace{-0.3cm}\\
	\multicolumn{7}{l}{\hbox{\textit{Household head's characteristics}}}\\
	\vspace{-0.4cm}\\
	\hbox{Male}	 & 	0.87	 & 	0.34	 & 	0.86	 & 	0.35	 & 	-0.03	 & 	0.03 	\\
	\hbox{Race: White}	 & 	0.87	 & 	0.34	 & 	0.88	 & 	0.32	 & 	0.04	 & 	-0.04	 	\\
	\hbox{Primary/Middle Education}	 & 	0.37	 & 	0.48	 & 	0.44	 & 	0.50	 & 	0.15	 & 	0.03 	\\
	\hbox{Occupation: Unemployed}	 & 	0.46	 & 	0.50	 & 	0.49	 & 	0.50	 & 	0.07	 & 	0.00	 	\\
	\vspace{-0.2cm}\\
	\hbox{Multivariate measure} & & & & &1.04\\
	\vspace{-0.3cm}\\
	\hline
\end{array}
$$		
\end{table}
\newpage

\clearpage
\section{Bayesian Causal Inference Conditional on an MSE-optimal Subpopulation: Computational Details} \label{sec:MSEUs0_BA}

\subsection{Likelihood Functions}

We can write the  observed likelihood function in terms of the observed data as follows:
\begin{eqnarray*}
\lefteqn{\mathscr{L}\left(\btheta\mid  \bX, \bS, \bZ, \bY; \Uset\right)=
	\mathscr{L}\left(\bgamma_0, \bgamma_1 \mid  \bX, \bS, \bZ, \bY; \Uset \right)\propto}\\&&
\prod\limits_{i\in \Uset: Z_i=0}   p(Y_i \mid \bX_i, i \in \Uset; \bgamma_0)\times
\prod\limits_{i\in\Uset: Z_i=1}   p(Y_i \mid \bX_i, i \in \Uset; \bgamma_1)
\end{eqnarray*}

In the Bolsa Fam\'ilia study, we have
$\bgamma_{0}=(\gamma_{0,0}, \bgamma_{X,0})$;
$\bgamma_{1}=(\gamma_{0,1}, \bgamma_{X,1})$;
and we impose $\bgamma_{X,0} =  \bgamma_{X,1} \equiv  \bgamma_{X}$. Therefore, the observed-data likelihood is
\begin{eqnarray*}
\lefteqn{\mathscr{L}\left(\btheta\mid  \bX, \bS, \bZ, \bY; \Uset\right)=
	\mathscr{L}\left(
	\gamma_0^{+}, \gamma_1^{+}, \bgamma_X,
	\mid  \bX, \bS, \bZ, \bY;\Uset \right)\propto}\\& &
\prod\limits_{i\in \Uset: Z_i=0}  \Phi\left(\gamma_{0,0}+ \bX_i'\bgamma_{X}\right)^{Y_i}
\left[1-\Phi\left(\gamma_{0,0}+ \bX_i'\bgamma_{X}\right)\right]^{1-Y_i}\times\\&&
\prod\limits_{i \in \Uset: Z_i=1}   \Phi\left(\gamma_{0,1}+ \bX_i'\bgamma_{X}\right)^{Y_i}
\left[1-\Phi\left(\gamma_{0,1}+ \bX_i'\bgamma_{X}\right)\right]^{1-Y_i}
\end{eqnarray*}

\subsection{Prior Distributions}

In the Bolsa Fam\'ilia study, we assume that parameters are a priori independent and use proper Normal prior distributions. 
Specifically, the prior distributions for the parameters of the outcome models are:  $\gamma_{0,z}  \sim N(0, \sigma^{2}_{\gamma_{0,z}})$, $z=0,1$, and      $\bgamma_X \sim N_{p}(\boldsymbol{0}, \sigma^{2}_{\bgamma_X} \mathbf{I}_{p})$,  
where $\sigma^{2}_{\bgamma^{-}}$,  $\sigma^{2}_{\bgamma^{+}}$  $\sigma^{2}_{\gamma_{0,z}}$, $z=0,1$, and  $\sigma^{2}_{\bgamma_X}$ are hyperparameters set at $1$.

\subsection{MCMC Algorithm for a selected subpopulation, $\Uset$, for the Bolsa Fam\'ilia study}

\begin{enumerate}
\item  Sample the coefficients $\gamma_{0,0}$ and $\gamma_{0,1}$  

\begin{enumerate}
	\item  Sample the latent variable $\bY^\ast$
	
	\begin{enumerate}
		\item  For $i \in \Uset$ with $Z_i=0$, sample the latent variable $Y_i^\ast$
		from $N(\gamma_{0,0} + \bX_i'\bgamma_X, 1)$ truncated to $[0, +\infty)$ if $Y_i=1$ and to $(-\infty, 0]$ if $Y_i = 0$;
		\item  For $i \in \Uset$ with $Z_i=1$, sample the latent variable $Y_i^\ast$
		from $N(\gamma_{0,1} + \bX_i'\bgamma_X, 1)$ truncated to $[0, +\infty)$ if $Y_i=1$ and to $(-\infty, 0]$ if $Y_i = 0$;
		
	\end{enumerate}
	\item    Sample the coefficient $\gamma_{0,0}$ from $N\left(\mu(\gamma_{0,0}), \sigma^2(\gamma_{0,0})\right)$ where
	$$
	\sigma^2(\gamma_{0,0}) =  \left[\dfrac{1}{\sigma^2_{\gamma_{0,0}}} + \sum_{i \in \Uset} (1-Z_i) \right]^{-1} $$
	and $$
	\mu(\gamma_{0,0})= \sigma^2(\gamma_{0,0}) \left[
	\dfrac{0}{\sigma^2_{\gamma_{0,0}}}+
	\sum_{i \in\Uset}(Y_i^{\ast}- \bX_i' \bgamma_X) (1-Z_i)\right]=   \sigma^2(\gamma_{0,0}) 
	\sum_{i \in\Uset}(Y_i^{\ast}- \bX_i' \bgamma_X) (1-Z_i)   
	$$ 
	
	\item    Sample the coefficient $\gamma_{0,1}$ from $N\left(\mu(\gamma_{0,1}), \sigma^2(\gamma_{0,1})\right)$ where
	$$
	\sigma^2(\gamma_{0,1}) =  \left[\dfrac{1}{\sigma^2_{\gamma_{0,1}}} + \sum_{i \in \Uset} Z_i \right]^{-1} $$
	and $$
	\mu(\gamma_{0,1})= \sigma^2(\gamma_{0,1}) \left[
	\dfrac{0}{\sigma^2_{\gamma_{0,1}}}+
	\sum_{i \in\Uset}(Y_i^{\ast}- \bX_i' \bgamma_X) Z_i\right]=   \sigma^2(\gamma_{0,1}) 
	\sum_{i \in\Uset}(Y_i^{\ast}- \bX_i' \bgamma_X) Z_i   
	$$ 
\end{enumerate}

\item  Sample the coefficients $\bgamma_X$  

\begin{enumerate}
	\item  Sample the latent variable $\bY^\ast$
	
	\begin{enumerate}
		\item  For $i \in \Uset$ with $Z_i=0$, sample the latent variable $Y_i^\ast$
		from $N(\gamma_{0,0} + \bX_i'\bgamma_X, 1)$ truncated to $[0, +\infty)$ if $Y_i=1$ and to $(-\infty, 0]$ if $Y_i = 0$;
		\item  For $i \in \Uset$ with $Z_i=1$, sample the latent variable $Y_i^\ast$
		from $N(\gamma_{0,1} + \bX_i'\bgamma_X, 1)$ truncated to $[0, +\infty)$ if $Y_i=1$ and to $(-\infty, 0]$ if $Y_i = 0$;
	\end{enumerate} 
	\item Let  $\widetilde{Y}^\ast$ be a $N_{\Uset}-$dimetional   vector with $N_{\Uset}=\sum_{i=1}^N \mathbb{I}\{i \in \Uset\}$ with $i$th element equal to
	$$\widetilde{Y}_i^\ast= \begin{cases}
		Y_i^\ast - \gamma_{0,0} & \hbox{if } i \in \Uset, Z_i=0\\
		Y_i^\ast - \gamma_{0,1} & \hbox{if } i \in \Uset, Z_i=1
	\end{cases} 
	$$
	and let $\bX_{\Uset}$ the $N_{\Uset} \times p$ sub-matrix of $\bX$ stacking information on the covariates for units in $\Uset$
	\item    Sample the coefficients $\bgamma_X$ from $N_{p}\left(\bmu(\bgamma_X), \Sigma(\bgamma_X)\right)$ where
	$$
	\Sigma(\bgamma_X) = \left[\dfrac{1}{\sigma^2_{\bgamma_X}}\mathbf{I}_{p} +  \bX_{\Uset}'\bX_{\Uset} \right]^{-1} $$
	and 
	$$
	\bmu(\bgamma_X)=\Sigma(\bgamma_X)  \left[
	\dfrac{1}{\sigma^2_{\bgamma_X}}\mathbf{I}_{p} \boldsymbol{0} +  \bX_{\Uset}'\widetilde{Y}^\ast \right]= \Sigma(\bgamma_X)   \bX_{\Uset}'\widetilde{Y}^\ast  
	$$ 
\end{enumerate}
\end{enumerate}
    
\subsection{Results on the causal risk difference}
Table~\ref{tab:baRDUset} shows summary statistics of the posterior distributions of the causal risk difference $RD_{\Uset}$ in each of the four  MSE-optimal subpopulations. 
Results are consistent with those we found for the causal relative risk $RD_{\Uset}$: The posterior medians of the causal risk difference are positive, but the 95\% posterior credible intervals always cover zero. The posterior probability that the causal risk difference is negative, that is, that eligibility for Bolsa Fam\'ilia benefits decreases leprosy rate, ranges between $18.4\%$ and $26.0\%$.  

\begin{table*}[t]	 
\caption{Bolsa Fam\'ilia study: 
Summary statistics of the posterior distributions of the finite sample causal risk difference, $RD_{\Uset}$, for specific MSE-optimal subpopulations.}\label{tab:baRDUset}
$$
 \hspace*{-1cm}
\begin{array}{lc ccccc}
\hline
\vspace{-0.35cm}\\
&&&\multicolumn{3}{c}{95\% \hbox{ HDI}}  \\ 
\cline{4-6}
\vspace{-0.35cm}\\
\hbox{Kernel function } (p) & \hbox{Subpopulation: }  \Uset  & \hbox{Median}  & \hbox{LB} & \hbox{UB}& \hbox{(width)} &Pr\left(RD_{\Uset}  < 0\right)\\
\vspace{-0.35cm}\\
\hline
\vspace{-0.35cm}\\
\hbox{Uniform kernel } (p=1) & \{i \in \mathcal{U}:  43.5 \leq S_i\leq 163.9\}  & 0.0005 & -0.0009 & 0.0017& (0.0026)
& 0.256\\ 
\vspace{-0.15cm}\\
\hbox{Triangular kernel } (p=1) & \{i \in \mathcal{U}:  39.0 \leq S_i\leq 174.5\} &  0.0005 & -0.0009 & 0.0016& (0.0025) & 0.260\\ 
\vspace{-0.15cm}\\
\hbox{Uniform kernel } (p=2) & \{i \in \mathcal{U}: \ \ 0.0 \leq S_i\leq 166.0\}&0.0006& -0.0007 & 0.0017& (0.0024) & 0.184\\
\vspace{-0.15cm}\\
\hbox{Triangular kernel } (p=2) & \{i \in \mathcal{U}: \ \ 0.0 \leq S_i\leq 165.6\}&0.0006& -0.0007 & 0.0017& (0.0024) & 0.194\\
\vspace{-0.35cm}\\
\hline
\end{array}
 \hspace*{-1cm}
$$
\end{table*}

\section{Results based on the Continuity Assumption}\label{sec:sa}
Under the assumption that the conditional regression or distribution functions of the potential outcomes given the forcing variable, $Y_i(0) \mid S_i=s$ and $Y_i(1) \mid S_i=s$
are continuous in $s$, causal effects at the threshold can be estimated using
local polynomial Regression Discontinuity point estimators \cite[e.g.][]{ImbensLemieux2008, LeeLemieux2010, CalonicoCattaneoTitiunik2014, CalonicoCattaneoFarrell2018, CalonicoCattaneoTitiunik2014, CalonicoEtAl:2019}.

We use uniform and triangular kernel functions to construct the local-polynomial estimator with order of the local-polynomial equal to 1 and 2 and two-sided Mean Square Error-optimal bandwidth selectors  (below and above the cutoff)  for the RD treatment effect estimator. 

The causal estimand which the approach based on continuity assumptions usually focuses on is the causal risk difference at the threshold. In the Bolsa Fam\'ilia study, the causal estimand of primary interest is the causal relative risk. We show results for both causal estimands. Estimates of the causal relative risk at the threshold and the 95\% confidence intervals are obtained as follows. Let $\widehat{\Pr}(Y_i(0)=1\mid S_i=s_0)$ and $\widehat{\Pr}(Y_i(1)=1\mid S_i=s_0)$ denote the local polynomial RD point estimators of $\Pr(Y_i(0)=1\mid S_i=s_0)$ and $\Pr(Y_i(1)=1\mid S_i=s_0)$ and let $\widehat{\hbox{var}}(\widehat{\Pr}(Y_i(0)=1\mid S_i=s_0))$ and $\widehat{\hbox{var}}(\widehat{\Pr}(Y_i(1)=1\mid S_i=s_0))$ denote the local polynomial RD point estimators of their variance. Then, 
$$
\widehat{RR}_{s_0}= \dfrac{\widehat{\Pr}(Y_i(1)=1\mid S_i=s_0)}{\widehat{\Pr}(Y_i(0)=1\mid S_i=s_0)}
$$
and the $(1-\alpha)\%$ CI for $R_{s_0}$ is given by
\begin{align*}
\Bigg(&\exp\left\{ \log(\widehat{RR}_{s_0})  - z_{1-\alpha/2} \sqrt{\widehat{\hbox{var}}( \log(\widehat{RR}_{s_0}) )}\right\}; \\
&\exp\left\{ \log(\widehat{RR}_{s_0}) + z_{1-\alpha/2} \sqrt{\widehat{\hbox{var}}( \log(\widehat{RR}_{s_0}) )}\right\}\Bigg)
\end{align*}
where, by the delta method, 
$$
\widehat{\hbox{var}}( \log(\widehat{RR}_{s_0}) ) = 
\dfrac{\widehat{\hbox{var}}(\widehat{\Pr}(Y_i(1)=1\mid S_i=s_0))}{\widehat{\Pr}(Y_i(1)=1\mid S_i=s_0)^2} + 
\dfrac{\widehat{\hbox{var}}(\widehat{\Pr}(Y_i(1)=1\mid S_i=s_0))}{\widehat{\Pr}(Y_i(1)=1\mid S_i=s_0)^2}
$$ 

Table~\ref{StandardApproach} shows the results. The point estimates of the causal relative risk and the causal risk difference suggest that eligibility for the Bolsa fam\'ilia benefits reduces leprosy incidence, but results are not statistically significant, with the 95\% confidence intervals covering null effect values. 

\begin{table}[t]\caption{Bolsa Fam\'ilia study: Local polynomial RD point estimators of the causal risk difference at the threshold, $RD_{s_0}$, and the causal relative risk at the threshold, $RR_{s_0}$.} \label{StandardApproach}
 $$	
 \hspace*{-1cm}
 \begin{array}{lcc c ccc c ccc}
 	\hline
\hbox{Kernel ($p$)} & h_r & h_l  & & RR_{s_0}&  \multicolumn{2}{c}{\hbox{95\% CI for }RR_{s_0}} & & ATE_{s_0} & \multicolumn{2}{c}{\hbox{95\% CI for }ATE_{s_0}} \\ 
 	\cline{1-3} \cline{5-7} \cline{9-11}	\vspace{-0.2cm}\\
\hbox{Uniform Kernel $(p=1)$} & 43.9&  76.5 &&
0.78402  &  (0.33548; & 1.83228) & & -0.00077 &   (-0.00366; & 0.00211)\\ 
	\vspace{-0.2cm}\\
\hbox{Triangular Kernel $(p=1)$} & 54.5 & 81.0 && 
0.86901 & (0.37025;  & 2.03964) & & -0.00047 &  (-0.00344; & 0.00250)\\  
 	\vspace{-0.2cm}\\
\hbox{Uniform Kernel $(p=2)$} &46.0&120.0&&
0.65865 & (0.20849; & 2.08077) & &-0.00147; &  (-0.00610; & 0.00317) \\
 	\vspace{-0.2cm}\\
\hbox{Triangular Kernel $(p=2)$}& 45.6& 120.0&&
0.78989 & (0.15574; & 4.00615) && -0.00087 &   (-0.00749; & 0.00575)\\ 
 	\hline
	\multicolumn{11}{l}{\hbox{\footnotesize{$p=$ Order of the local polynomial; $h_l=$ Left bandwidth ($Z_i=1$); $h_r=$ Right bandwidth ($Z_i=1$)}}}
 \end{array}
  \hspace*{-1cm}
 $$
\end{table}

 \end{appendices}

\bibliographystyle{apalike}
\bibliography{RDDUSelection} 
\end{document}